\documentclass[superscriptaddress,groupedaddress,nofootnoteinbib,11pt]{article}
\pdfoutput=1 
\usepackage{jheppub}

\usepackage{amsmath, amsfonts, amsthm, amssymb, graphicx, xcolor, hyperref, mathrsfs}
\usepackage{subfigure}
\usepackage{caption} %

\usepackage[utf8]{inputenc}

\allowdisplaybreaks[3]

\def\({\left(}
\def\){\right)}
\def\[{\left[}
\def\]{\right]}
\def\d{\mathrm{d}}
\newcommand{\n} {\nabla}

\newcommand{\f}[2]{\frac{#1}{#2}}
\def \bal#1\eal  {\begin{align} #1 \end{align}}
\newcommand{\eref}[1]{{Eq.~(\ref{#1})}}
\newcommand{\be} {\begin{equation}}
\newcommand{\ee} {\end{equation}}
\newcommand{\bc}{\begin{center}}
\newcommand{\ec}{\end{center}}
\newcommand{\bim} {\begin{itemize}[noitemsep]}

\newcommand{\eim} {\end{itemize}}

\newcommand{\ud} {\mathrm{d}}

\newcommand{\pd} {\partial}
\newcommand{\mc} {\mathcal}

   \newcommand{\bfk} {{\bf k}}

      \newcommand{\bfx} {{\bf x}}

\newcommand{\ai}{{\alpha}}

\newcommand{\ri}{{\rho}}
\newcommand{\si}{{\sigma}}

\newcommand{\oi}{\omega}

\newcommand{\Li}{\Lambda}

\title{Charge-Swapping Q-balls and Their Lifetimes}
\author[a,b]{Qi-Xin Xie}
\author[c]{, Paul M. Saffin}
\author[a,b]{and Shuang-Yong Zhou}

\affiliation[a]{Interdisciplinary Center for Theoretical Study, University of
Science and Technology of China, Hefei, Anhui 230026, China}
\affiliation[b]{Peng Huanwu Center for Fundamental Theory, Hefei, Anhui 230026, China}
\affiliation[c]{School of Physics and Astronomy, University Park, University of Nottingham,\\ Nottingham NG7 2RD, United Kingdom}

\emailAdd{xqx2018@mail.ustc.edu.cn}
\emailAdd{paul.saffin@nottingham.ac.uk}
\emailAdd{zhoushy@ustc.edu.cn}
\preprint{\small USTC-ICTS/PCFT-21-05}

\date{\today}

\abstract{
For scalar theories accommodating spherically symmetric Q-balls, there are also towers of quasi-stable composite Q-balls, called charge swapping Q-balls (CSQs). We investigate the properties, particularly the lifetimes, of these long-lived CSQs in 2+1D and 3+1D using numerical simulations with efficient second order absorbing boundary conditions. We find that the evolution of a CSQ typically consists of 4 distinct stages: initial relaxation, first plateau (CSQ stage), fast decay and second plateau (oscillon stage). We chart the lifetimes of CSQs for different parameters of the initial conditions and of the potential, and show the attractor behavior and other properties of the CSQs. 
}

\begin{document}

\maketitle
\flushbottom

\section{Introduction and summary}

Non-perturbative configurations, such as solitons, together with their fascinating properties are integral parts of understanding field theories. Apart from topological defects, which are stable due to the presence of topological charges, there also exist non-topological solitons, such as Q-balls \cite{Friedberg:1976me, Coleman:1985ki}. Q-balls are spatially localized and stationary but non-static, and are stable due to the presence of Noether charges. For example, Q-balls exist in U(1) symmetric scalar field theories with a potential that grows slower than the quadratic term away from its minimum. The shallow potential creates some sort of ``attractive forces between particles'' in the theory, so particles prefer to condense to form a localized lump rather than dissipate to infinity. In other words, the Q-ball condensate is the energetically preferred state for such a system. The properties and dynamics of Q-balls have been extensively studied 
\cite{Safian:1987pr, Kusenko:1997ad, Laine:1998rg, Axenides:1999hs, Multamaki:1999an, Battye:2000qj, Paccetti:2001uh, Volkov:2002aj,  Gleiser:2005iq, Campanelli:2007um,  Sakai:2007ft, Bowcock:2008dn,  Tsumagari:2008bv,  Copeland:2009as, Mai:2012yc, Gulamov:2015fya,  Bazeia:2015gkq, Bazeia:2016wco, Levkov:2017paj, Smolyakov:2017axd, Loiko:2018mhb, Hasegawa:2019bbo}. Crucially, they may play an important role in the early universe (see, e.g., 
\cite{Frieman:1988ut, Enqvist:1997si, Kusenko:1997zq, Kasuya:1999wu, Enqvist:2000cq, Fujii:2001xp, Postma:2001ea, Fujii:2002kr, Kawasaki:2002hq,  Allahverdi:2002vy,  Palti:2004is, Berkooz:2005rn, Pearce:2012jp, Krylov:2013qe, Zhou:2015yfa, Hasegawa:2018yuy, Cotner:2019ykd}) and can be  candidates of dark matter (see, e.g.,~\cite{Kusenko:1997si,Enqvist:1998ds,Kasuya:2000sc,Banerjee:2000mb,Kusenko:2004yw,Roszkowski:2006kw,Kasuya:2011ix,Kasuya:2014ofa, Kawasaki:2019ywz}). They can also be prepared in cold atom systems \cite{Enqvist:2003zb, Bunkov:2007fe}.

The most stable form of Q-balls are spherically symmetric, as other forms of spatial configurations increase the gradient energy. Recently, non-spherically symmetric, composite Q-balls have been identified \cite{Copeland:2014qra}. In the theories where the spherically symmetric Q-balls exist, there are also a tower of composite Q-balls with different multipoles, within which both positive and negative charges co-exist and swap with time (see Fig.~\ref{fig:evolution in one period} for a dipole CSQ and see Fig.~\ref{fig:four_and_eight_Q_balls} for higher multipole CSQs). They are thus dubbed charge-swapping Q-balls (CSQs) \cite{Copeland:2014qra}. Nevertheless, their energy densities remain mostly spherically symmetric (see Figs.~\ref{fig:evolution in one period_ham} and \ref{fig:four_and_eight_Q_balls}). They have been shown to exist in 2+1D and 3+1D, and can be prepared by simply placing positive and negative charge elementary Q-balls (or in general simply lumps) tightly together such that their nonlinear cores overlap, and then the configurations, after initially emitting a burst of radiation, can quickly relax to CSQs. 

As expected, CSQs are only quasi-stable and will ultimately decay. In this sense, although much more complex, they are spiritually like oscillons \cite{Bogolyubsky:1976yu, Copeland:1995fq} which are spatially localized and quasi-stationary and exist even in real scalar field theories. (Indeed, CSQs decay to oscillons, as will see in this paper.) Oscillons do not contain any Noether charges but nonetheless live for an extended period of time (see, e.g.,~\cite{Honda:2001xg, Adib:2002ff, Fodor:2006zs, Saffin:2006yk, Farhi:2007wj, Gleiser:2008ty, Fodor:2009kf, Gleiser:2009ys, Amin:2010dc, Amin:2011hj, Salmi:2012ta, Amin:2013ika, Zhou:2013tsa, Mukaida:2016hwd, Lozanov:2017hjm, Gleiser:2018kbq, Amin:2018xfe, Ibe:2019vyo, Kou:2019bbc, Zhang:2020bec} for more details). While the existence of CSQs has been identified in \cite{Copeland:2014qra}, their detailed properties have not been sufficiently explored. In particular, Ref.~\cite{Copeland:2014qra} makes use of periodic boundary conditions, which are not suitable for extracting the lifetimes of the CSQs.

In this paper, we shall investigate the properties and evolution of the CSQs in more detail, and determine their lifetimes. We will focus on dipole CSQs in 2+1D and 3+1D in the simplest $\phi^6$ potential, with quadrupole and octupole CSQs also briefly touched on. To be able to determine the lifetimes of CSQs, we shall utilize absorbing boundary conditions (ABCs). This is crucial as our method of obtaining the CSQs simply involves superimposing elementary Q-balls, which is, of course, not the configuration of a CSQ. Thus, in the initial relaxing phase, the configuration radiates a substantial amount of energy (see, e.g., Fig.~\ref{fig:energy_vs_period}). With the previous periodic boundary conditions, this radiation would travel back and echo around the periodic space, continuously perturbing the CSQ. So, while the periodic boundary conditions of \cite{Copeland:2014qra} are sufficient to determine the existence of CSQs,  the lifetimes of the CSQs can not be reliably determined in that setup. Therefore, for an accurate determination of the lifetimes of CSQs, it is essential to employ effective ABCs. 

We will survey the effectiveness of a few ABCs for our particular problem: Sommerfeld’s ABCs \cite{alcubierre2008introduction}, Engquist-Majda’s ABCs \cite{engquist1977absorbing} and Hidgon's ABCs \cite{higdon1994radiation, higdon1986absorbing}. Sommerfeld’s ABCs are first order, designed to absorb spherically symmetric radiation, while Engquist-Majda’s ABCs and Hidgon's ABCs can be implemented at higher orders. For our applications, we find that the second order ABCs are sufficient to determine the lifetimes within a couple of percent, and Hidgon's second order ABCs with a judicial choice of the $c_1$ and $c_2$ generally give the best absorbing effects, which are used to produce most of the results in the paper. Another way to damp radiation in the far field regions would be to add a Kreiss-Oliger term. However, in our simulations, we do not find the Kreiss-Oliger term significantly increases the accuracy, and therefore this artificial term is not adopted for the results presented in the paper.

We find that the evolution of a CSQ can be divided into 4 distinct stages (see Figs.~\ref{fig:energy_vs_period} and \ref{fig:charge_vs_period}): {(1)} Initial relaxation, {(2)} First plateau (CSQ stage), {(3)} Fast decay and {(4)} Second plateau (oscillon stage). The presence of the initial relaxation stage is, as mentioned, because the CSQ is not precisely prepared. Indeed, we find that the CSQ is an attractor solution, as a quasi-stable configuration should be, and can be formed with relatively general initial configurations (see Figs.~\ref{ec_same_value}, \ref{fig:other_phenomena_of_charge} and \ref{fig:deformed_EQ}).  After settling down, the CSQ stage is characterized by a mostly spherical and slightly oscillating energy density profile (see Fig.~\ref{fig:evolution in one period_ham}) with charges swapping with time (see Fig.~\ref{fig:evolution in one period}). The total energy and charge decrease slowly with time, and the swapping period remains mostly constant (see Fig.~\ref{fig:freq_vs_period_zeros}), which is a few times the oscillating period of an elementary Q-ball. We find that long-lived dipole CSQs in 2+1D can be achieved in a diagonal strip of the parameter space of the initial Q-ball frequency $\oi$ and the initial separation between them $d$ (see Fig.~\ref{parameter space}). We also find the the lifetime of the dipole CSQ has an exponential dependence on the $\phi^6$ coupling $g$ (see Fig.~\ref{fig:gLifetimeAndFitting}). The CSQ stage is followed by a short stage of fast decay of both energy and charges, the end result of which is, interestingly, an oscillon with roughly half of the total energy (and with very small charge densities). In this oscillon stage, the charge of the Q-ball components comprising the CSQ, $Q_s$, decays exponentially. Although within the time limits of our simulations we have not seen the decay of the second oscillon plateau, the oscillons will ultimately decay as they are not absolutely stable.

For the higher multipole CSQs in 2+1D, their lifetimes are shorter than those of the dipoles but nevertheless remain at the same order, while their total energies are about twice those of the dipoles.  On the other hand, the charges of the Q-ball components, $Q_s$, decays faster than that of the dipoles, unless the coupling $g$ is tuned to be a smaller value (see Fig.~\ref{fig:plateaus_of_4_CSQ}). Also, for high multipoles, it appears that the lifetimes converge to the same value for different $g$ (see Fig.~\ref{fig:g_and_complex_CSQ}). In 3+1D, the lifetimes of the CSQs are much shorter (see Figs.~\ref{fig:swapCSQ_3D} and \ref{fig:EandQ_3D}), presumably because there are more possible decay modes with three spatial directions. However, this is also largely to do with the potential of the theory. For example, for the logarithmic potential, which is the fiducial example of \cite{Copeland:2014qra}, the 3+1D CSQs are also long-lived, even for the higher multipoles \cite{HSXZ}. 

The paper is organized as follows.  In Section \ref{sec:model}, we first introduce the fiducial field model we consider in this paper, define a few quantities that will be used later, and review elementary Q-balls and CSQs; then we specify the numerical implementations and introduce the Higdon's ABCs that are used to produce the results in the paper; additional ABCs, which are used to cross-check some of the results, are introduced in the Appendix \ref{sec:otherABCs}. In Section \ref{sec:csq2D}, we investigate the evolution histories, lifetimes and attractor behavior of  the dipole CSQs in 2+1D; the different stages of a CSQ evolution are detailed, and various properties of the dipole CSQs are explored, in particular the lifetimes of the dipole CSQs are surveyed for different parameters $\oi$, $d$ and $g$. In Section \ref{sec:multipleCSQ}, we briefly study higher multipole CSQs in 2+1D. In Section \ref{sec:csq_in_3d}, we briefly study CSQs in 3+1D.

\section{Model and setup}
\label{sec:model}

The existence of Q-balls in a complex scalar field theory requires a potential with an attractive nature that condenses field perturbations (or, loosely speaking, ``particles'' in the language of quantum field theory), rather than dissipates them.
The simplest $\mathbb{Z}_2$ symmetric ($\varphi\to -\varphi$), polynomial potential that supports Q-balls is given by
\be
S = \int\ud^{d+1}  \tilde x \Big[-\Big|\frac{\pd\varphi}{\pd\tilde x^\mu}\Big|^2-V(|\varphi|) \Big],~~{\rm with} ~~ V(|\varphi|) = m^2|\varphi|^2 -\lambda |\varphi|^4 + \tilde g |\varphi|^6 ,
\label{Lagrangian0}
\ee
where $d$ is the number of the spatial dimensions. Defining dimensionless variables $ x^\mu = m \tilde x^\mu$, $\phi=\lambda^{1/2}\varphi/m$ and $g=\tilde g m^2/\lambda^2$,  the action can be re-written as
\be
S = \lambda^{-1}m^{3-d}\int\ud^{d+1}   x \Big[-|\pd_{\mu}\phi|^2-V(|\phi|) \Big], ~~{\rm with}~~V(|\phi|) = |\phi|^2 - |\phi|^4 + g |\phi|^6 ,
\label{Lagrangian}
\ee
which only contains one dimensionless free parameter $g$. In other words, we are expressing the coordinates in the units of the particle mass, $m^{-1}$, and the field value in the units of $m\lambda^{-1/2}$. Unless otherwise stated, in the following, we will consider the potential with $g=1/2$ as a fiducial model,
\be
\label{Veffphi}
V(|\phi|) = |\phi|^2 - |\phi|^4 + \frac12 |\phi|^6.
\ee
The energy density and conserved energy for a spatial volume are given respectively by
\be 
{\cal H}  = |\dot{\phi}|^2+|\n{\phi}|^2+V \label{hamiltonian density},~~~~~~
E = \int \ud^d x {\cal H}  .
\ee
Since the action is invariant under a global symmetry $\phi \rightarrow e^{i\ai} \phi$ ($\ai$ being constant), by Noether's theorem, there is a conserved current and thus a conserved charge for this symmetry. The associated charge density and conserved charge are given respectively by
\be
\ri = i(\phi \dot{\phi}^* - \phi^* \dot{\phi})= -2{\rm Im}(\phi \dot{\phi}^*)  = 2(\phi_1\dot\phi_2-\dot\phi_1\phi_2) ,
 \label{charge density}
~~~~~~~~
Q = \int \ud^d x \rho  ,
\ee
where we have defined
\be
\phi=\phi_1+i \phi_2  .
\ee
Numerically, we are simulating the system in a large but finite box, so the total energy $E$ and total charge $Q$ will refer to the total energy and charge in the simulation box.

Note that for this potential the $|\phi|^4$ term is negative, which provides the attractive force mentioned above, and thus Q-balls can form. Technically, this means that $V/|\phi|^2$ has a minimum away from $|\phi|=0$. Since $V/|\phi|^2$ at $|\phi|=0$ gives the mass of the particle in the free theory, $V/|\phi|^2$ having a minimum at $|\phi|\neq 0$ means in the interacting theory there are condensate configurations where the mass of the particle is smaller than that of the free theory. The minimum of these configurations, {\it i.e.}, elementary Q-balls, are the minima of the energy functional \cite{Coleman:1985ki}. That is, because of the leading interacting potential being shallower than the free quadratic potential, particles tend to condense rather than propagate away from each other.

\begin{figure}[tbp]
\centering
\includegraphics[width=.45\textwidth,trim=90 260 110 280,clip]{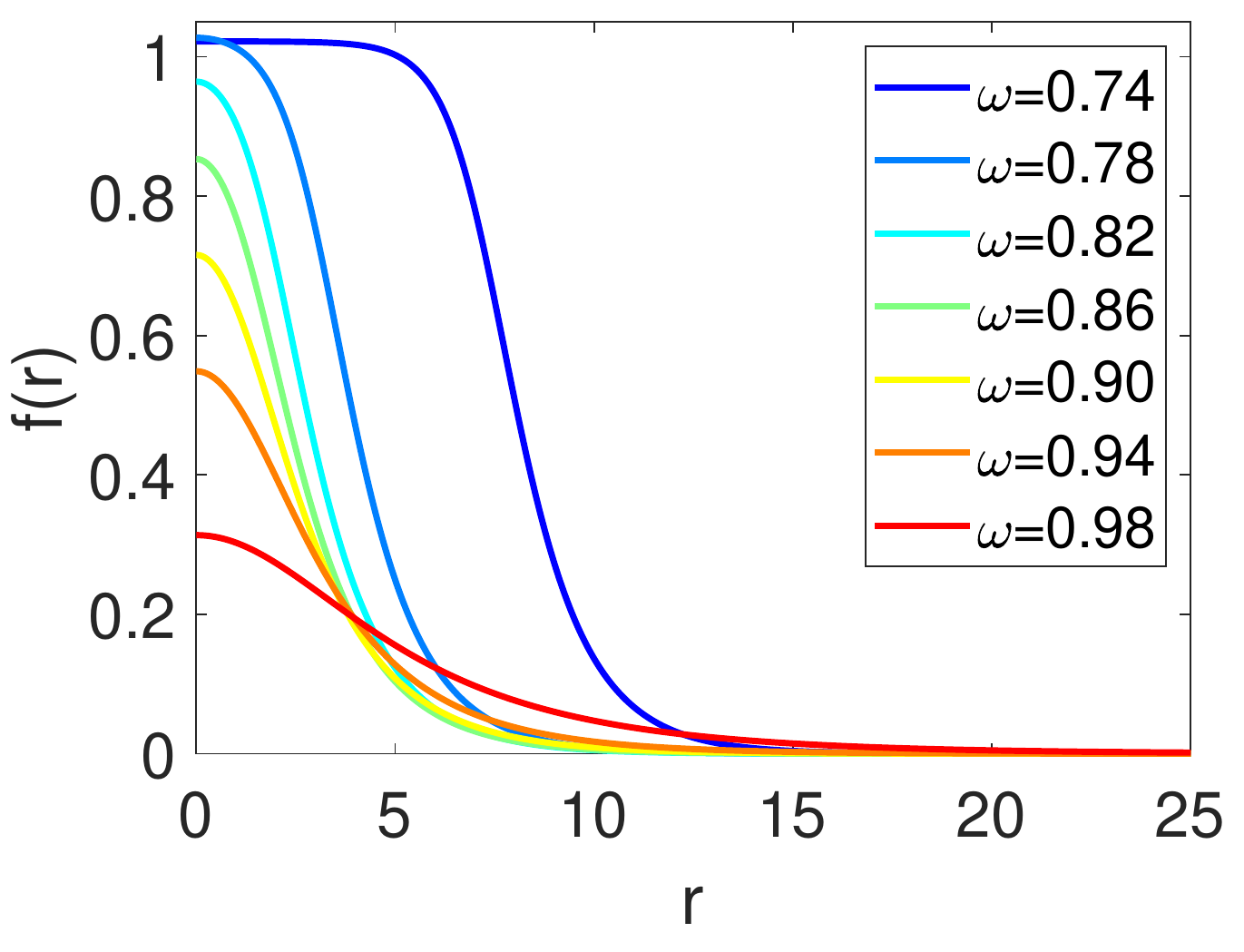}
\hfill
\includegraphics[width=.45\textwidth,trim=90 260 100 280,clip]{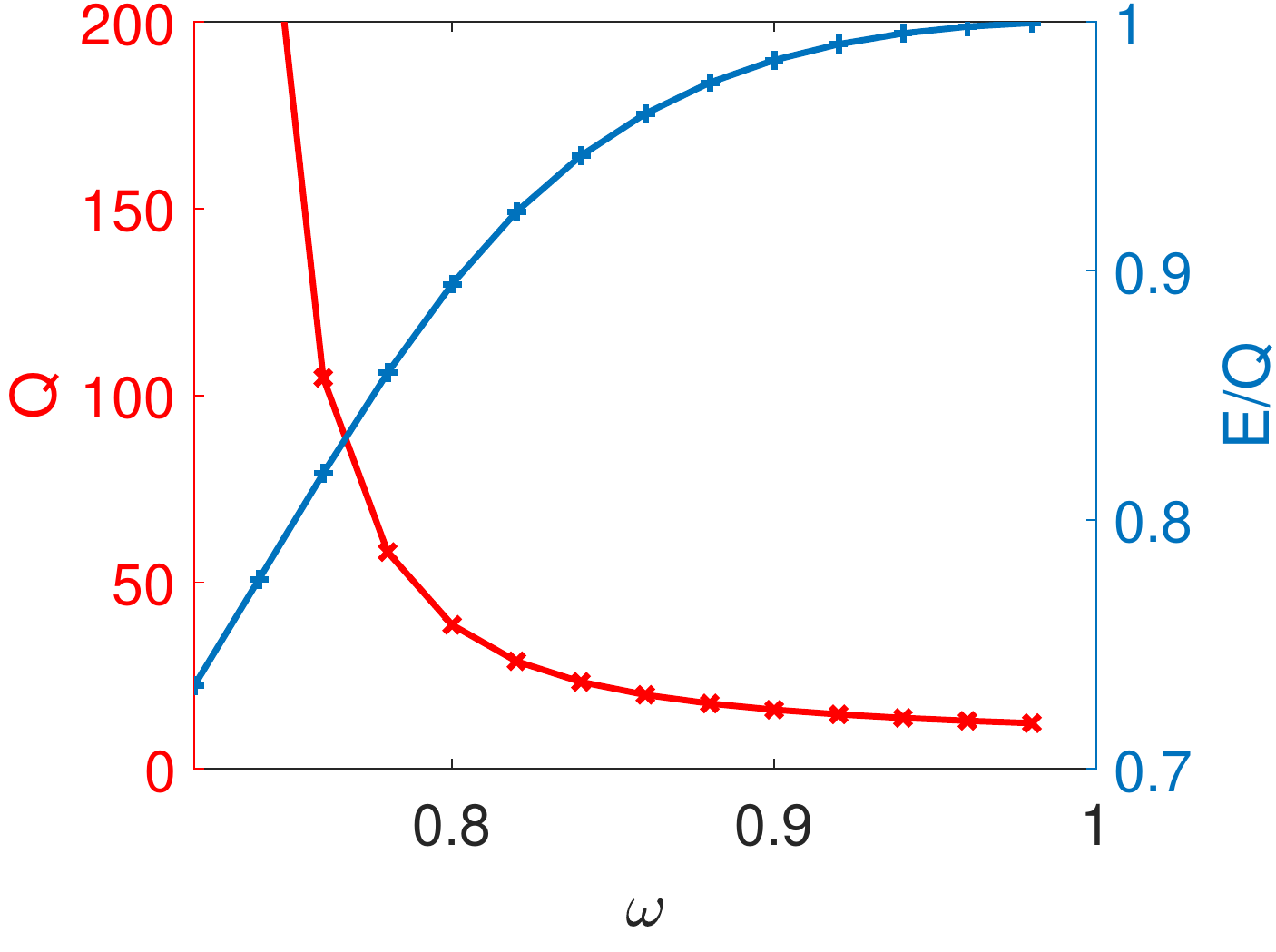}
\caption{Profiles (left plot), total charges and ratios of total charge to total energy (right plot) of elementary Q-balls with different frequencies for potential (\ref{Veffphi}). The fact that $E/Q<1$ ensures the stability of elementary Q-balls. Dimensionful quantities are in units of the scalar mass $m$.}
\label{fig:profile charge vs freq}
\end{figure}

Elementary Q-balls are stable stationary solutions that are spherically symmetric. To obtain the radial profile of an elementary Q-ball, we take the following ansatz
\be
\phi(t,r) = f(r)e^{i \oi t}   ,
\label{stationary ansatz}
\ee
where $\oi$ is the Q-ball frequency or the angular velocity in the $\phi$ field space. We shall call it an anti-Q-ball if $\oi$ is negative.  Substituting this ansatz to the equation of motion for $\phi^*$, we get an equation for $f(r)$
\be
\f{d^2 f}{dr^2} = \f{1-d}{r}\f{df}{dr}-\oi^2f+\f{1}{2} \f{\pd V}{\pd f} , \label{profile equ}  
\ee
which is subject to the boundary conditions ${df(0)}/{dr}=0$ and $f(\infty)=0$. Viewing $r$ as ``time'', \eref{profile equ} along with its boundary conditions can be viewed as a problem where a point mass, initially at rest  ${df(0)}/{dr}=0$, moves with a time-dependent friction term and in the effective potential $V_{\rm eff}=\f{1}{2}\oi^2 f^2 -\f{1}{2} V$, and eventually comes to stop in the infinite future at $f(\infty)=0$. For an appropriate $\oi$, we can find a unique solution that interpolates between $f(0)\neq 0$ and $f(\infty)=0$ without oscillations. If $\oi^2$ is greater than the perturbative mass of the free theory, $f=0$ is a local minimum of $V_{\rm eff}$, such a non-oscillating solution is impossible. To get an elementary Q-ball solution, the upper bound of the frequency is given by $\oi_+^2 =\f{1}{2} {\ud^2 V}/{\ud f^2}|_{f=0}={V}/{f^2}|_{f=0}  = 1$. If $\oi^2$ is smaller than the minimum of $V/f^2$ at some $f_0\neq 0$, $V_{\rm eff}$ does not have a maximum away from $f=0$ that is greater than $V_{\rm eff}(f=0)$, such a Q-ball solution is again impossible, so the lower bound of the frequency is $\oi_-^2= {V}/{f^2} |_{f=f_0}$. For our fiducial potential, we have $\oi_-=1/\sqrt{2}$.  Solving \eref{profile equ} numerically by the shooting method, we can get the Q-ball profile and total charge for the corresponding $\oi$; see Fig.~\ref{fig:profile charge vs freq}. We see that the total charge $Q$ is larger for smaller $\oi$ and the peak of $f(r)$ is also higher except when $\oi$ is close to $\oi_-$. Also, for an elementary Q-ball, the ratio between the total energy and the total charge $E/Q$ is, as expected, smaller than 1, meaning that the Q-ball will not decay into free particles. 

Apart from the elementary Q-ball solutions, the theory (\ref{Lagrangian}) also admits nonlinear, quasi-stationary, real solutions called oscillons \cite{Bogolyubsky:1976yu, Copeland:1995fq}, which take the form 
\be
\phi = g(t,r) ,
\ee
where the imaginary part of the field vanishes. As a very crude approximation, $g(t,r)$ goes like 
\be
g(t,r)\sim g_0(r)\cos  \oi t .
\ee
These are also localized lumps, very similar to the elementary Q-ball. Oscillons can be supported by merely a real scalar field theory. For a complex field, if the initial configuration is real, the field remains real through out its time evolution. Indeed, for oscillons to exist, we also need a potential that supports attractive forces, and thus a theory that supports Q-balls also has oscillon solutions. But different from the elementary Q-balls, without the protection of any exact symmetry, oscillons are only quasi-stable solutions, although in 2+1D their life time can be long \cite{Salmi:2012ta}. 

\subsection{Charge-swapping Q-balls (CSQs)}

\label{sec:CSQs}

Recently, it has been observed that, apart from the elementary Q-balls, which have been proven to be stable \cite{Coleman:1985ki}, there are also composite, quasi-stable solutions in the theory where elementary Q-balls exist \cite{Copeland:2014qra}. These composite Q-balls are not spherically symmetric. They can form when Q-balls and anti-Q-balls are placed closely together with their cores overlapping. Remarkably, the positive and negative charges of the composite Q-balls swap with time, and thus they are dubbed charge-swapping Q-balls (CSQs).

\begin{figure}[tbp]
\centering
\includegraphics[width=1\textwidth,trim=40 170 0 140,clip]{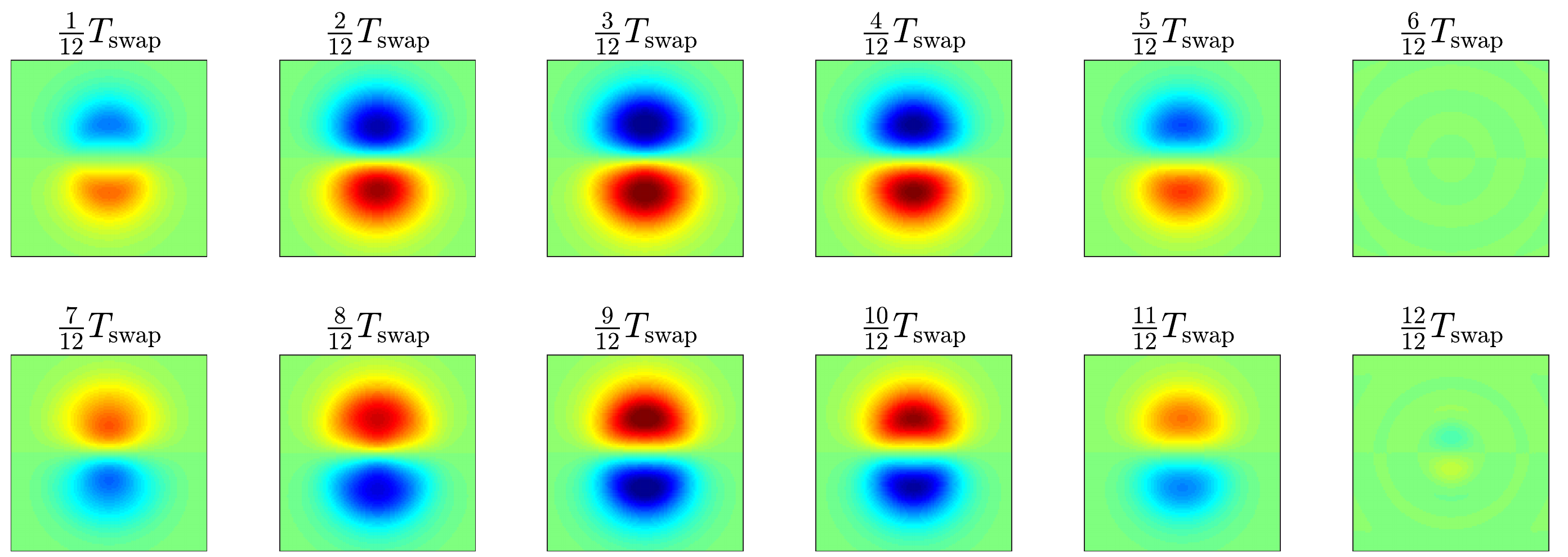}
\caption{\label{fig:evolution in one period}Evolution sequence of the charge density of the dipole CSQ in one charge-swapping period $T_{\rm swap}$. The red color depicts positive charge density and the blue color depicts negative charge density. $T_{\rm swap}$ is usually a few times the oscillation period of the field, which is roughly $T_0\equiv 2\pi/m=2\pi$.}
\end{figure}

\begin{figure}[tbp]
\centering
\includegraphics[width=1\textwidth,trim=40 170 0 140,clip]{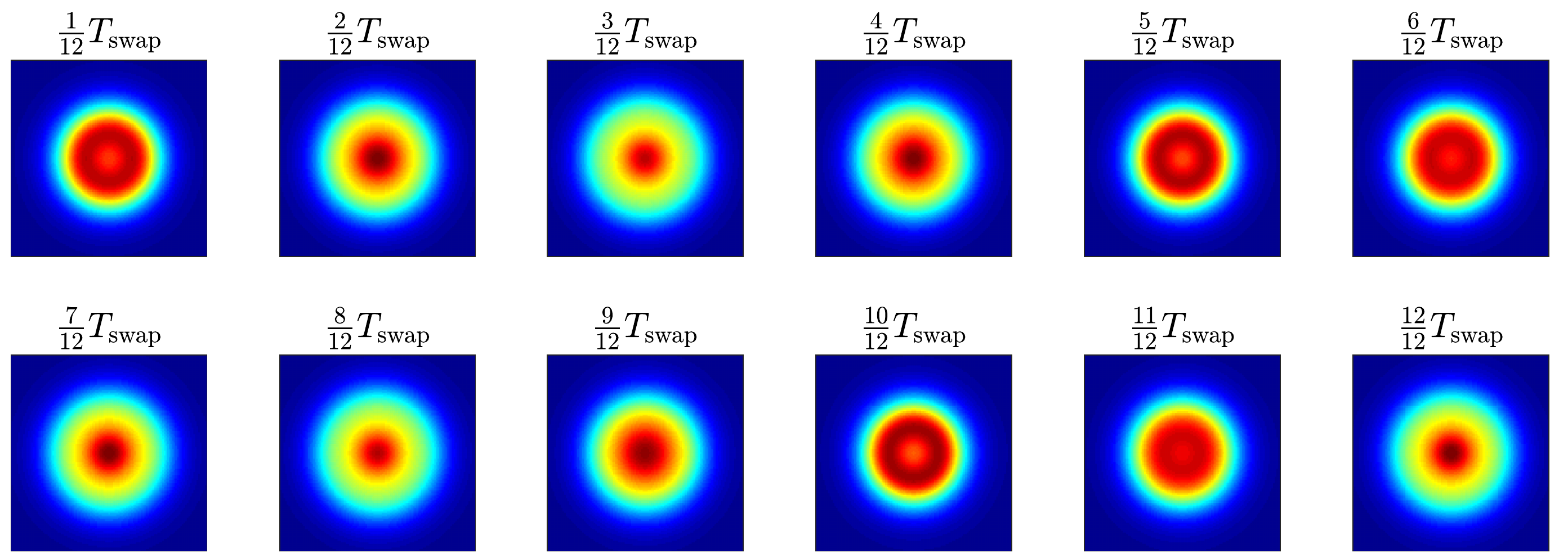}
\caption{\label{fig:evolution in one period_ham}Evolution sequence of the energy density of the dipole CSQ in the same charge-swapping period $T_{\rm swap}$ as Fig.~\ref{fig:evolution in one period}, with the blue depicting lower density and the red depicting higher density.}
\end{figure}

\begin{figure}[tbp]
\centering
\includegraphics[width=.5\textwidth,trim=90 260 110 280,clip]{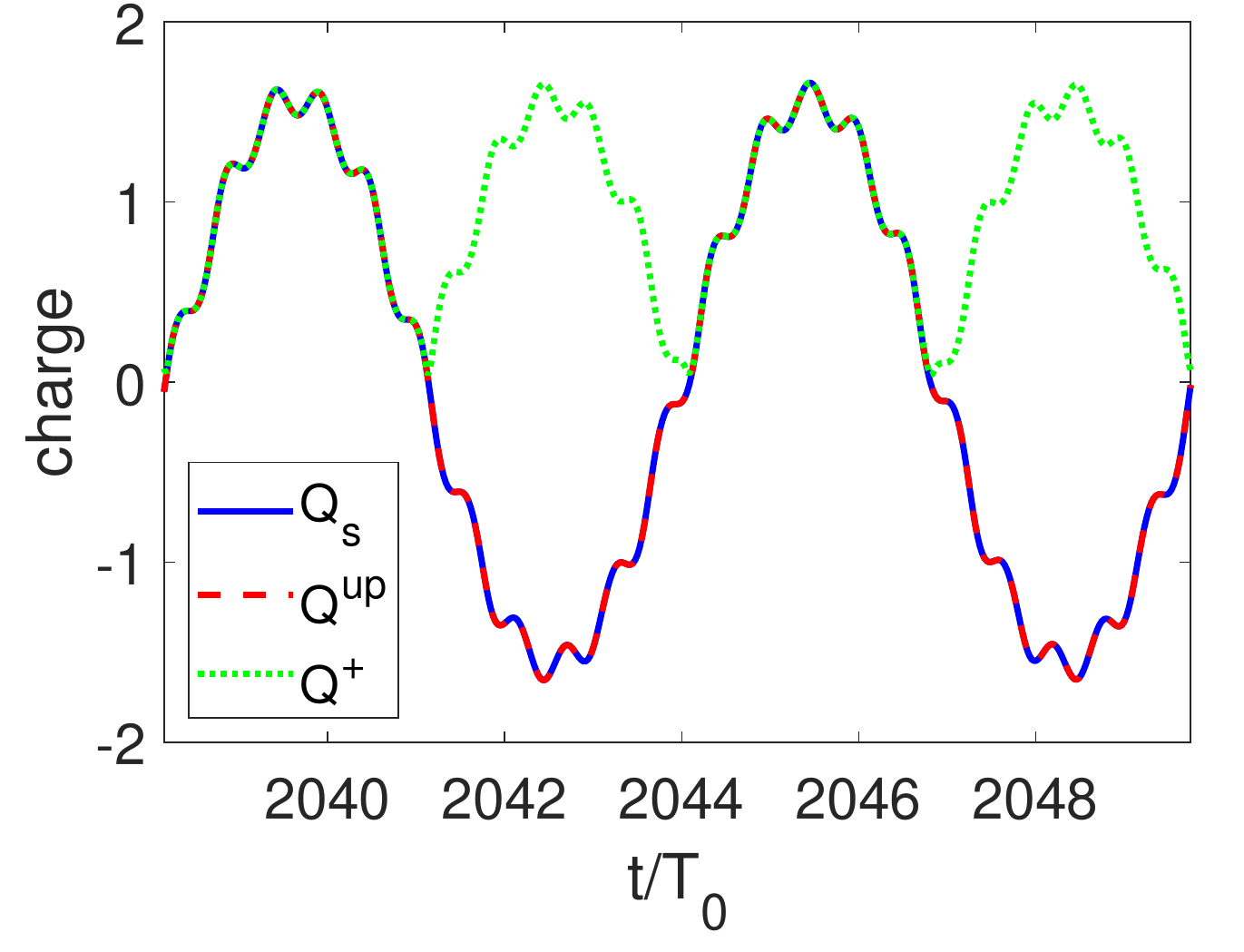}
\caption{\label{fig:charge1}Evolutions of three different charge integrals $Q_s$, $Q^{\rm up}$ and $Q^+$ with time, the absolute values of which are mostly the same. $T_0=2\pi/m$ is roughly the oscillation period of the field $\phi$.}
\end{figure}

The simplest CSQ can be prepared by superposing a Q-ball and an anti-Q-ball. For example, we can place a Q-ball on the positive $y$-axis and an anti-Q-ball with equal but negative charge on the negative $y$-axis, i.e. a system with reflection symmetry about $x$-axis. The initial relative phase difference between the two Q-balls can be chosen to be zero. Because of this placement, the real component of the scalar field is symmetric about the $x$-axis and $y$-axis, while the imaginary component is symmetric about the $y$-axis and antisymmetric about the $x$-axis. The charges will swap along the $y$ axis as the system evolves. Of course, this superposed configuration is not the quasi-stable CSQ, but it will quickly relax to a CSQ, as we shall see later. In fact, it is not essential to superpose exact Q-balls and anti-Q-balls in the initial preparation; we may as well initially superpose oscillating lumps that resemble Q-balls and anti-Q-balls. That is, CSQs are attractor solutions of the theory. See Figs.~\ref{fig:evolution in one period} and \ref{fig:evolution in one period_ham} for the sequence of one charge-swapping period for the already relaxed dipole CSQ, and see Fig.~\ref{fig:charge1} for the evolution of the different charge integrals defined shortly below. On the other hand, the energy density of a CSQ is mostly spherically symmetric. There are also more complex CSQs with more Q-balls and anti-Q-balls, as we shall see in Section \ref{sec:multipleCSQ}. Their lifetime is shorter in 3+1D; see Section \ref{sec:csq_in_3d}. We are focusing on the polynomial potential in this paper, but they also exist in other models \cite{Copeland:2014qra}.

\begin{figure}[tbp]
\centering
\includegraphics[width=.45\textwidth,trim=170 290 160 280,clip]{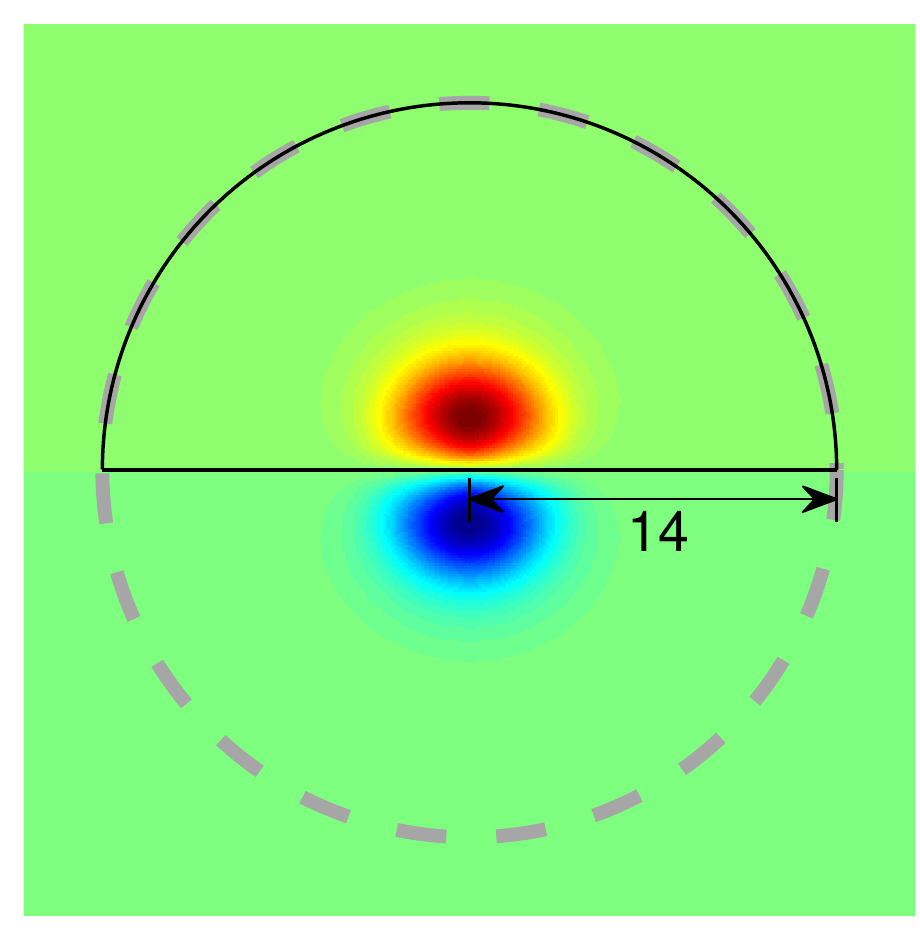}
\caption{Circular disk region used to evaluate $E_c$ and $Q_c$ and semi-circular disk region to evaluate energy $E_s$ and charge $Q_s$. The radius of the semi-circle is $14$, in units of $1/m$. (In 3D, we will use a corresponding semi-ball with radius of 20.) }
\label{fig:integration range}
\end{figure}

For later convenience, we shall define a few energy and charge quantities denoting integrations over various different regions of the dipole CSQ (the center of the CSQ is placed at the origin of the coordinate system):

\begin{itemize}

\item $E$ and $Q$: the total energy and charge in the simulation box respectively

\item $Q^+$: the positive charge over the whole simulation box

\item $E^{\rm up}$ and  $Q^{\rm up}$:  the energy and charge obtained by integrating over the upper half space ($y>0$ for 2D or $z>0$ for 3D) respectively

\item $E_{s}$ and  $Q_{s}$:  the energy and charge obtained by integrating over an upper semi-circular disk (2D) or an upper semi-ball (3D) with a radius of 14 (for 2D) or 20 (for 3D) around the CSQ respectively (see the black solid line in Fig.~\ref{fig:integration range} for 2D)

\item $E_{c}$ and  $Q_{c}$:  the energy and charge obtained by integrating over a circular disk (2D) or a ball (3D) with a radius of 14 (for 2D) or 20 (for 3D) around the CSQ respectively (see the thick, dashed line in Fig.~\ref{fig:integration range} for 2D)

\end{itemize}

While the existence of CSQs has been firmly shown in \cite{Copeland:2014qra}, the properties of CSQs are yet to be studied in more detail. In particular, the periodic boundary conditions used in the simulations of \cite{Copeland:2014qra} are not appropriate to determine the lifetimes of these CSQs, as CSQs radiate perturbations, which propagate back to affect the CSQs in a periodic box. In this paper, we shall set up lattices that have absorbing boundary conditions that can absorb radiation   effectively from the CSQs, which allows us to investigate CSQs in much more detail and determine the lifetimes of various CSQs.

\subsection{Numerical setup}

\begin{figure}[tbp]
\centering
\includegraphics[width=.5\textwidth,trim=90 260 90 260,clip]{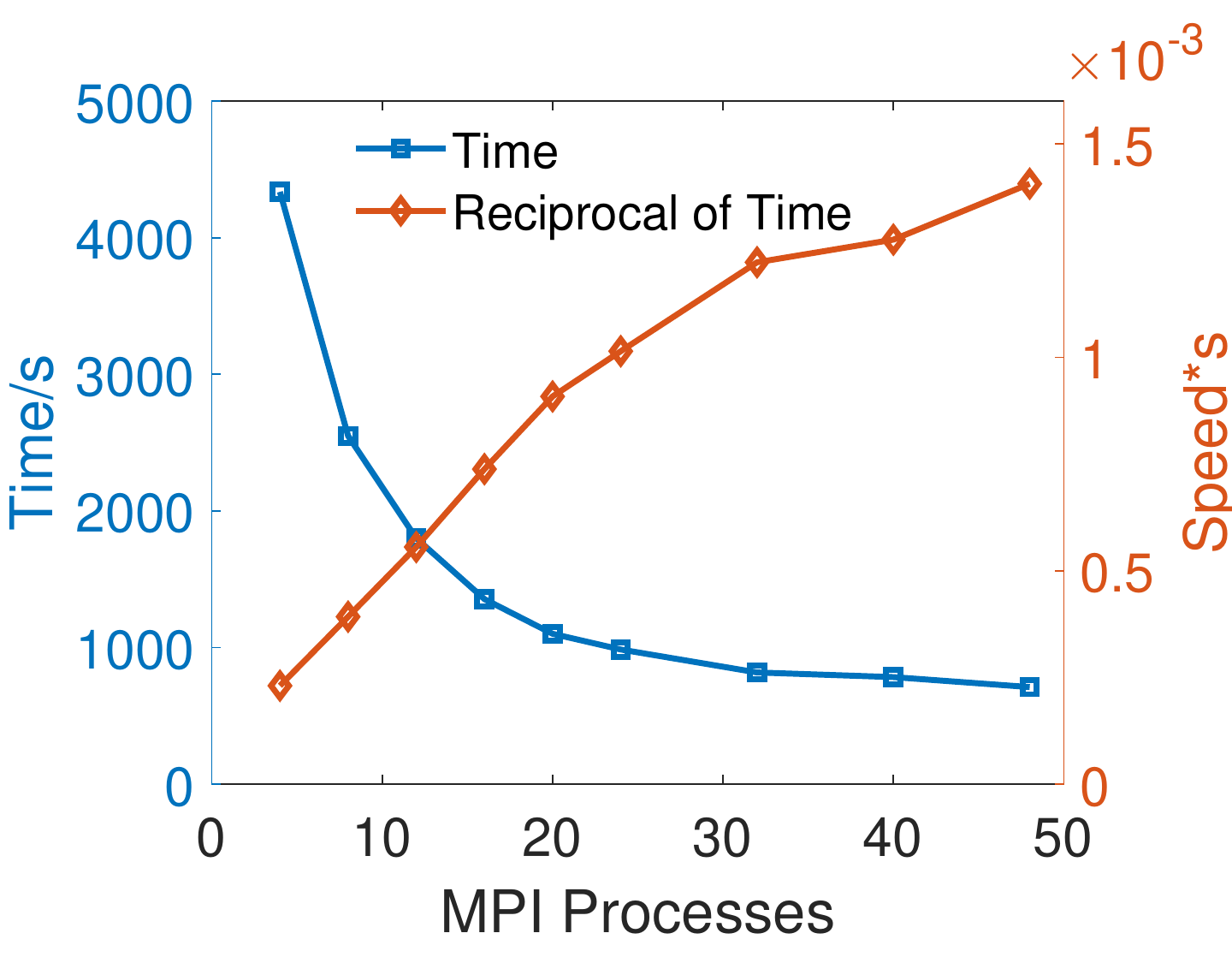}
\caption{Times used to run simulations with a $256^2$ grid for 145600 time steps on a workstation with dual CPUs (two Intel Xeon Platinum 8280 CPUs) and shared memory. The ``speed'' is defined as the reciprocal of the time duration.}
\label{fig:cores speedup}
\end{figure}

Our lattice code makes use of the open-source LATfield2 C++ library \cite{David:2015eya}, which defines objects such as Lattice, Site and Field, allowing for fast and easy implementations of classical field simulations.  However,  LATfield2 uses periodic boundary conditions, which is not suitable for our purposes. We modified the library to incorporate several absorbing boundary conditions, which will be introduced in Section \ref{sec:Higdon BC} and Appendix \ref{sec:otherABCs}. We use a 4th order finite difference stencil for spatial derivatives and evolve in time with the classical Runge-Kutta 4th order method. The code has excellent parallel speedups for many CPU cores with MPI as seen in Fig.~\ref{fig:cores speedup}.

We shall superimpose elementary Q-ball solutions, both for $\phi(0,\bfx)$ and $\dot\phi(0,\bfx)$,  as the initial configuration and let it relax to obtain CSQs. The internal frequency $\oi$ of the elementary Q-ball and the initial distance $d$ between the elementary Q-balls are the free parameters we choose, and we will chart the lifetime of the CSQ in this two dimensional parameter space. 

The Courant-Friedrichs-Lewy (CFL) factor $\d t/\d x$ is set to be 0.1 both for 2D and 3D simulations. Unless otherwise stated, in 2D simulations, we will use a $512^2$ lattice and the grid spacing is $\d x=0.2$, with a $1024^2$ lattice frequently used to check for convergence, and in 3D we will use a $256^3$ lattice and the grid spacing is  $\d x=0.4$, with a $512^3$ lattice to check for convergence. As we will see ({\it e.g.}, in Figs.~\ref{convergence_study} and \ref{bnd_cfl}) that these numerical settings are sufficient for our purposes.

\subsection{Higdon's absorbing boundary conditions}
\label{sec:Higdon BC}

\begin{figure}[tbp]
\centering
\includegraphics[width=.45\textwidth,trim=90 260 110 280,clip]{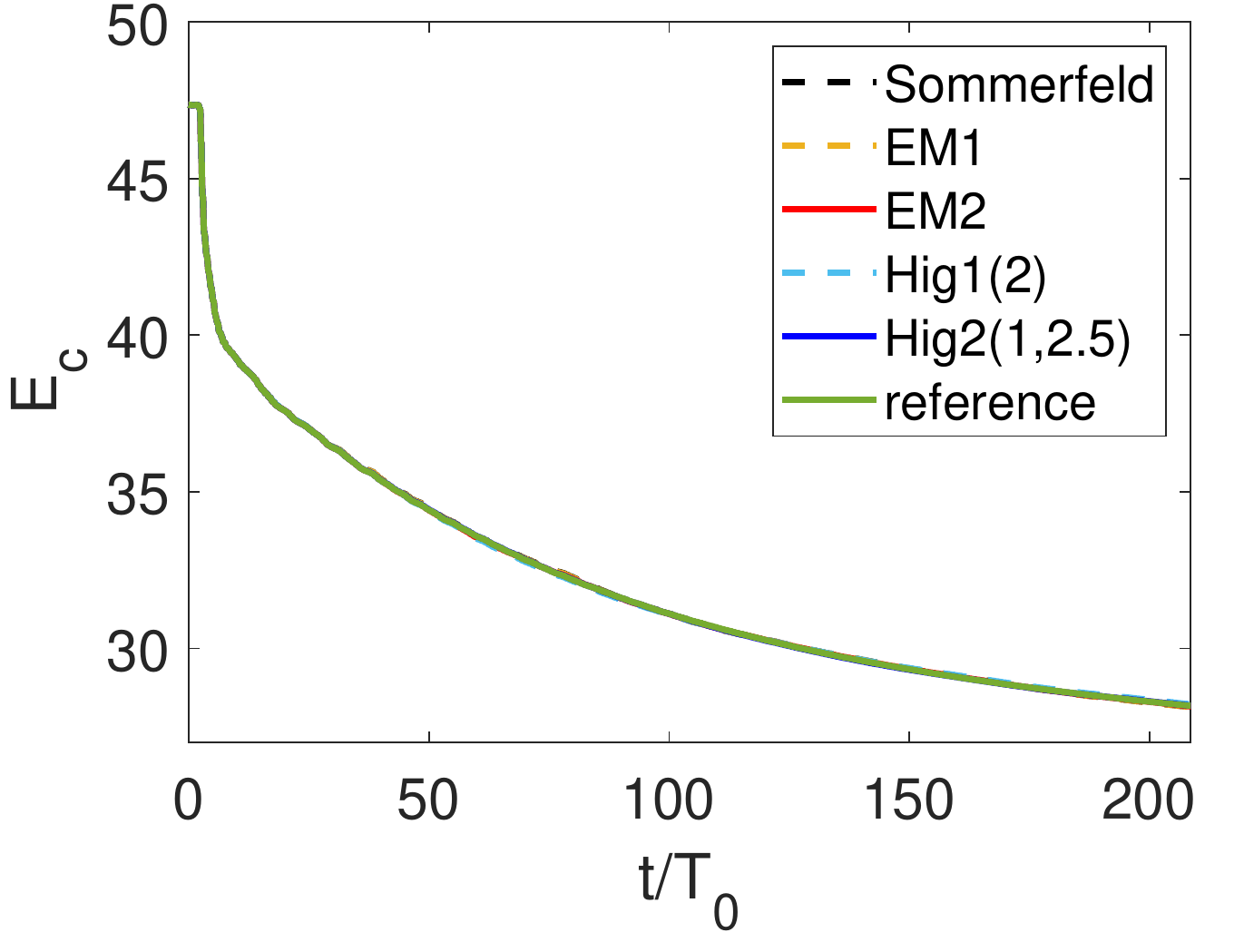}
\hfill
\includegraphics[width=.45\textwidth,trim=90 260 110 280,clip]{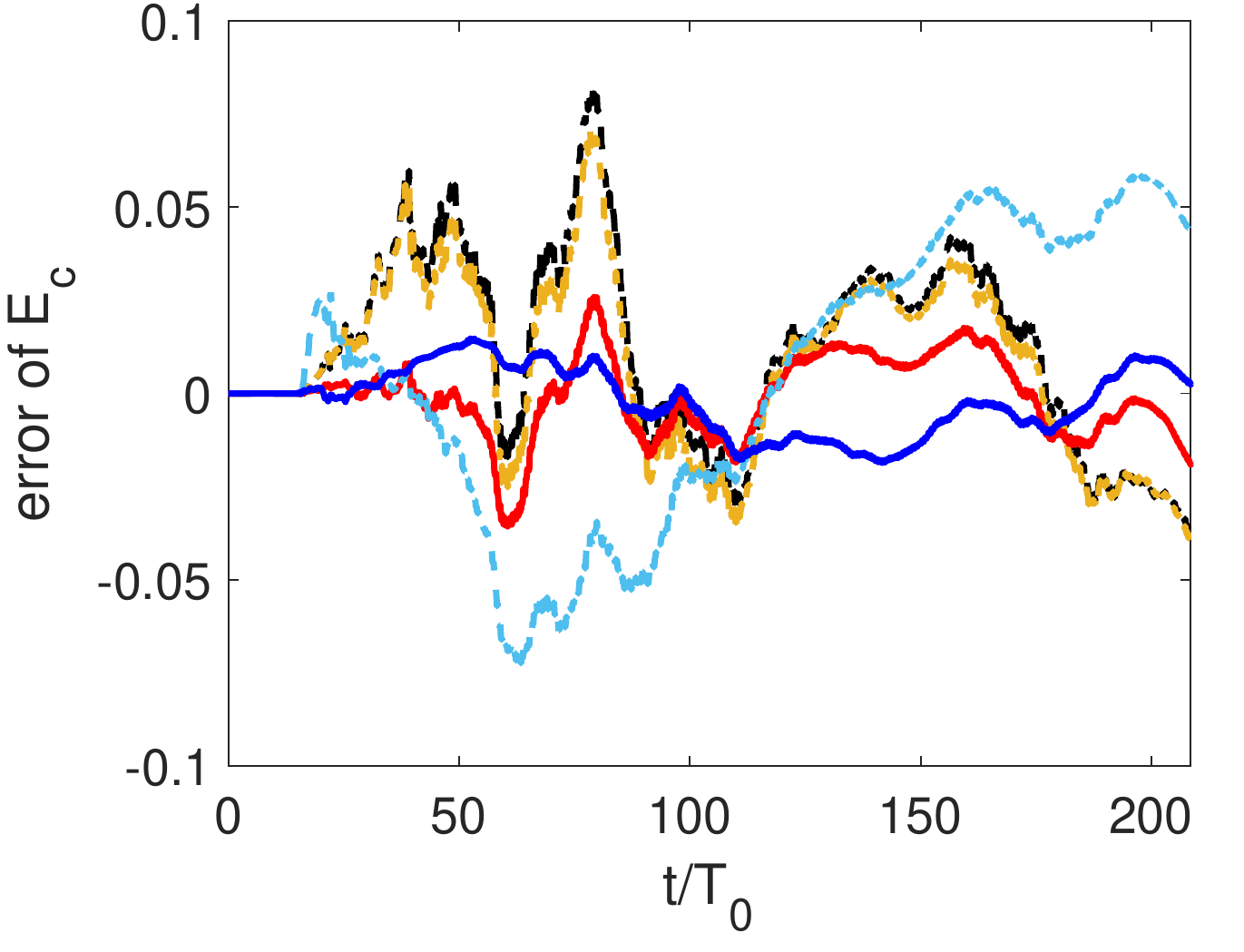} 
\includegraphics[width=.45\textwidth,trim=90 260 110 280,clip]{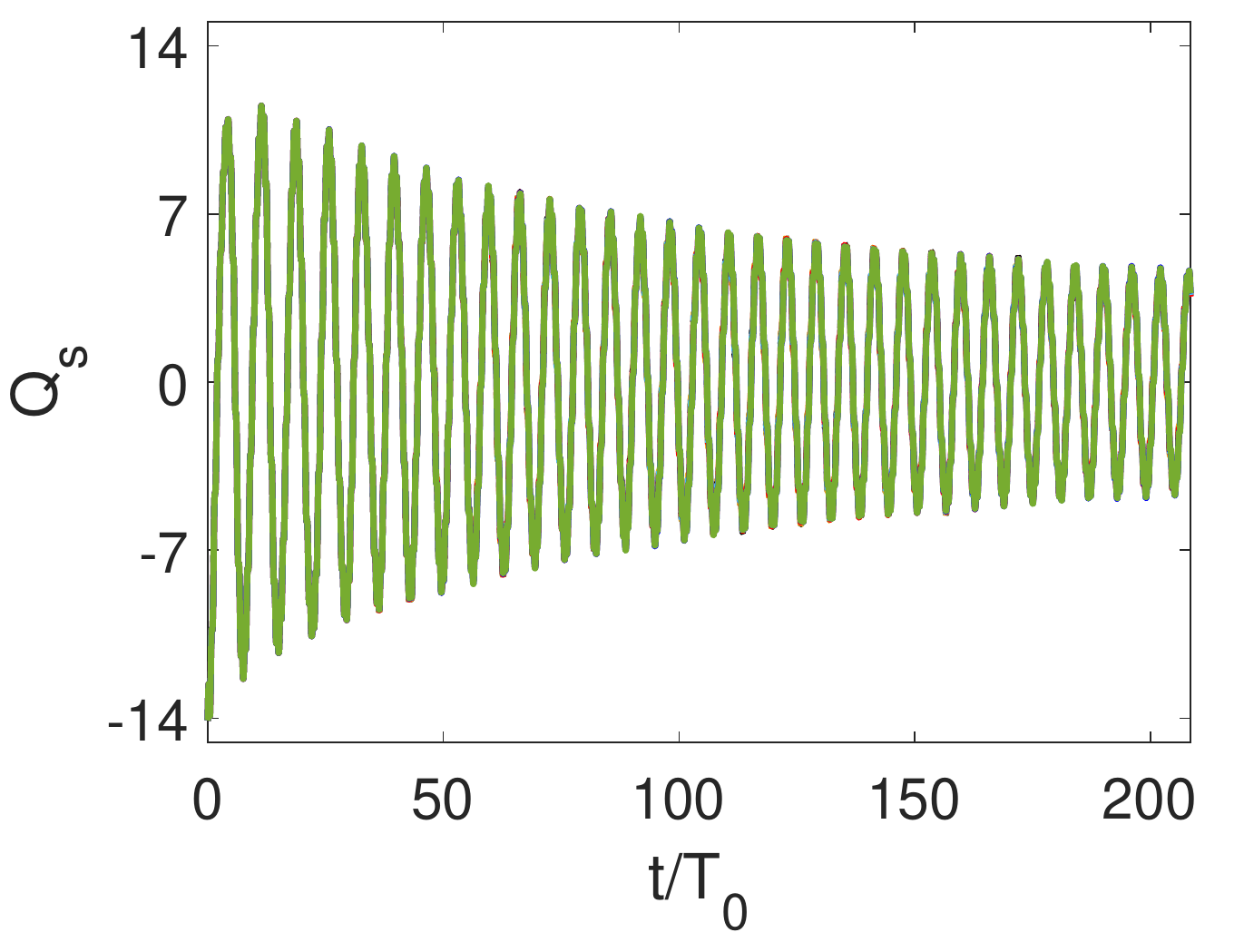} 
\hfill
\includegraphics[width=.45\textwidth,trim=90 260 110 280,clip]{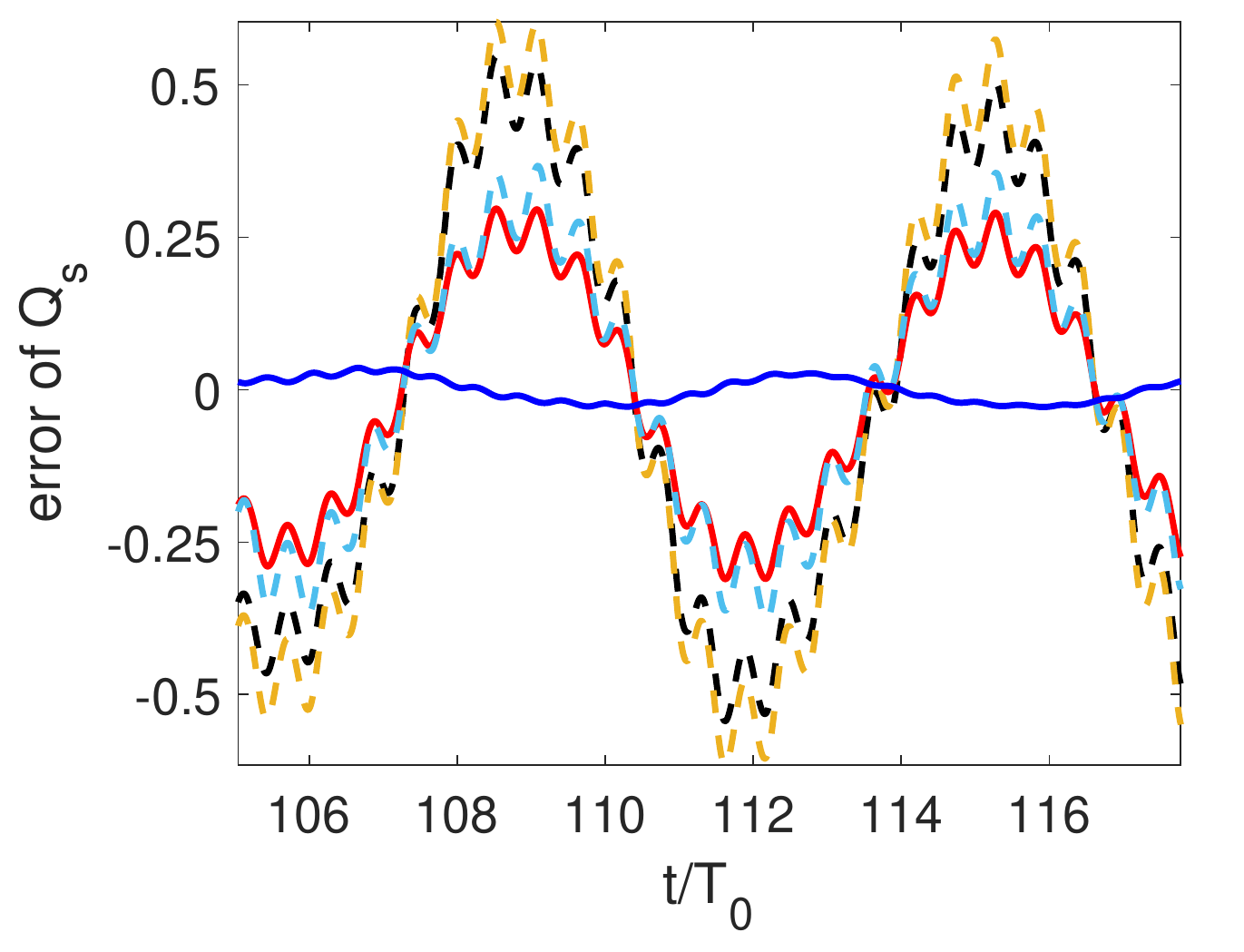}
\caption{\label{bcs_energy charge} Comparisons of simulations with various absorbing boundary conditions (ABCs) in the initial relaxation phase. The left two plots are from the reference simulation, and the right two plots are deviations of simulations with various ABCs from the reference simulation. The various ABCs are defined in the text below \eref{Hig2}, and energy $E_c(t)$ and charge $Q_s(t)$ are defined in Section \ref{sec:CSQs}. The reference simulation is obtained by using a large lattice such that the radiation has not reached the boundaries in the run. Dashed lines represent first-order ABCs and solid lines represent second-order ones. We see that second order ABCs generally perform better than the first order ABCs, and the Hig2(1,2.5) ABC has the best behavior with regard to eliminating the initial relaxing radiation. $T_0=2\pi/m$ is roughly the oscillation period of the field $\phi$.}
\end{figure}

As mentioned above, to determine the lifetime of a CSQ, we need to run the code for an extended period of time, and the periodic boundary conditions, while sufficient for determining the existence of CSQs \cite{Copeland:2014qra}, is unsuitable for this purpose. This is because CSQs emit radiation (or waves), especially during the initial relaxation phase, and this radiation travels back to interfere with the CSQs when using periodic boundary conditions. In reality we are interested in the lifetimes of CSQs in Minkowski space where radiation should simply propagate to infinity. To use a finite computation region to solve an infinite domain problem, we may make use of suitable absorbing boundary conditions (ABCs). Unless the problem is highly symmetric (say spherical symmetry) or the outgoing waves are very simple, ABCs are usually not perfect. However, a good ABC should let the majority of outgoing waves go through the boundary transparently and only incur minor reflections. We will explore several ABCs in this paper, namely Sommerfeld's ABC \cite{alcubierre2008introduction}, Engquist-Majda's ABC \cite{engquist1977absorbing} and Higdon's ABC \cite{higdon1994radiation,higdon1986absorbing}. We find that Higdon's 2nd order ABCs usually produce the best accuracy, which will be adopted in the majority of the simulations in this paper, while the other ABCs are used for sanity checks. We will introduce the Higdon ABCs below and the other ABCs are introduced in Appendix \ref{sec:otherABCs}.

The Higdon ABCs are a set of easily implementable conditions at boundary $x^i=a$ \cite{higdon1994radiation}:
\be
\prod_{j=1}^{M} \( \f{\pd}{\pd t}\pm c_j \f{\pd}{\pd x^i} \)\phi|_{x^i=a}=0 , \label{Higm}
\ee
where the $+$ ($-$) sign is for a right (left) boundary, $t$ and $x^i$ are Cartesian coordinates and $c_j$, to be chosen by the user for specific problems, are the phase velocities of the normal outgoing plain waves that can be absorbed exactly. For the massive Klein-Gordon equation, $c_j$ should be chosen to be no less than 1. It is clear that $c_j$ should be chosen to annihilate the dominant wavenumbers near the boundary, which can be obtained by performing Fourier transforms of the field near the boundary. A rough guide is that $c_j$ is to be chosen to minimize the reflection rate between the amplitudes of the incoming waves and the outgoing waves
\begin{equation}
{\mc R[k^i]}=\prod_{j=1}^{M}\left|\frac{-\sqrt{(k^i)^2+m^2}+c_{j} k^i}{-\sqrt{(k^i)^2+m^2}-c_{j} k^i}\right|  ,
\end{equation}
for the dominant wavenumbers, where $k^i$ is the wavenumber and $m$ is the perturbative mass of field. From the reflection rate formula, we can see that even for the outgoing waves whose phase velocities are not $c_j$, the Higdon ABC can still often absorb much of them. In practice, the most commonly used Higdon ABCs are when $M=1,2$,  which are simply given by
\bal
\( \f{\pd}{\pd t}\pm c_1\f{\pd}{\pd x^i} \)\phi|_{x^i=a} &= 0 , \label{Hig1}
\\
\( \f{\pd^2}{\pd t^2}\pm (c_1+c_2)\f{\pd}{\pd x^i \pd t}+c_1 c_2\f{\pd^2}{\pd (x^i)^2} \)\phi|_{x^i=a} &= 0 , \label{Hig2}
\eal
where the $+$ ($-$) sign is for a right (left) boundary.
Note that the 1st and 2nd order Higdon ABC with $c_1=1$ and $c_1=c_2=1$ are actually equivalent to the 1st and 2nd order Engquist-Majda ABC respectively, upon using the equations of motion. It is advised that, without prior knowledge of the spectrum of the outgoing waves, the default 2nd Higdon parameters can be chosen as $c_1=c_2=1$ \cite{higdon1994radiation}.

As we find CSQs by superimposing elementary Q-balls and letting them relax, the initial relaxation phase of the CSQ evolutions produces the largest amount of radiation, and we shall choose our ABCs to maximize the absorption of the radiation at the boundary from this phase. We shall compare the absorbing effects of different ABCs and select the best one among them. Since the relaxation phase is short, we can actually simulate this phase with a sufficiently large lattice such that by the end of this phase the radiation has not reached the boundary yet. We compare the effects of different ABCs with this reference run, which allows us to pick the one with the smallest deviations. The ABCs that we have tested against the reference run are:
\begin{itemize}

\item {\it Sommerfeld}: Sommerfeld boundary condition with $v=1$, $\phi_0=0$ and no non-wavelike term $h(t)$

\item {\it EM1}: Engquist-Majda's 1st order condition, which is equivalent to Higdon's 1st order condition with $c_1=1$

\item {\it EM2}: Engquist-Majda's 2nd order condition, which is equivalent to Higdon's 2nd order condition with $c_1=c_2=1$

\item {\it Hig1(2)}: Higdon's 1st order condition with $c_1=2$

\item {\it Hig2(1,2.5)}: Higdon's 2nd order condition with $c_1=1$, $c_2=2.5$ (this is the best ABC for our problem, and we will mostly use this method in the following, with other methods used for sanity checks.)

\end{itemize}

\begin{table}[tbp]
\centering
\resizebox{0.9\textwidth}{!}{
\begin{tabular}{c|cccccc}
\hline
\hline
errors &  Sommerfeld & EM1 & EM2 & Hig1(2) & Hig2(1,2.5) \\
\hline
$\si_{E_c}$ & $2.80\times 10^{-2}$ & $2.53\times 10^{-2}$ & $1.08\times 10^{-2}$ & $3.72\times 10^{-2}$ & $0.91\times 10^{-2}$ \\
$\si_{Q_s}$ & $1.93\times 10^{-1}$ & $2.19\times 10^{-1}$ & $1.18\times 10^{-1}$ & $1.38\times 10^{-1}$ & $0.20\times 10^{-1}$ \\
\hline
\hline
\end{tabular}
}
\caption{\label{2-norm of error}Standard deviations of the various  ABCs (see the text below \eref{Hig2}) against the reference simulation for energy $E_c(t)$ and charge $Q_s(t)$, which are defined in Section \ref{sec:CSQs}. The reference simulation is set up such that the lattice is sufficiently large so that the radiation has not reached the boundaries at the end of the run. The Hig2(1,2.5) ABC has the best accuracy and is the default ABC used below unless stated otherwise.}
\end{table}

Specifically, we test the different ABCs against the reference simulation in a $256^2$ grid for a run of 32769 steps. We use the fiducial model and prepare the consistent Q-balls with $\oi=\pm 0.84$ and place them together with a separation $d=2.8$. We compute energy $E_c(t)$ and charge $Q_s(t)$, which are defined in Section \ref{sec:CSQs}, as functions of time, and compute the standard deviations:
\bal
\si_{E_c} &= \[\sum^N_{i=1} \f{(E^{\rm ABC}_c(t_i)-E^{\rm ref}_c(t_i))^2}N\]^{1/2} ,
\\
\si_{Q_s} &= \[\sum^N_{i=1}\f{(Q^{\rm ABC}_s(t_i)-Q^{\rm ref}_s(t_i))^2}N\]^{1/2} .
\eal
From Table \ref{2-norm of error} and Fig.~\ref{bcs_energy charge}, we see that all these ABCs, especially the second order ABCs, are relatively good, and the Hig2(1,2.5) ABC appears to have the best accuracy in eliminating the outgoing radiation from the initial relaxation phase of the CSQ. The effectiveness of the Hig2(1,2.5) ABC can also be seen by looking at far-field regions of the CSQ, {\it i.e.}, at points  far from the CSQ, in which case the runs with the Hig2(1,2.5) ABC match the reference run much better than the other ABCs.

Among the properties of CSQs that will be explored in the later sections, the lifetimes of CSQs are probably most sensitive to the ABCs. Indeed, if we use different ABCs to find the lifetimes, we get slightly different results. However, the differences are small, about a couple of percent. Generally, the Hig2(1,2.5) ABC leads to the longest lifetimes. We will assume that a longer lifetime means that the ABC absorbs the radiation better at the boundaries, on which basis we will use the Hig2(1,2.5) ABC as our default ABC.

\section{Dipole CSQs in 2+1D}
\label{sec:csq2D}

In this section, we will explore the properties of the simplest dipole CSQ in 2+1D, derived from a figuration with a Q-ball and an equal (opposite) charge anti-Q-ball. We will see that they have distinct stages of evolution and can be formed from various different initial setups, confirming that they are attractor solutions. We will also chart the lifetimes of these CSQs for different initial conditions.

\subsection{Stages of a CSQ evolution}

\begin{figure}[tbp]
\centering
\includegraphics[width=.5\textwidth,trim=90 260 110 280,clip]{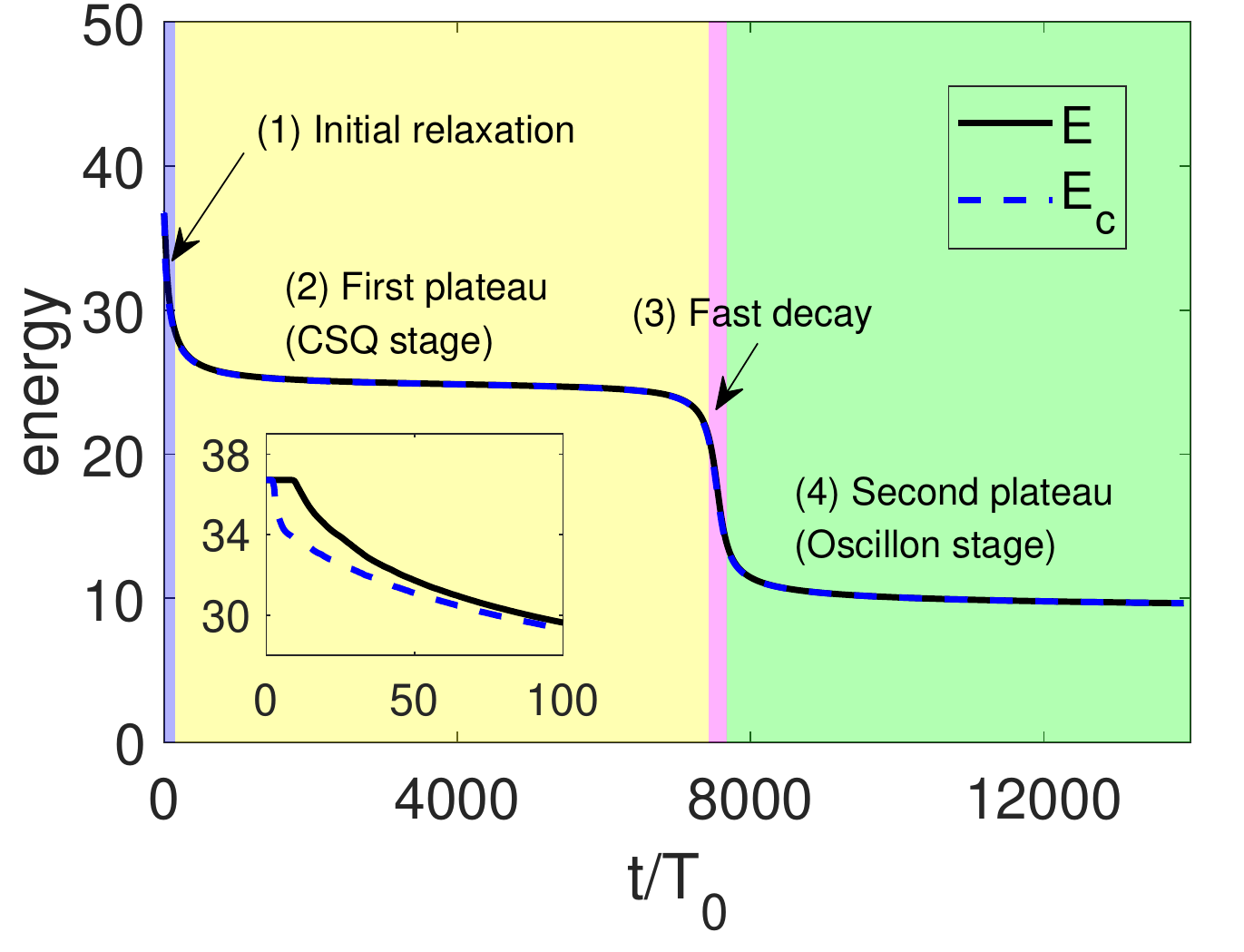}
\caption{\label{fig:energy_vs_period}Evolution of total energy $E$ and energy $E_c$, which are defined in Section \ref{sec:CSQs}. $t$ is in the units of the oscillation period of the field $T_0$, which is $2\pi$ in the units of ${1}/{m}$. Inset: an enlarged view of the initial relaxation phase where there are noticeable differences between $E$ and $E_c$. The constituent elementary Q-balls are initially placed $d=2.8$ apart from each other, with $\oi=\pm 0.86$.}
\end{figure}

\begin{figure}[tbp]
\centering
\includegraphics[width=.5\textwidth,trim=90 260 110 270,clip]{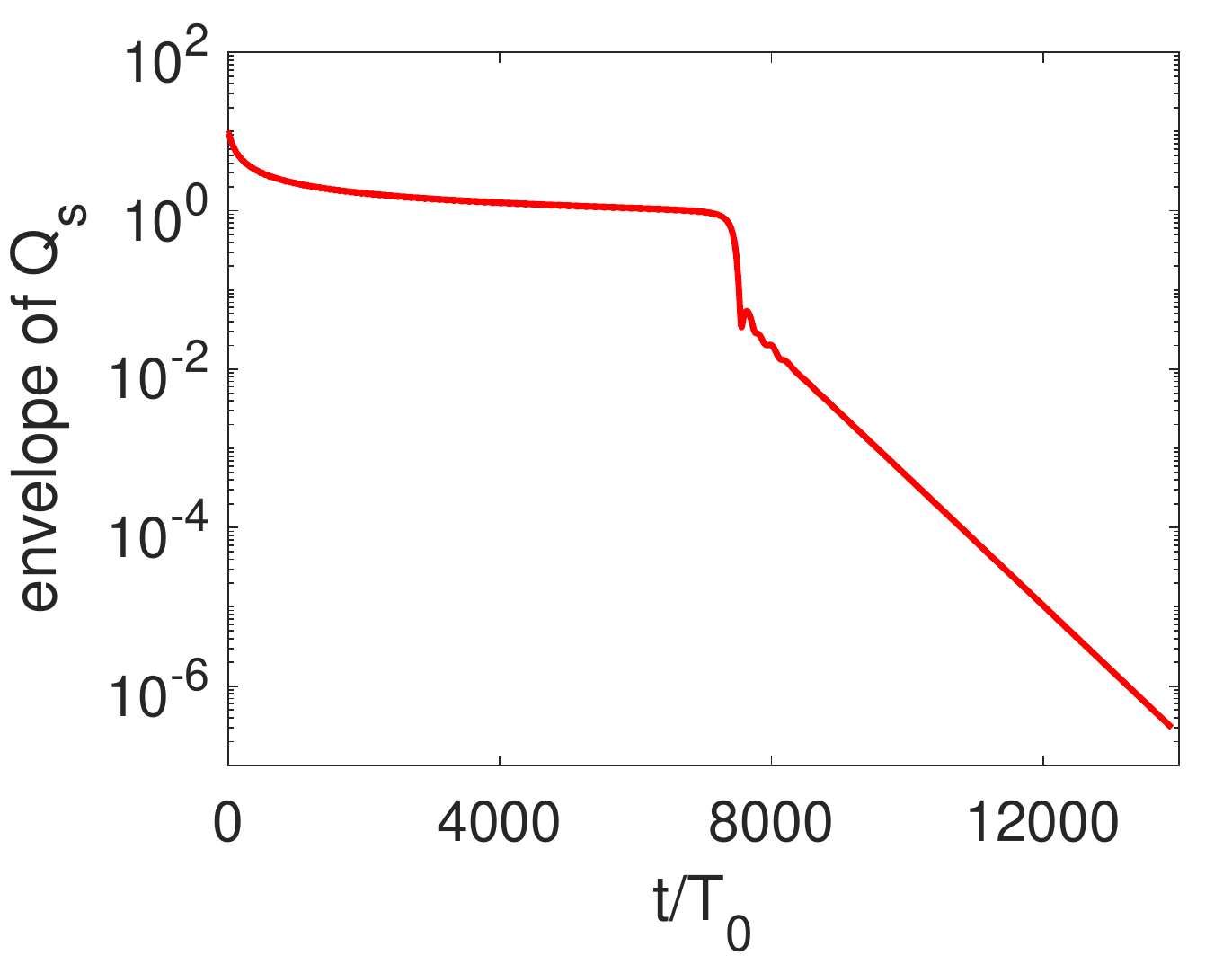}
\caption{\label{fig:charge_vs_period} Envelope of the $Q_s$ evolution, where $Q_s$ is defined in Section \ref{sec:CSQs}. $Q_s$ oscillates quickly according to the charge swapping frequency $T_{\rm swap}$ within the envelope. All the settings are identical to Fig.~\ref{fig:energy_vs_period}. $T_0=2\pi/m$ is roughly the oscillation period of the field $\phi$.}
\end{figure}

\begin{figure}[tbp]
\centering
\includegraphics[width=.5\textwidth,trim=90 260 90 280,clip]{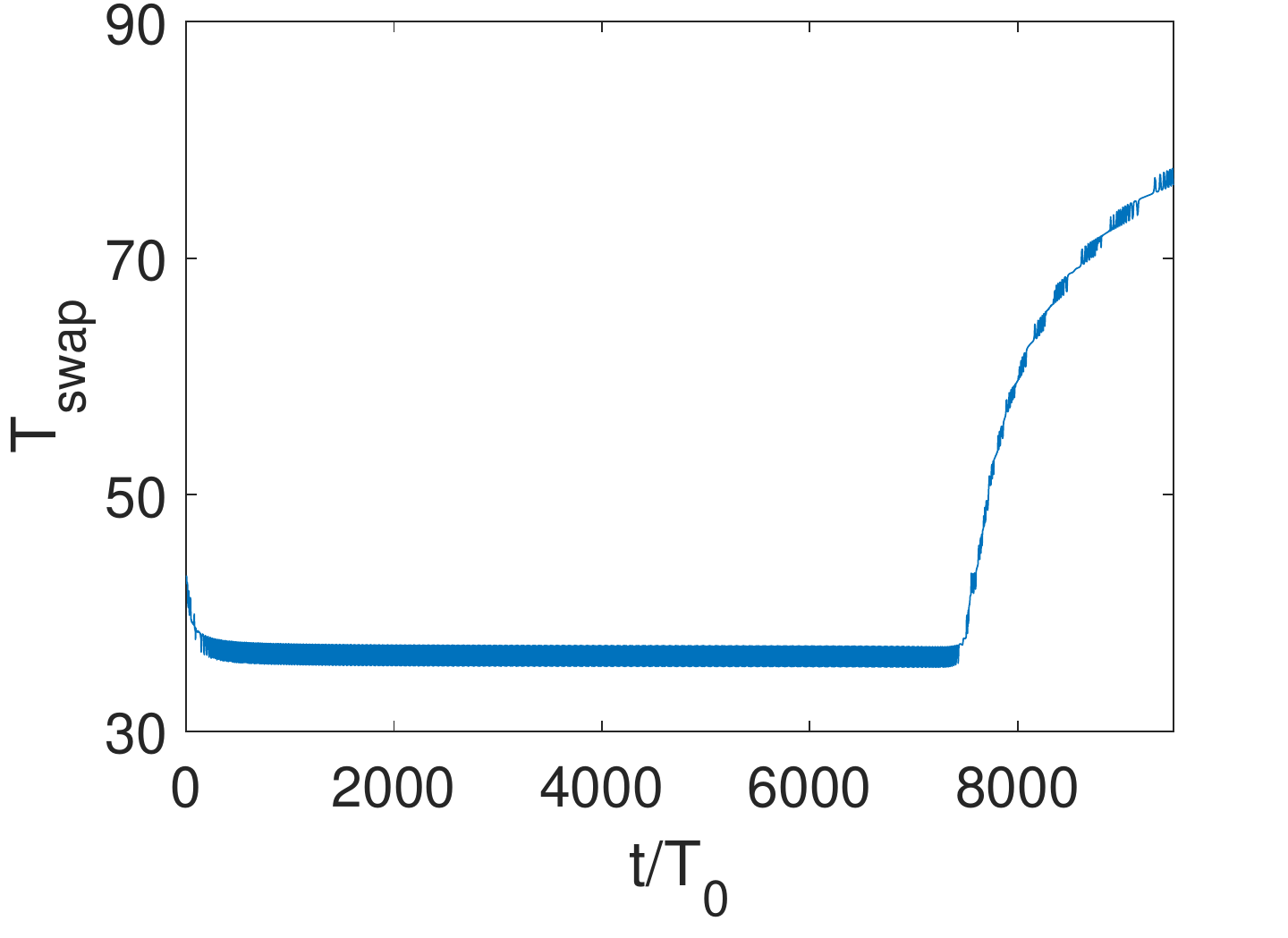}
\caption{\label{fig:freq_vs_period_zeros}Charge-swapping period $T_{\rm swap}$. $T_0=2\pi/m=2\pi$ is roughly the oscillation period of the field $\phi$. All the settings are identical to Fig.~\ref{fig:energy_vs_period}.}
\end{figure}

\begin{figure}[tbp]
\centering
\includegraphics[width=.5\textwidth,trim=90 260 110 280,clip]{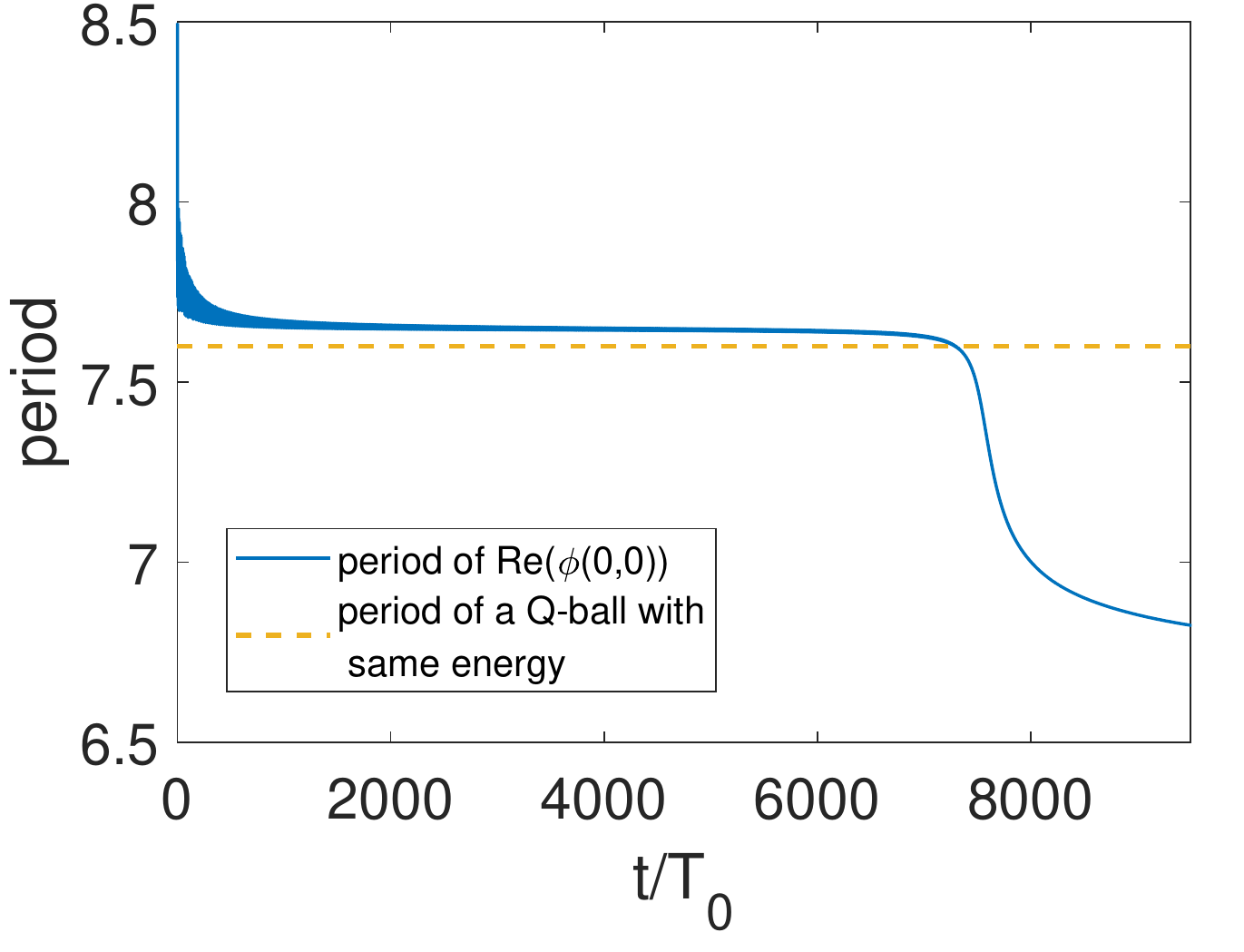}
\caption{\label{fig:origin_real_period}Evolution of the period of the real part of the field $\phi$ at the origin $(x=0,y=0)$. All the settings are identical to Fig.~\ref{fig:energy_vs_period}. The orange dashed line is the oscillation period of an elementary Q-ball with the same energy as the CSQ. Note that if we prepare the CSQ with elementary Q-balls with different frequencies, the period of the resulting CSQ remains the same.}
\end{figure}

As we mentioned above, we prepare the dipole CSQ by superimposing an elementary Q-ball and an equal charge elementary anti-Q-ball with a short separation. An elementary Q-ball or anti-Q-ball is totally specified by $\oi$ in \eref{stationary ansatz}, that is, $f(r)$ is determined by $\oi$. For a typical CSQ prepared in this way, there are four distinct stages of its evolution, as can be seen in Figs.~\ref{fig:energy_vs_period}, \ref{fig:charge_vs_period}, \ref{fig:freq_vs_period_zeros} and \ref{fig:origin_real_period}: {\bf (1)} Initial relaxation, {\bf (2)} First plateau (CSQ stage), {\bf (3)} Fast decay and {\bf (4)} Second plateau (oscillon stage). In the following, we will discuss these four stages separately.\\

\noindent{\bf (1) Initial relaxation}

As we set up the initial configuration by simply superimposing two elementary Q-balls, which we emphasize are not the quasi-stable CSQs, this stage is characterized by fast decrease of energy and charge from the initial lump. For example, in Figs.~\ref{fig:energy_vs_period} and \ref{fig:charge_vs_period}, we see that, for a typical CSQ, about a third of the energy and about three quarters of the charge of the initial lump are shed during about a hundred oscillations, after which the initial configuration settles down to become a CSQ. The radiation of energy from the initial lump can be easily seen in the inset of Fig.~\ref{fig:energy_vs_period}, where the energy in the central circle $E_c$ initially decreases much faster than that of the total energy in the box $E$, and upon absorption at the boundaries the difference between $E_c$ and $E$ diminishes within about a hundred oscillations for this fiducial model. During this transition stage, the swapping period of the lump $T_{\rm swap}$, which is defined at zero points of $Q_s$ by summing the time durations to the previous and next zero point, decreases quickly to the plateau of the CSQ (see Fig.~\ref{fig:freq_vs_period_zeros}), and the oscillating period of the real part of the field $\phi$ at the origin, which is defined at zero points of Re$\phi(t,0,0)$ by summing the time durations to the previous and next zero point, quickly settles to a value that is slightly larger than the intrinsic period of the elementary Q-ball with the same energy(see Fig.~\ref{fig:origin_real_period}). From the trends in this initial stage of these figures, we might have expected that the CSQs are attractor solutions. We will look at this in more detail in the next subsection. \\

\noindent {\bf (2) First plateau (CSQ stage)}

In this stage, the energy and the charge of the CSQ have reached a plateau, decreasing very slowly with time; see  Figs.~\ref{fig:energy_vs_period} and \ref{fig:charge_vs_period}. While the average value of the swapping period of the CSQ $,T_{\rm swap}$, also remains mostly unchanged, its value oscillates noticeably around the average; see Fig.~\ref{fig:freq_vs_period_zeros}. This stage lasts for an extended period of time, the length of which depends on the oscillation frequency $\oi$ of the constituent elementary Q-balls and the initial separation $d$ between them, the parameter space of which will be charted in Section \ref{sec:lifeDiCSQ}. This stage is usually what we refer to as the CSQ, and we will refer to the duration of this stage as the lifetime of a CSQ.

\begin{figure}[tbp]
\centering
\includegraphics[width=.42\textwidth,trim=90 260 110 280,clip]{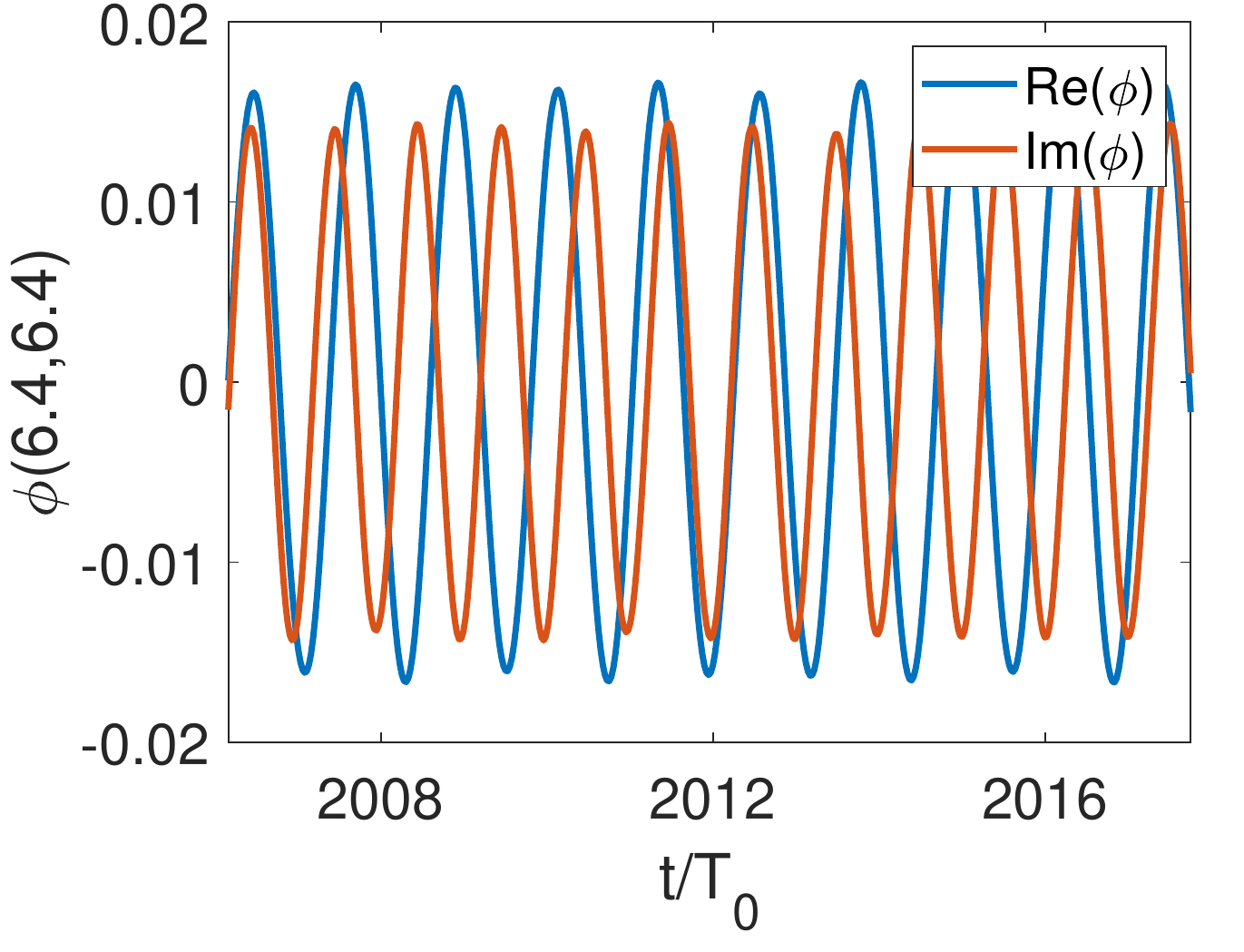}
~~~~~~~~~
\includegraphics[width=.42\textwidth,trim=90 260 110 280,clip]{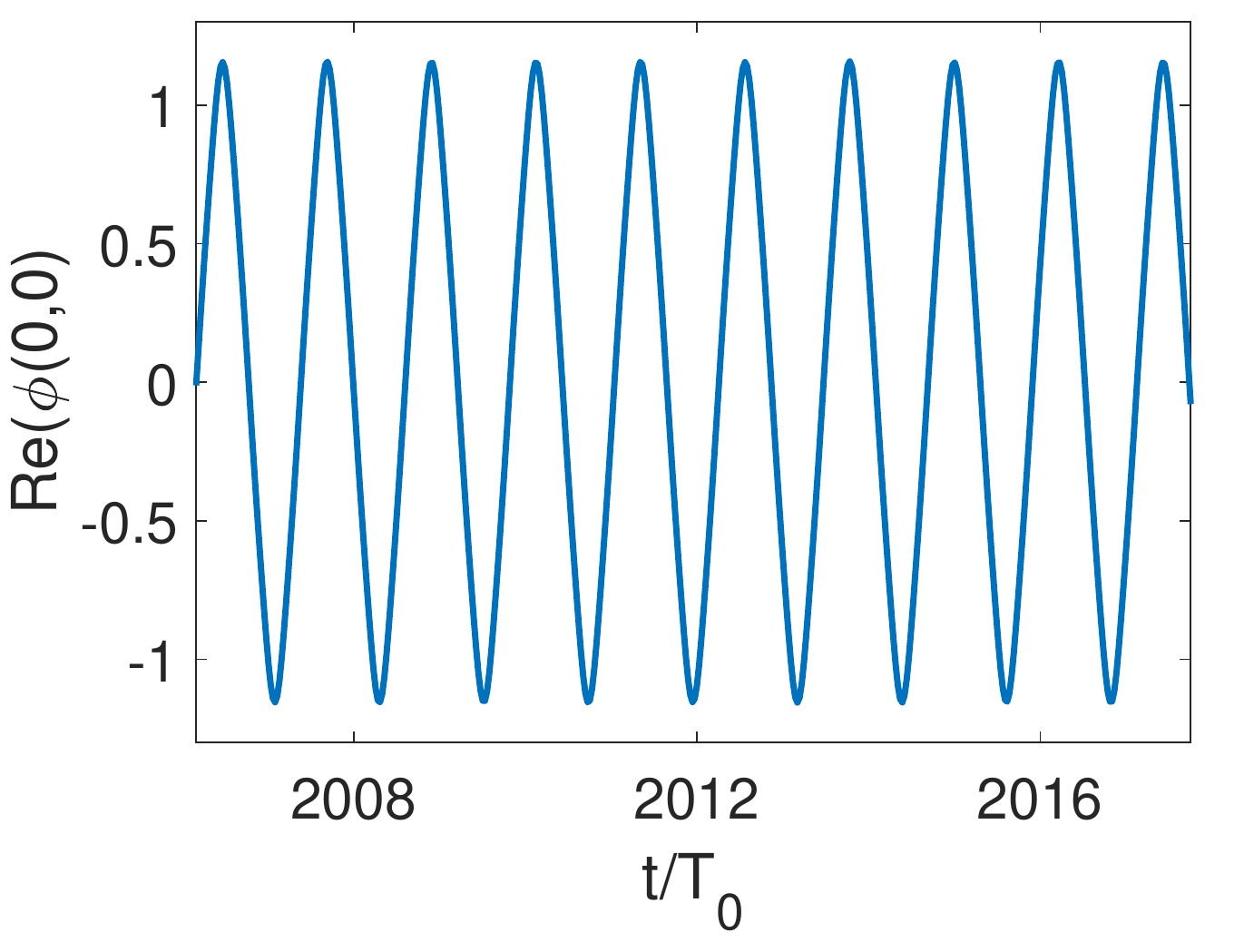}
\includegraphics[width=.42\textwidth,trim=90 260 110 270,clip]{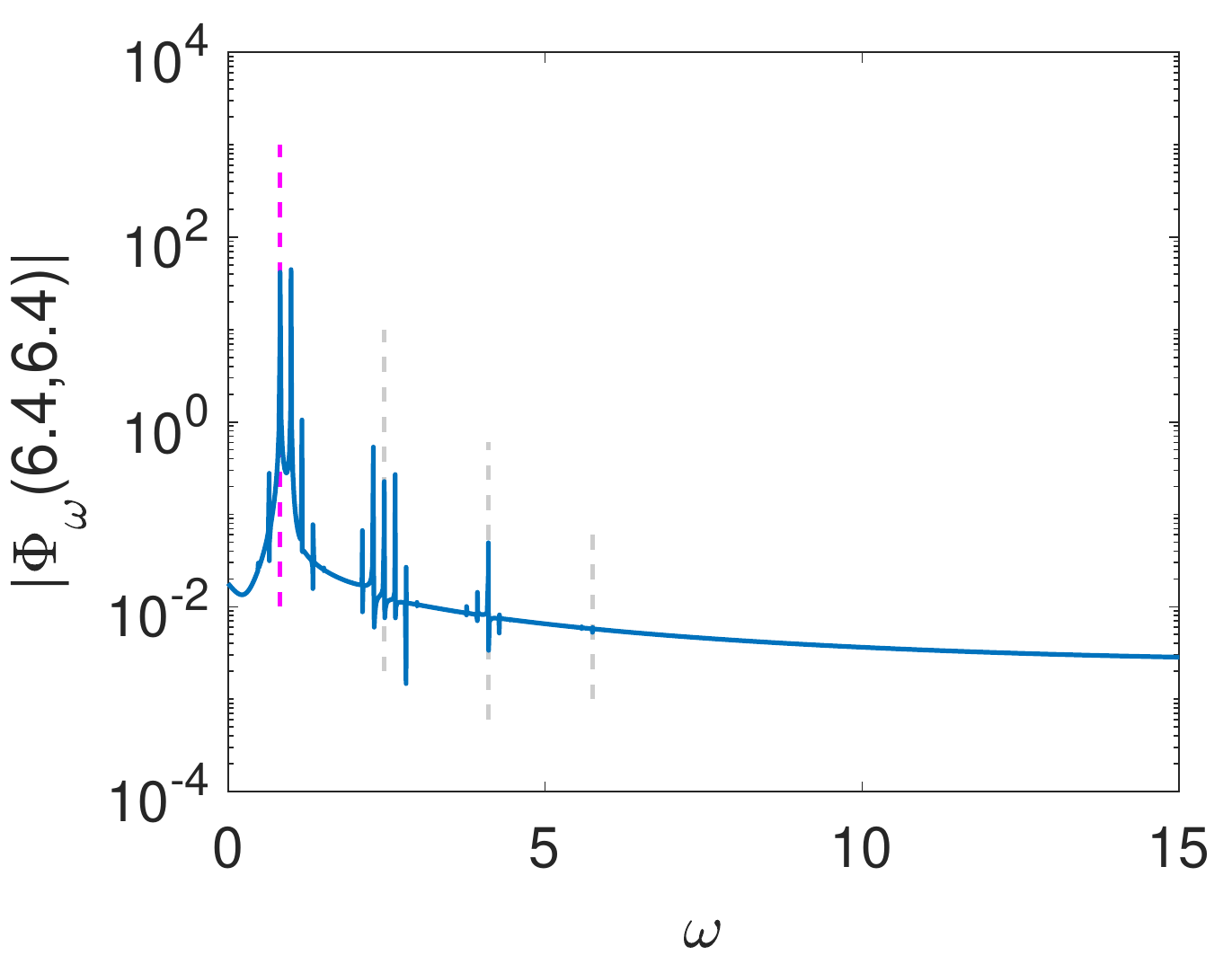}
~~~~~~~~~
\includegraphics[width=.42\textwidth,trim=90 260 110 270,clip]{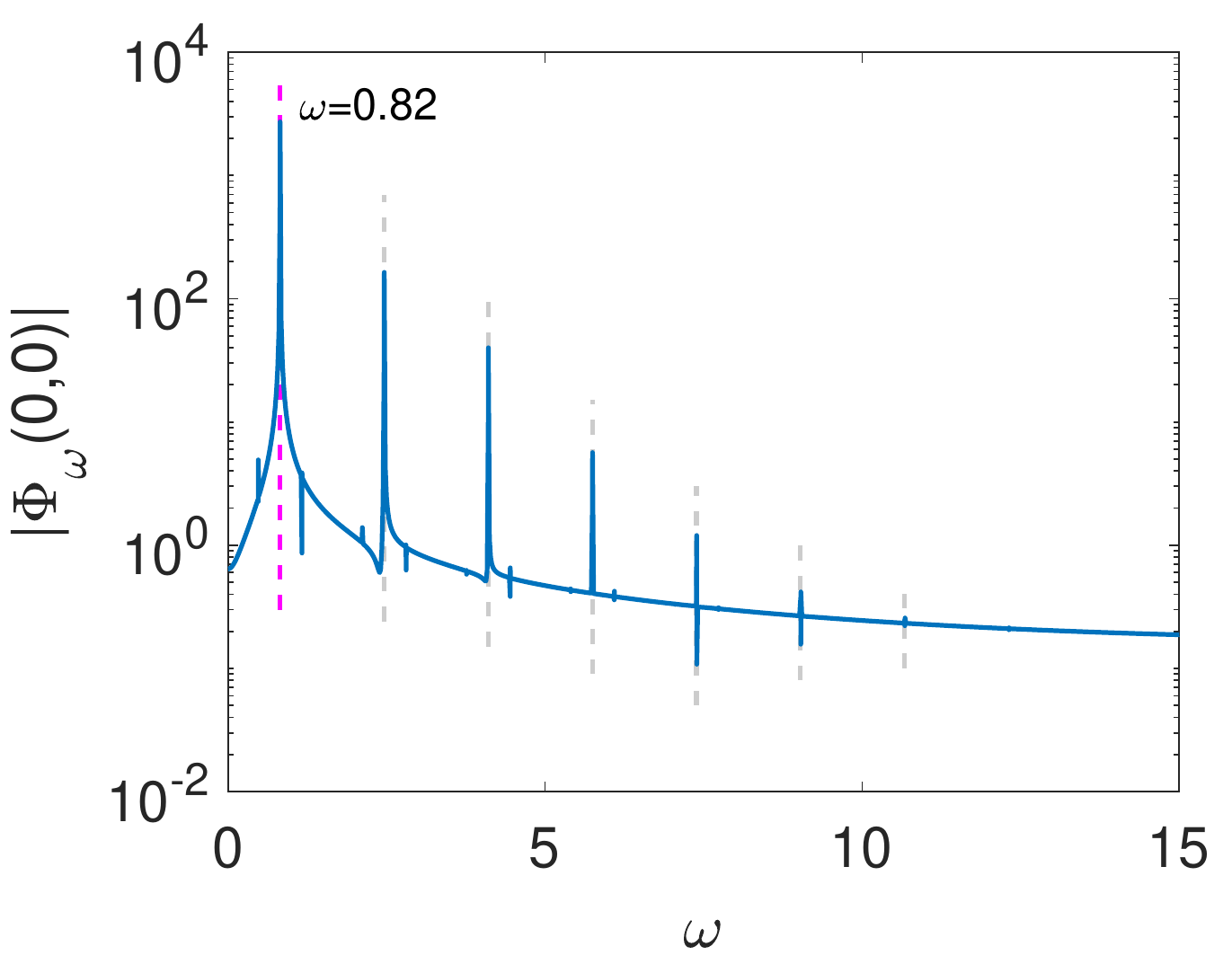}
\caption{\label{fig:points_components_and_spectrums}The field values at $(x=6.4,y=6.4)$ and $(x=0,y=0)$ and their temporal Fourier transforms in the CSQ stage. $\Phi_{\oi}(x,y)$ is the Fourier transform of $\phi(t,x,y)$. Dashed lines in the two bottom plots show the odd multiples of $\oi=0.82$, which match the spectral peaks. }
\end{figure}

\begin{figure}[tbp]
\centering
\includegraphics[width=.3\textwidth,trim=110 260 150 280,clip]{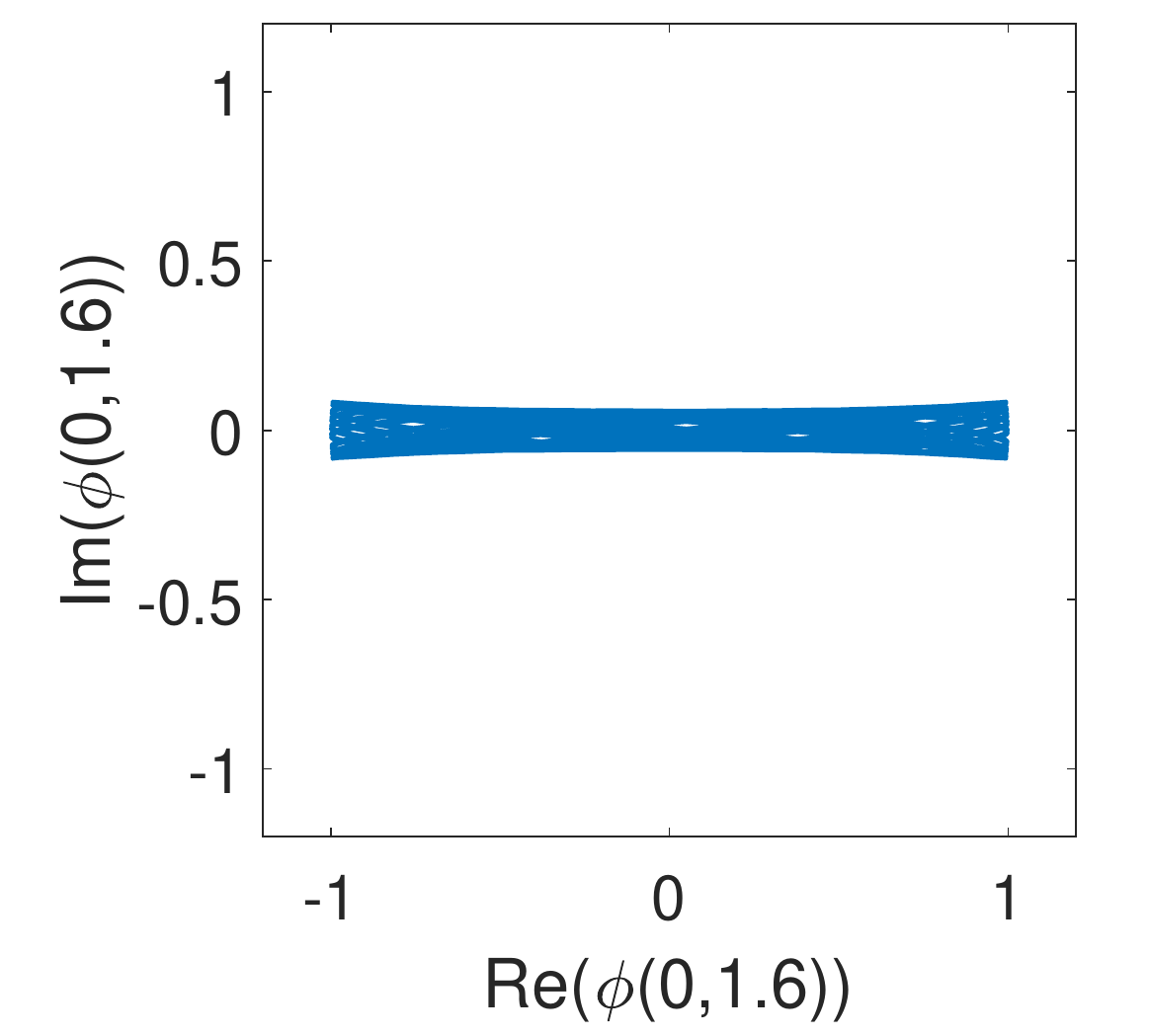}
\hfill
\includegraphics[width=.3\textwidth,trim=110 260 150 280,clip]{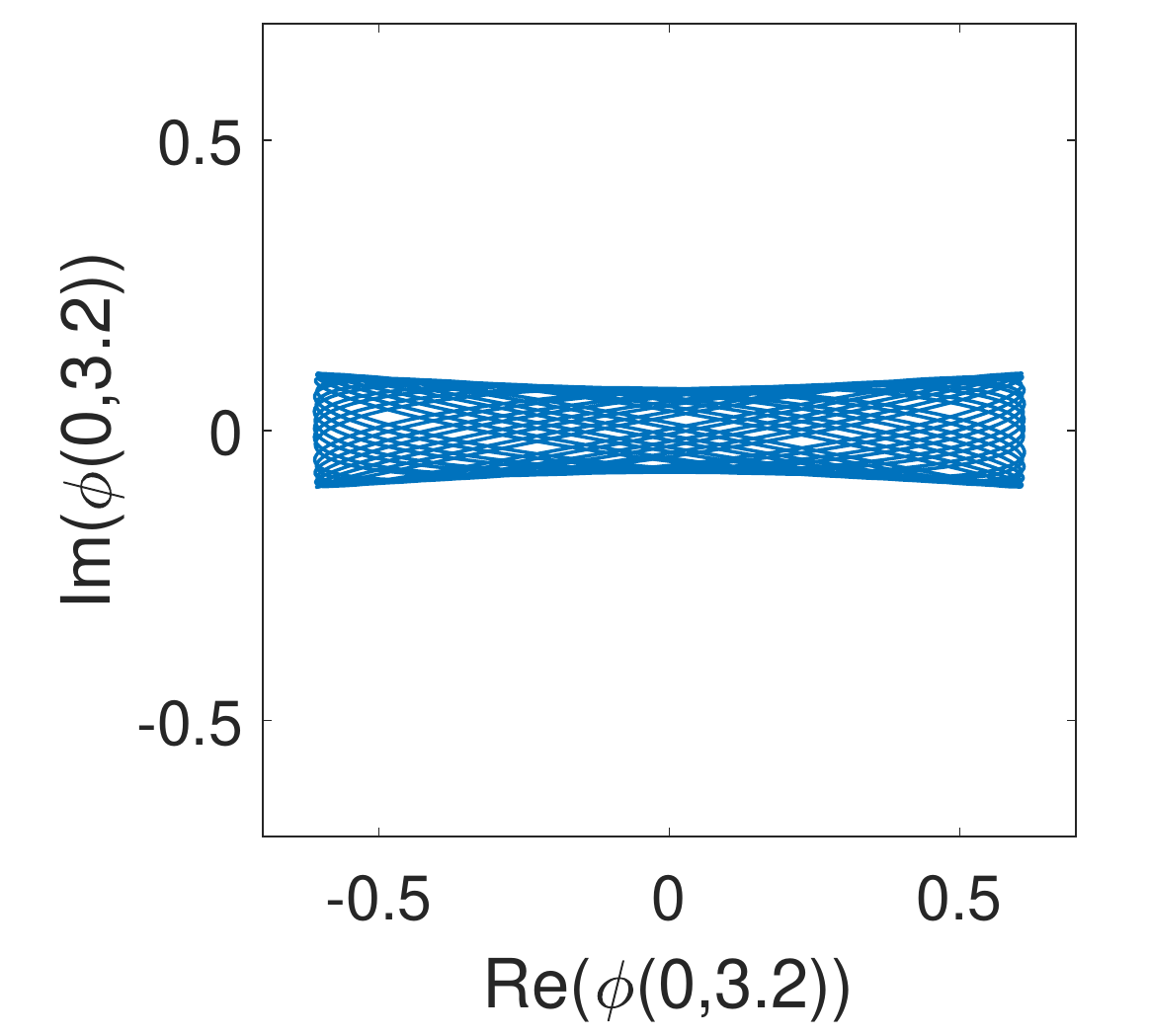}
\hfill
\includegraphics[width=.3\textwidth,trim=110 260 150 280,clip]{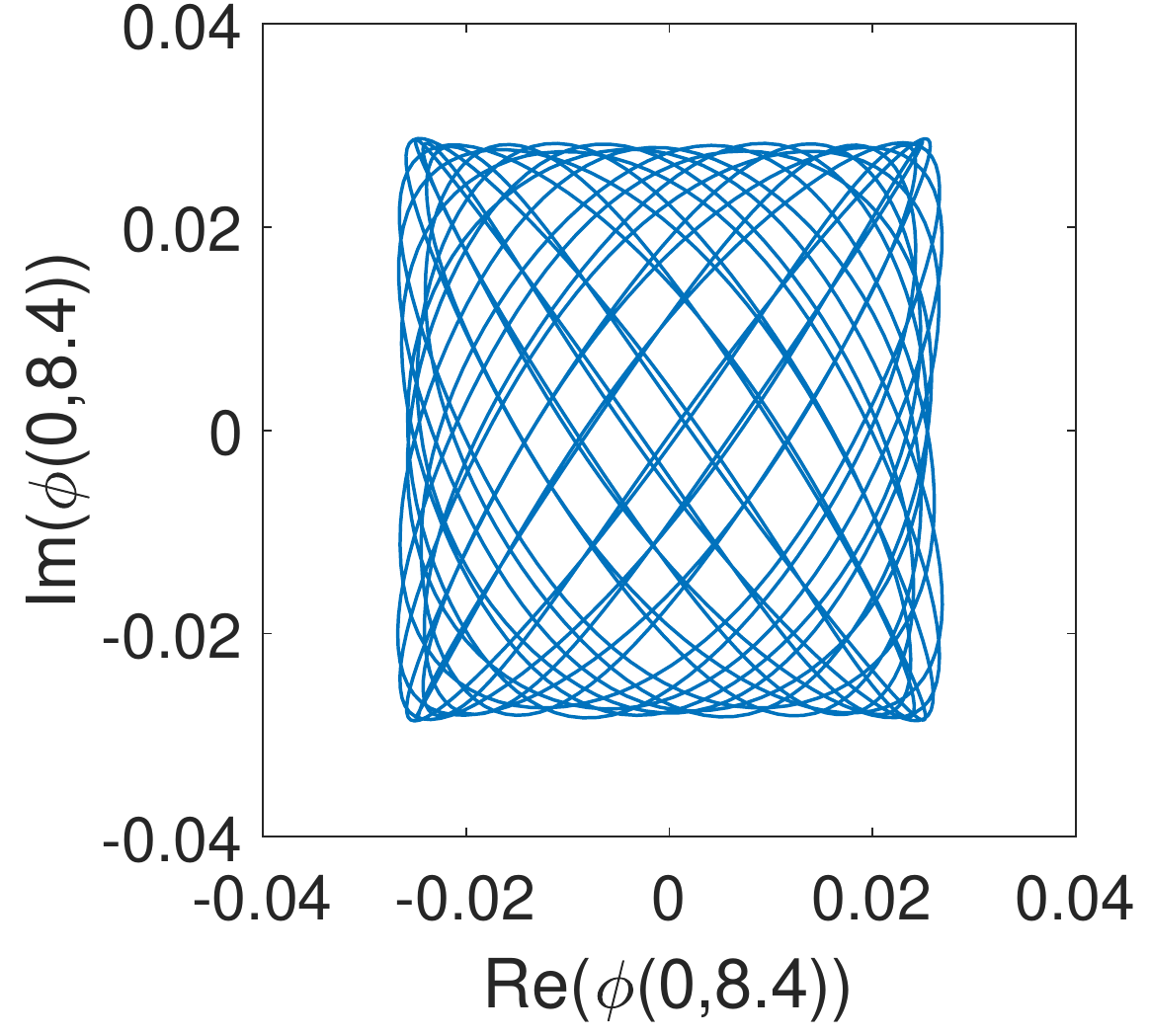}
\caption{\label{fig:phase_plot}
Phase portraits on three points on the $y$ axis in the CSQ stage. }
\end{figure}

In Fig.~\ref{fig:origin_real_period}, we see that if an elementary Q-ball and a dipole CSQ have the same energy, the oscillation period of the CSQ, defined as the period of Re$\phi(x=0,y=0)$, is greater than that of the elementary Q-ball. On the other hand, if an elementary Q-ball and a dipole CSQ have the same oscillation frequency, the energy of the elementary Q-ball is greater than that of the CSQ. In Fig.~\ref{fig:points_components_and_spectrums}, we plot the evolution of the field values and their Fourier transforms at point $(x=0,y=0)$ and point $(x=6.4,y=6.4)$. We see that most power in the spectra occurs around the first few odd times of the base frequency $\omega_0=0.82$. This is similar to the case of oscillons, the real scalar ``cousin'' of a Q-ball, where again only odd times of the base frequencies are significant in the spectra with a potential with the $\mathbb{Z}_2$ symmetry \cite{Salmi:2012ta}. In Fig.~\ref{fig:phase_plot}, we plot the phase portraits of various points on the $y$ axis within the CSQ, which carve out near-rectangle areas. When the elementary Q-balls are not prepared to be exactly in anti-phase, these portrait rectangles will be rotated around the origin by an angle proportional to the initial phase misalignment. For an elementary Q-ball, the corresponding phase portrait is a circle at every field point, and for an oscillon the corresponding phase portrait is a line at every point, so the CSQ may be considered a hybrid between the two, whose phase portraits are different at different points. \\

\noindent {\bf (3) Fast decay}

\begin{figure}[tbp]
\centering
\includegraphics[width=.45\textwidth,trim=80 260 90 260,clip]{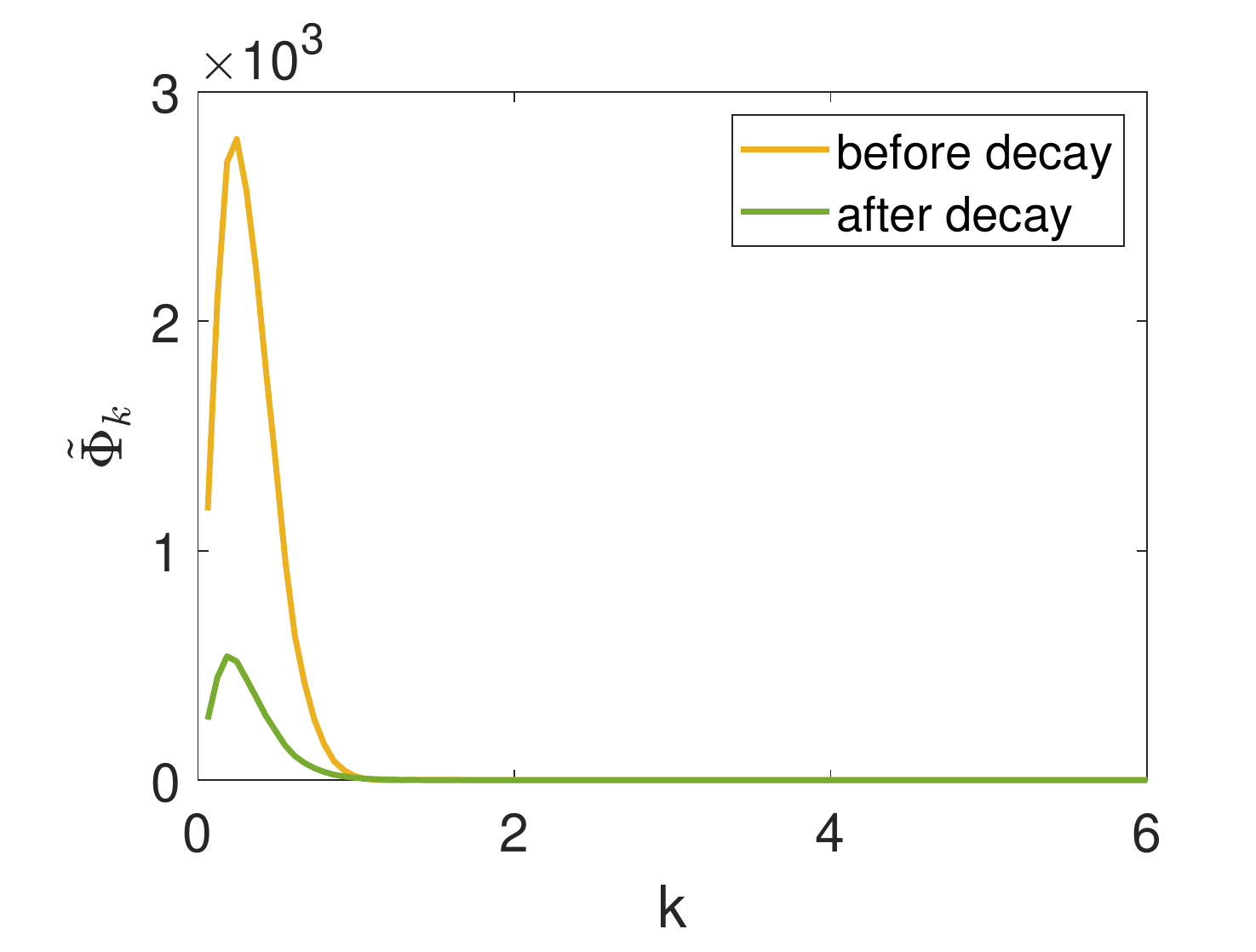}
\hfill
\includegraphics[width=.45\textwidth,trim=80 260 90 260,clip]{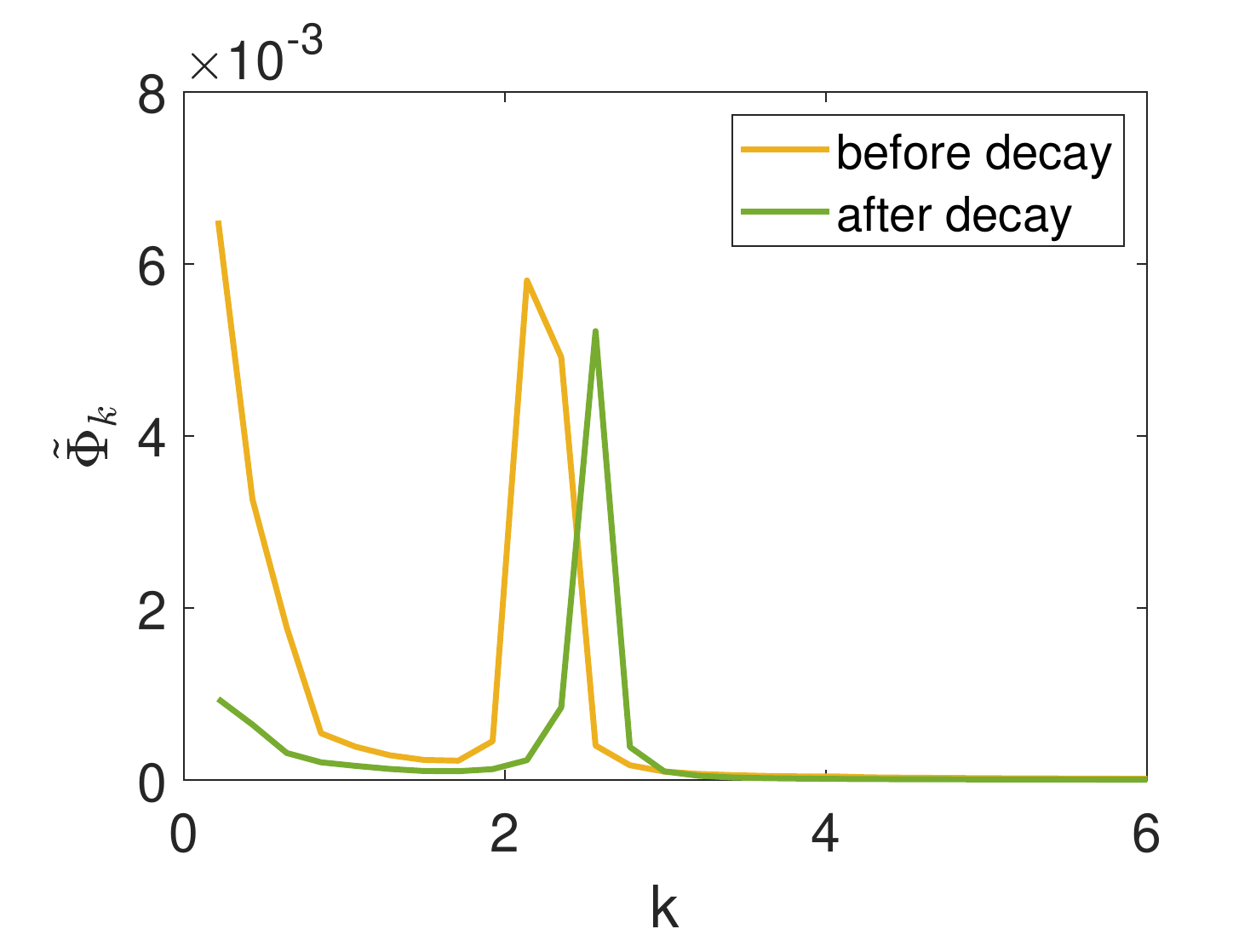}
\caption{\label{fig:space_wavenumber} Power spectrum of $\phi(\bfx)$ over the whole lattice (left plot) and over a small rectangular area outside the CSQ along the positive $x$-axis (right plot). Typical results before the decay (orange line, $t \approx 3766 T_0$) and after the decay (green line, $t \approx 5793 T_0$) are shown. The constituent elementary Q-balls with $\oi=\pm 0.86$ are initially placed at $d=4.0$ away from each other.
}
\end{figure}

In this short stage, the energy and charge of the CSQ decrease dramatically in a short period of time and the Q-ball swapping period starts to increase unboundedly, which may be taken as the end of the life of the CSQ. The significant changes during the decay process can also be seen from the power spectrum of the field $\tilde{\Phi}_k = k \int \ud k_\theta |\Phi(\bfk)|^2$ where $k=|\bfk|$ and $\Phi(\bfk) =\int \ud^2 x \phi(\bfx) e^{-i  \bfk \cdot \bfx}$. From the left plot of Fig.~\ref{fig:space_wavenumber}, we can see the CSQ power spectrum decreases significantly during the decay, while, from the right plot, we see that the second peak of the power spectrum of the outgoing waves shifts to the higher wave-numbers, and the power of the first peak at low wave-numbers decreases substantially. \\

\noindent {\bf (4) Second plateau (oscillon stage)}

In this stage, the energy becomes quasi-stable again at a second plateau, which is around half of the first plateau. After dropping a couple of orders of magnitudes in the fast decay stage, the charge in this stage decays exponentially with time. That is, in this stage, the charge has been mostly radiated away, and the remnant of the fast decay of the CSQ is essentially an oscillon, which only oscillates along one linear direction in the phase space of Re$\phi$ and Im$\phi$.

\subsection{Lifetimes of CSQs}

\label{sec:lifeDiCSQ}

\begin{figure}[tbp]
\centering
\includegraphics[width=.45\textwidth,trim=90 260 100 270,clip]{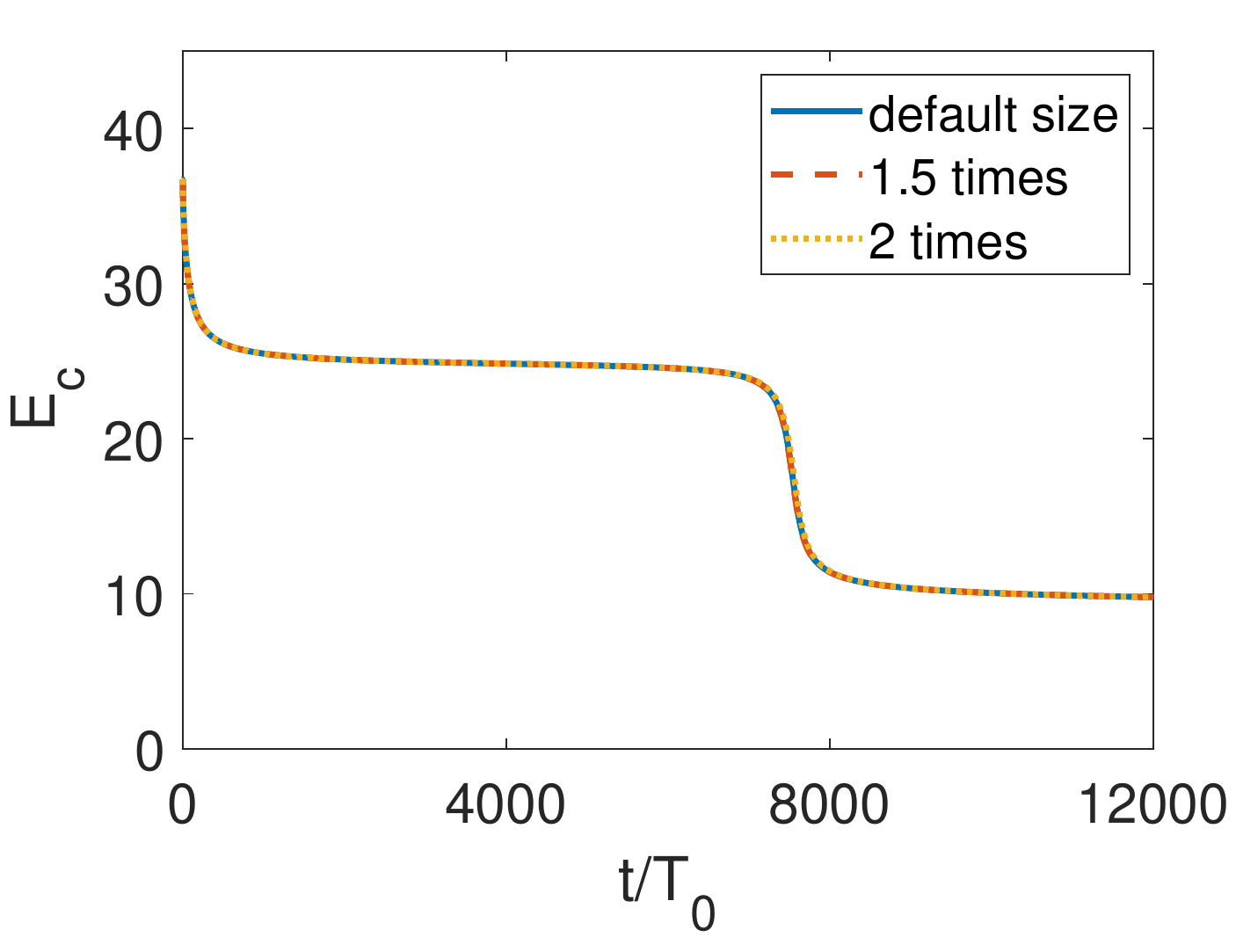}
\hfill
\includegraphics[width=.45\textwidth,trim=90 260 100 270,clip]{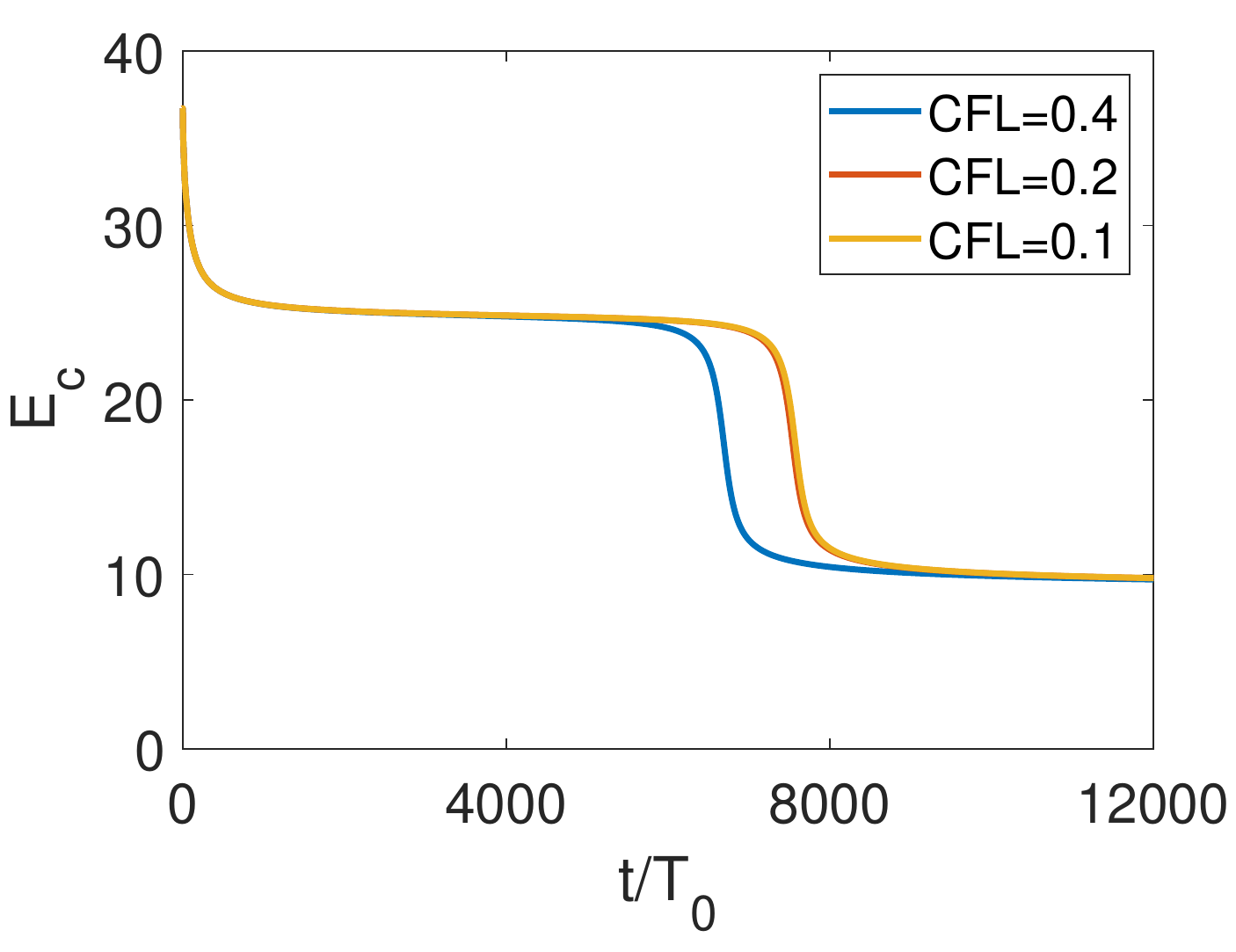}
\caption{\label{bnd_cfl}Convergence study with different physical box sizes (left) and different Courant-Friedrichs-Lewy (CFL) factors (right). In the left plot, the blue solid line is the physical box size we use for most of our simulations, which is 102.4 in each spatial direction, and the red and orange lines are $1.5$ and $2$ times of the default size respectively with $\d x$ kept at 0.2.}
\end{figure}

An important goal of this paper is to determine the lifespans of CSQs. In this subsection, we shall survey the lifespans of the dipole CSQs in 2D.

\begin{figure}[tbp]
\centering
\includegraphics[width=.5\textwidth,trim=90 260 100 270,clip]{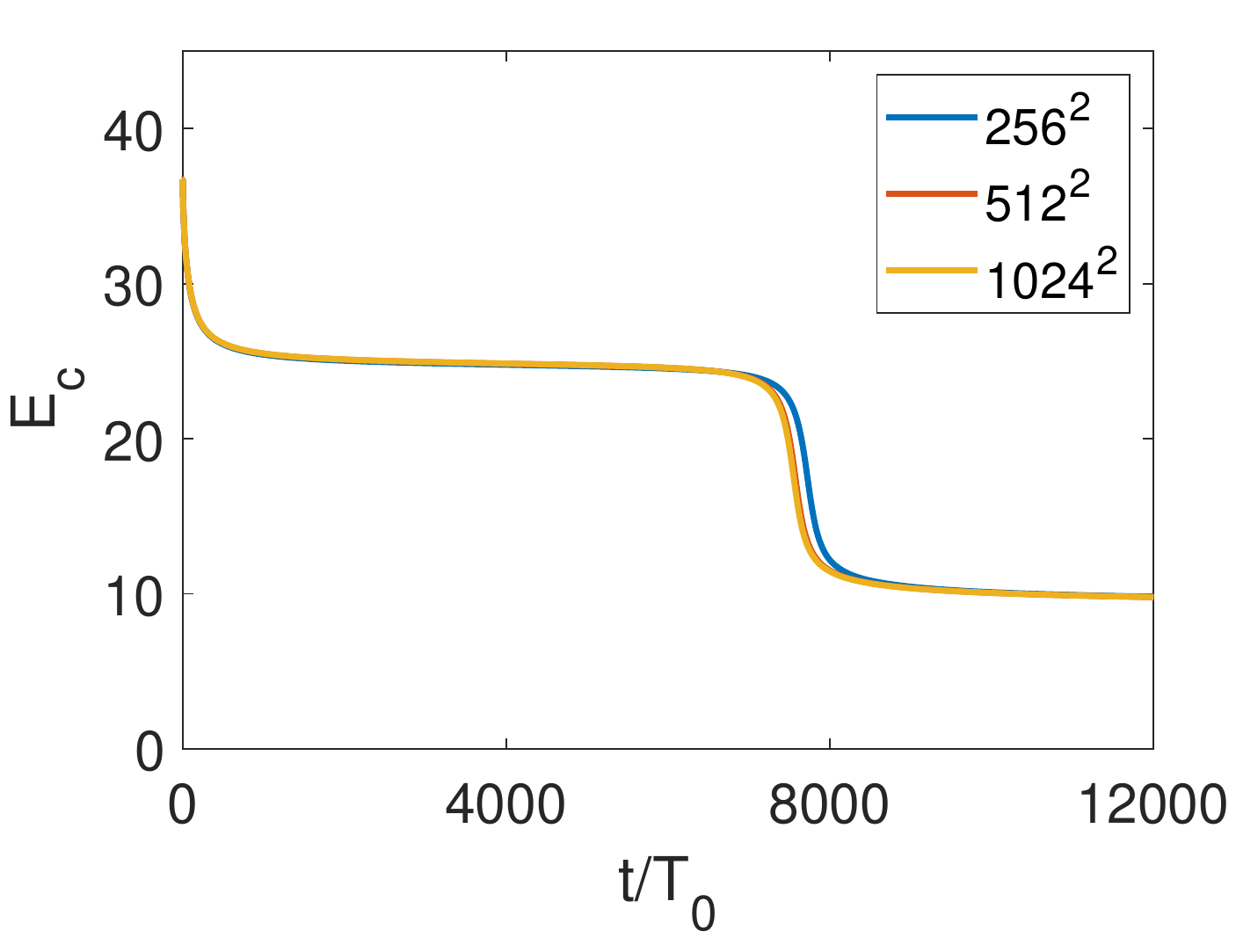}
\caption{\label{convergence_study}Convergence study with different spatial resolutions. The lifetime decreases slightly as the resolution increases.  We use a $512^2$ lattice to survey the lifespans of the CSQ, and the uncertainty due to the lattice resolution is within 1\%. The physical box size is $102.4^2$ and thus $\d x=0.2$.}
\end{figure}

Calculating the CSQ lifetimes is prone to accumulated numerical errors, as we need to run the code for very long time. So it is essential that we get all the numerical errors under sufficient control. Let us first check how the lifetimes vary with different numerical setups. From the left plot of Fig.~\ref{bnd_cfl}, we see that, thanks to the effective absorbing boundary condition, a box size of around $100/m$ is sufficient for our purposes. Also, from the right plot, we see that our choice of the CFL factor  $0.1$ is also very accurate to compute the lifetime. Our main numerical uncertainties for the lifetimes of the CSQs are from the lattice resolution $\d x$, or the number of lattice points for a fixed physical box size, which however are also under good control, as one may see  in Fig.~\ref{convergence_study}. If we define the lifetime of the CSQ stage by how long the charge-swapping period oscillates around a steady value, then the difference between the lifetimes of the $512^2$ and $1024^2$ runs is about 0.39\% in Fig.~\ref{convergence_study}. With the three resolutions, we find the convergence rate for the lifetime, due to the long evolution, is about second order. (At any given time, the differences between energies of the three resolutions always remain fourth order, as expected for the finite difference method we use.) Using the linear Richardson extrapolation, even for a second order convergence, we see that the difference between the extrapolated value and the $512^2$ run is about 0.55\%. In the following, we will survey the lifetimes in the parameter space of $\oi$ and $d$ with a $512^2$ lattice, and thus the accuracy of this survey is expected to be around 1\%.

We also need to adopt a reasonable measure to define the lifetime of a CSQ. One way, as we just used above, is to extract the length of the plateau of the CSQ stage in the $T_{\rm swap}$ evolution plot (such as in Fig.~\ref{convergence_study}). One can similarly define the lifetime from other plots, but the easily recognizable feature of the swapping period plateau in Fig.~\ref{fig:freq_vs_period_zeros} arguably provides the best measure for this purpose. After all, those objects have been dubbed charge-swapping Q-balls. So, for definiteness, we define the lifetime of a CSQ as the length of the thick bar in Fig.~\ref{fig:freq_vs_period_zeros}.

\begin{figure}[tbp]
\centering
\includegraphics[width=.5\textwidth,trim=90 260 100 260,clip]{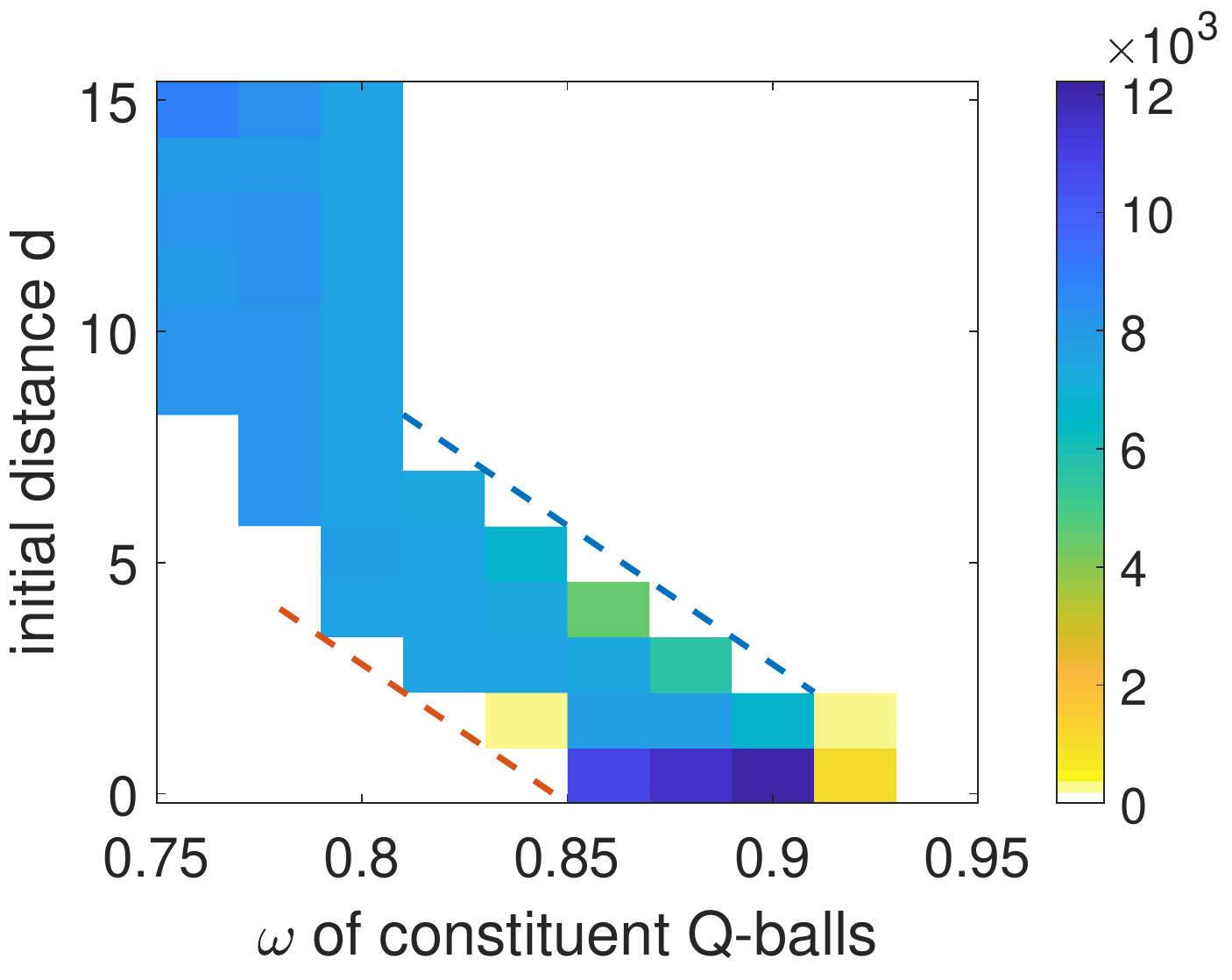}
\caption{Lifespans of a dipole CSQ with different initial parameters. The parameter space is spanned by the internal frequency $\oi$ of constituent Q-balls and the initial distance $d$ between the Q-ball centers. Lifetimes are shown in units of $T_0= 2\pi/m$. The white regions are where CSQ can not form. The two dashed guiding lines are $d+60\oi=50.8$ and $d+60\oi=56.8$ respectively.}
\label{parameter space}
\end{figure}

Fig.~\ref{parameter space} is the survey of the lifespans of the dipole CSQ in 2+1D in the parameter space of $\oi$ and $d$, where $\oi$ is the oscillating frequency of the constituent elementary Q-balls and $d$ is the initial separation between them.  We see that they can form in the diagonal strip of the parameter space, roughly between the lines of $d+60\oi=(50.8,56.8)$. That is, if the constituent Q-ball has lower frequencies, the separation between the constituents should be greater. When CSQs do form, as we see in Fig.~\ref{ec_same_value}, they evolve to some universal values of $E_c$ and $Q_s$, and their lifetimes are typically of the order of a few thousands of swapping periods, or tens of thousands in terms of $1/m$. The longest lifetime is observed at when $\oi\simeq 0.9$, $d\simeq 0.4$. Of course, not every set of parameters leads to the formation of a CSQ. When trying to prepare a CSQ with initial elementary Q-balls placed too close to each other, they will violently repel each other and scatter. On the other hand, if they are placed too far away, they will attract and pass through each other and scatter. 

\begin{figure}[tbp]
\centering
\includegraphics[width=.5\textwidth,trim=90 260 110 280,clip]{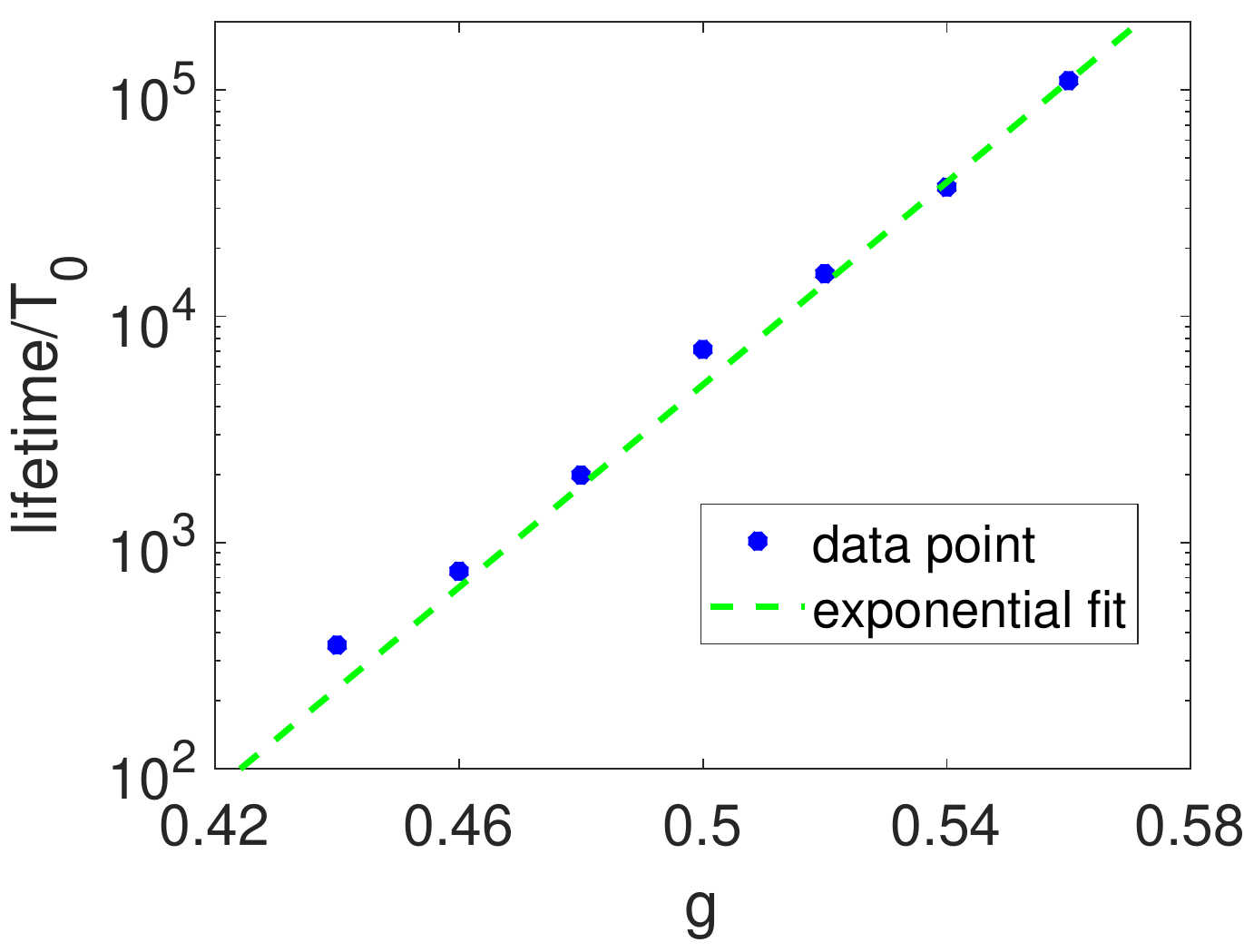}
\caption{\label{fig:gLifetimeAndFitting}
Dependence of the dipole CSQ lifetime on the coupling constant $g$. The green dashed line is the  exponential fit $a_0e^{b_0g}$ with $a_0=3.23\times10^{-8}, b_0=51.5$. The data points are obtained by superposing two Gaussian lumps with an oscillating frequency $\oi=0.84$ and with the same peak amplitude and width as those of an elementary Q-ball with $\oi=0.84$ and $g=0.5$, for more equal comparison. We vary distance $d$ to get the maximum lifetime for a given $g$.}
\end{figure}

If one wants to prolong the lifetimes of CSQs, changing the Q-ball frequency $\oi$ and the initial distance $d$ is not the most efficient way. In our fiducial model (\eref{Lagrangian}), the only theory parameter is the coupling of the $\phi^6$ term $g$. In Fig.~\ref{fig:gLifetimeAndFitting}, we show how the lifetime will vary with $g$. We see that the lifetime increases at least exponentially when $g$ is increased. There are also some other effects caused by increasing $g$: 1) the plateau CSQ energy $E_c$  decreases slightly; 2) the CSQ charge $Q_s$ decreases; 3) the CSQ swapping period increases.

We also find that the initial relative phase of the two constituent Q-balls does not significantly affect the lifetime of CSQs. It mainly changes the relative magnitude of the real and imagery part of the field $\phi$. For this reason, in the above we set the initial relative phase of the two constituent Q-balls to be zero.

\subsection{Attractor behaviors of CSQs}
\label{sec:CSQisAttractor}

\begin{figure}[tbp]
\centering
\includegraphics[width=.45\textwidth,trim=90 260 110 270,clip]{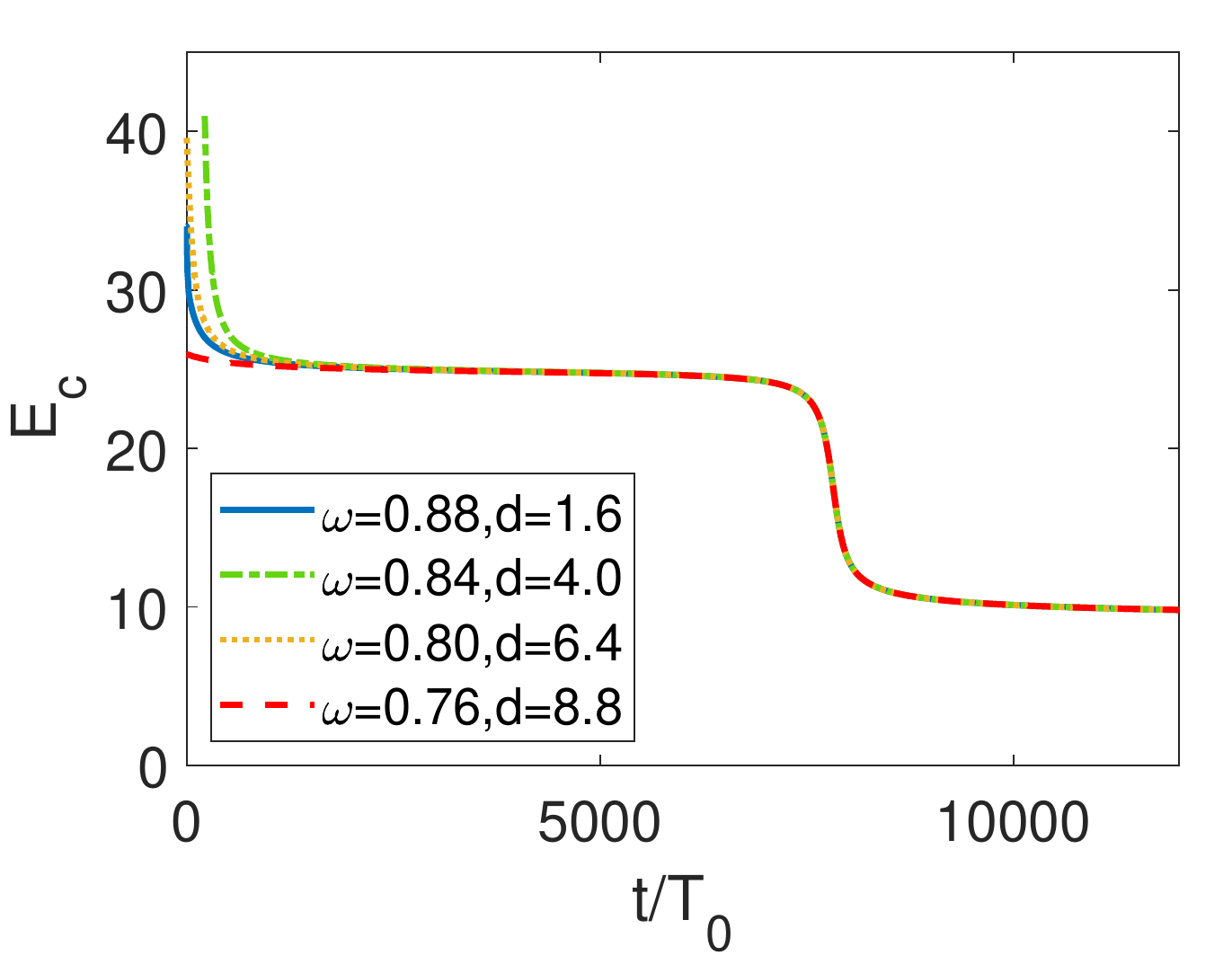}
\hfill
\includegraphics[width=.45\textwidth,trim=90 260 110 270,clip]{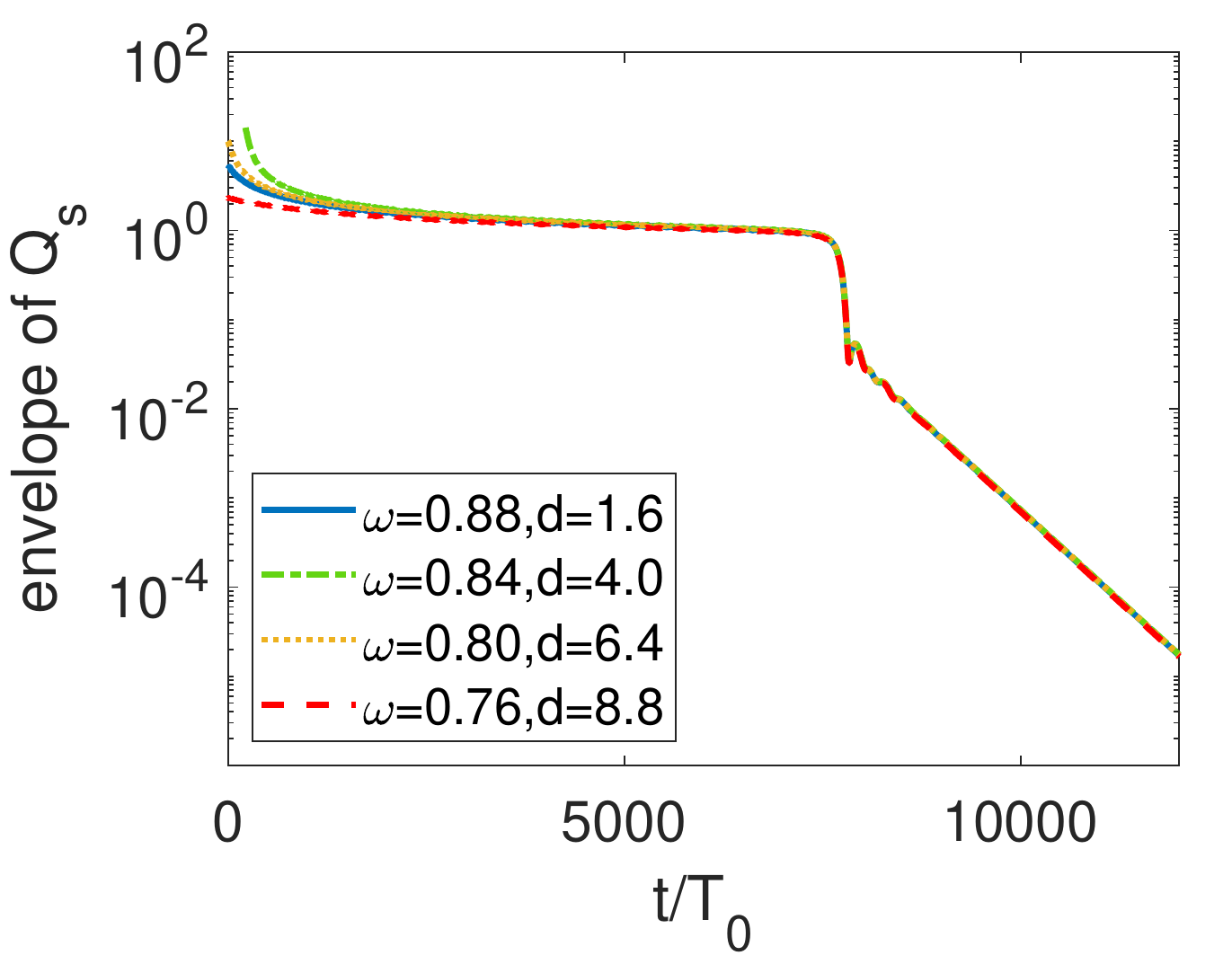}
\caption{\label{ec_same_value}Attractor behaviors of the dipole CSQ in the $E_c$ and envelope of $Q_s$ plot. The initial configurations are constructed by superimposing elementary Q-balls.}
\end{figure}

As quasi-stable solitons, CSQs must be attractor solutions, that is, their formation should be insensitive to initial conditions: if there are favorable but otherwise quite generic initial conditions, they can form spontaneously. Arguably, this is what makes them relevant in many physical circumstances. In this section, we will explore this aspect of the dipole CSQs.

\begin{figure}[tbp]
\centering
\includegraphics[width=.5\textwidth,trim=90 260 110 270,clip]{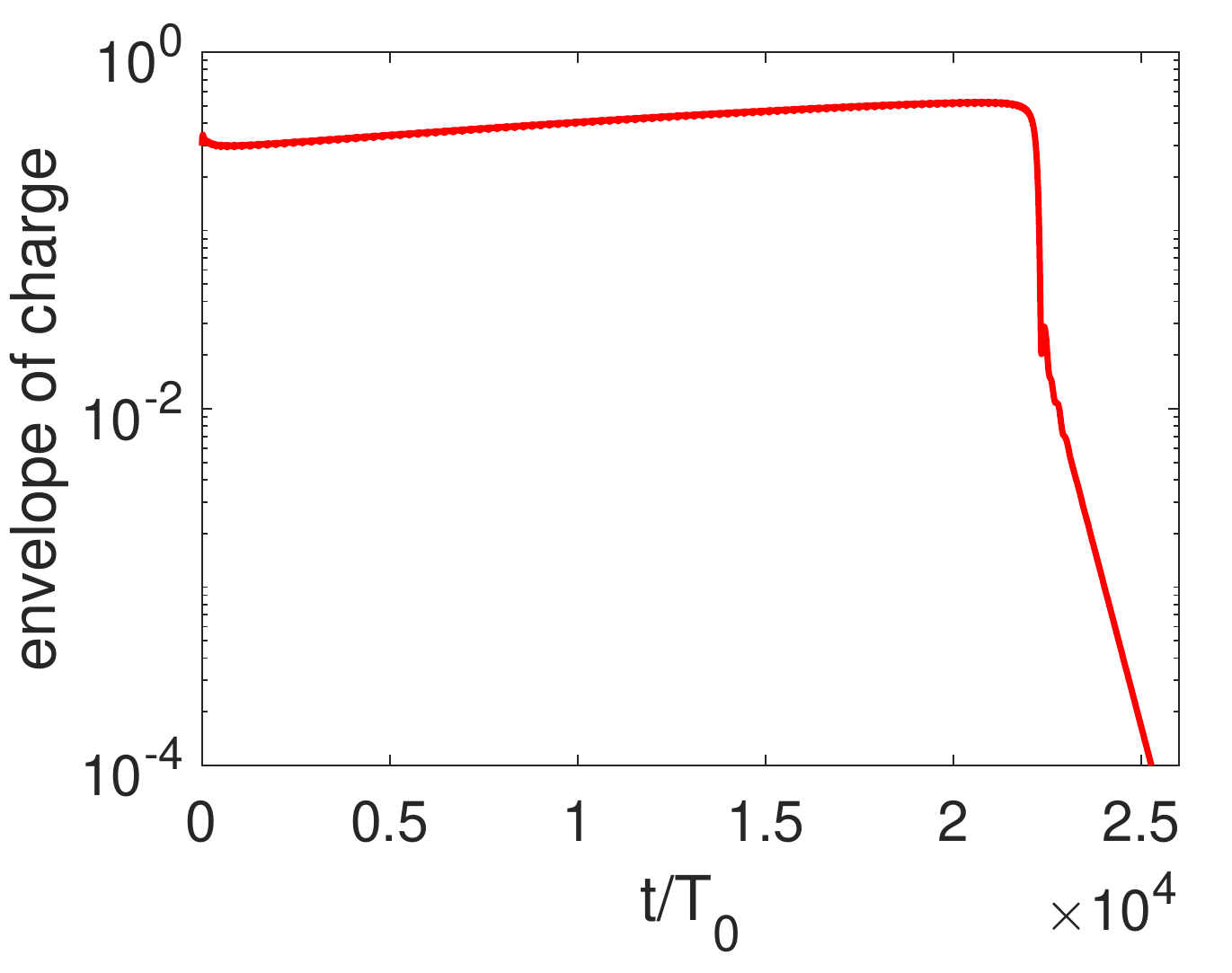}
\caption{\label{fig:other_phenomena_of_charge}Evolution of the envelope of $Q_s$ for the case where $\oi=0.88$, $d=0.1$. In this case, the initial amplitude of $Q_s$ is smaller than the steady value of the CSQ, and the amplitude of $Q_s$ will increase from $0.29$ to $0.53$ before the decay of the CSQ. This leads to a longer lifetime.}
\end{figure}

In Fig.~\ref{ec_same_value}, we see that for different initial $\oi$ and $d$ the CSQ energy $E_c$ and charge $Q_s$ evolves to the same trajectories after the short relaxation. This is why in Fig.~\ref{parameter space} we see that once CSQs are formed, their lifetimes are usually quite similar to each other. On the other hand, if the initial constituent Q-balls are (slightly) too close to each other, for $0.86\lesssim\omega\lesssim0.92$, the envelope of $Q_s$ will steadily increase to that of the quasi-stable CSQ, which leads to longer lifetimes; see Fig.~\ref{fig:other_phenomena_of_charge}.

\begin{figure}[tbp]
\centering
\includegraphics[width=.45\textwidth,trim=90 260 100 270,clip]{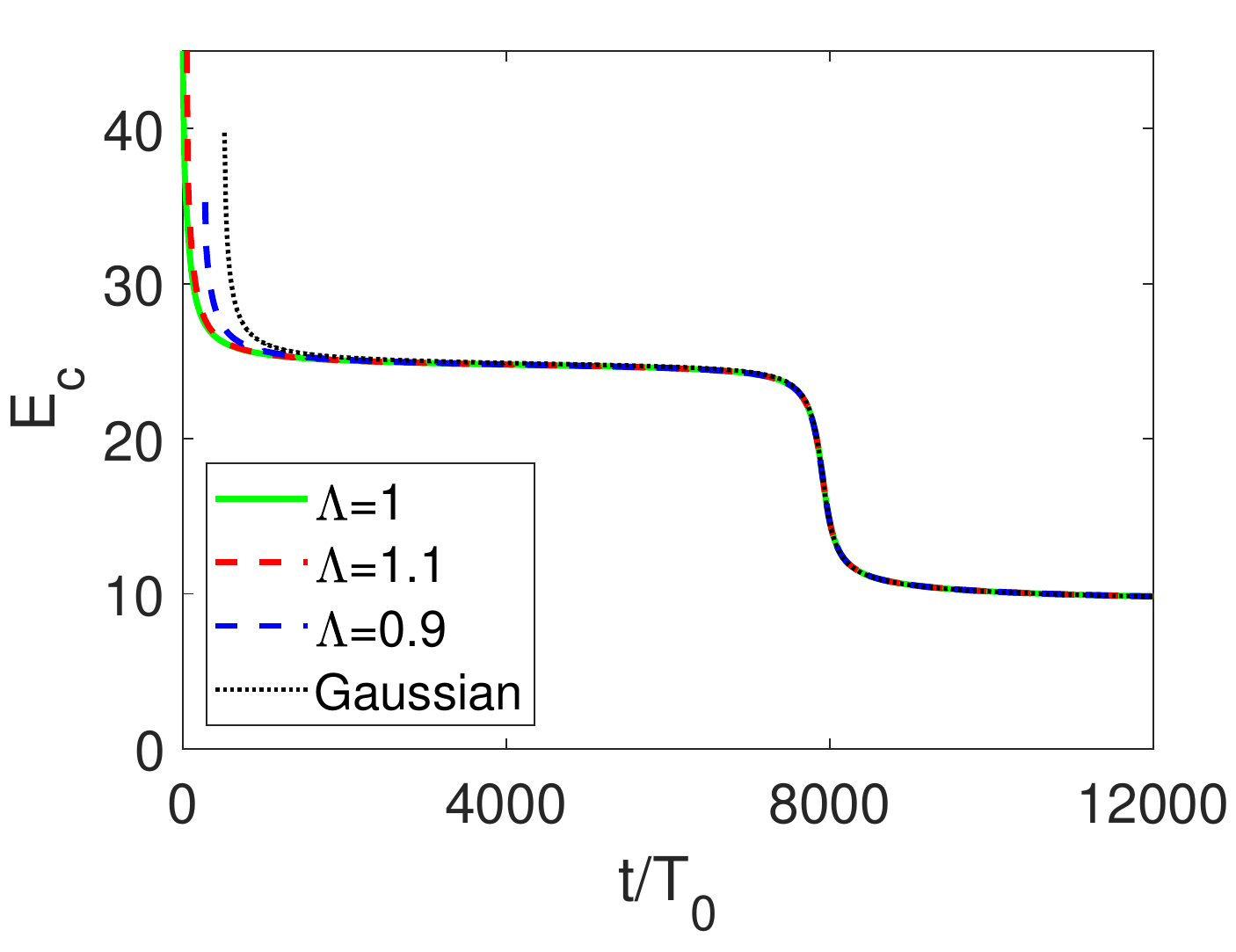}
\hfill
\includegraphics[width=.45\textwidth,trim=90 260 100 270,clip]{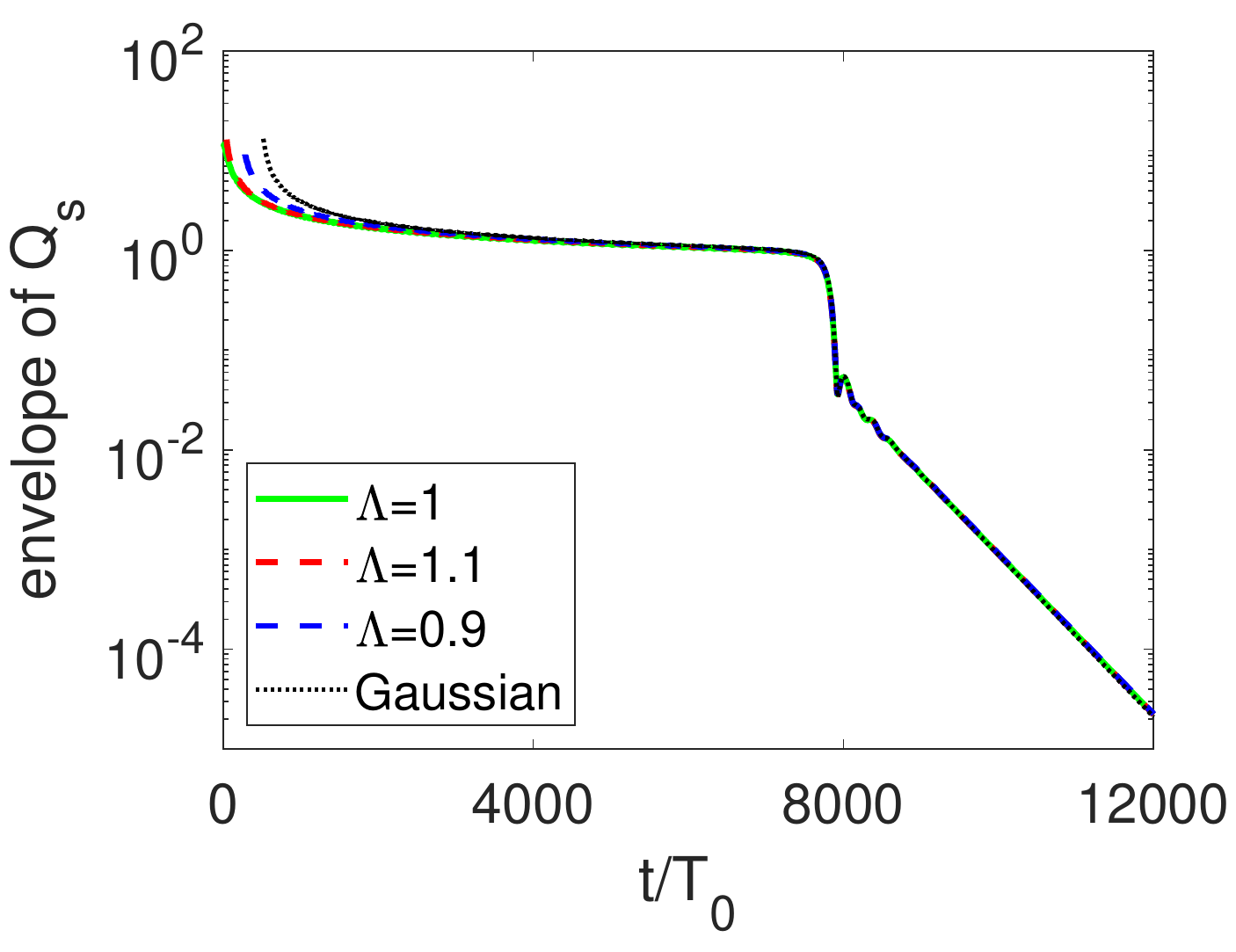}
\caption{\label{fig:deformed_EQ}
Evolution of energy $E_c$ and charge $Q_s$ for deformed CSQs. The deformation is done by multiplying the profile $f(r)$ of the constituent elementary Q-ball by a factor of $\Lambda$, $\Li=1$ being the un-deformed elementary Q-ball. Different cases are shifted to match the fast decay stages. We see that different initial configurations are attracted to the same quasi-stable CSQ configuration. }
\end{figure}

We have so far only superimposed elementary Q-balls to prepare the CSQs. The attractor nature of the CSQs means that one should also be able to use other configurations to prepare CSQs. For example, we can prepare the CSQs with oscillating Gaussian lumps
\be
\phi(t,\bfx)=A e^{-{(\bfx-\bfx_0)^2}/{\si^2}}e^{i \oi t}  ,
\ee
or deformed Q-balls
\be
\phi(t,\bfx) = \Lambda f(|\bfx-\bfx_0|)e^{i \oi t}  ,
\ee
where $A, \si, \Lambda$ are constants and $f(|\bfx-\bfx_0|)$ is the elementary Q-ball profile. We find that if they are sufficiently close to the elementary Q-balls, CSQs can form. Indeed, these configurations evolve to the same trajectory as those prepared with elementary Q-balls, although their lifespans are slightly shorter; see Fig.~\ref{fig:deformed_EQ}. 

Of course, if they are too different from Q-balls, CSQs can not form. Take the deformed Q-balls above for example. If $\Lambda$ is too big, the two deformed Q-balls will repel without forming a CSQ, while if $\Lambda$ is too small, the configuration will quickly shed away energy and charge and then decay to an oscillon without having the CSQ plateau.  

Here we have focused on the evolution of $E_c$ and $Q_s$ for different initial conditions, and we demonstrated an attractor behavior. One can also verify that similar attractor behavior can be found in the evolution of the charge-swapping frequencies, field-oscillating frequencies inside the CSQs and so on.

\section{Higher multipole CSQs in 2+1D}

\label{sec:multipleCSQ}

\begin{figure}[tbp]
\centering
\includegraphics[width=0.8\textwidth,trim=40 290 10 260,clip]{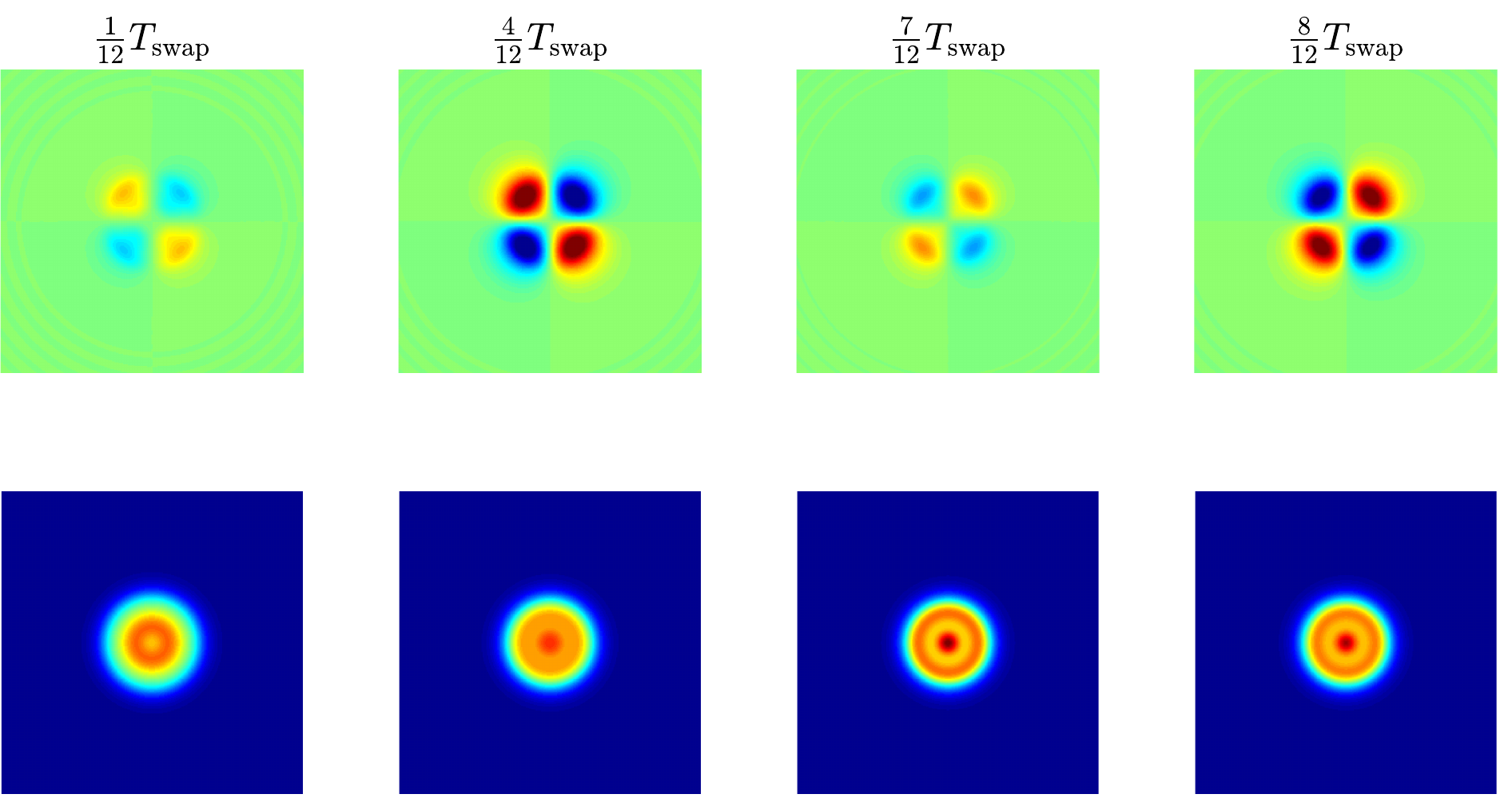}
\includegraphics[width=.8\textwidth,trim=40 290 10 260,clip]{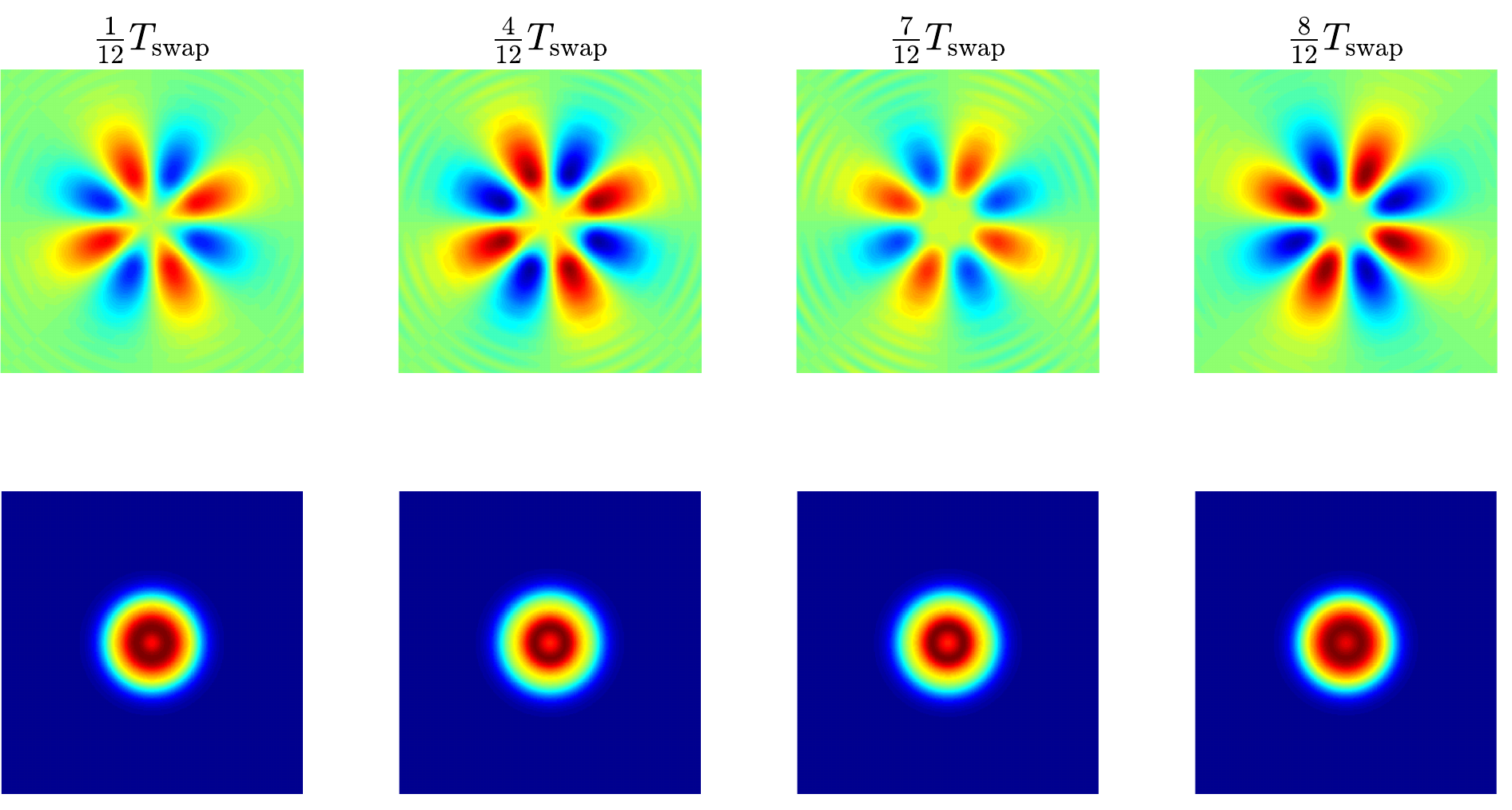}
\caption{\label{fig:four_and_eight_Q_balls}Evolution sequences of the charge and energy density of a quadrupole and octupole CSQs in 2+1D. In the first and third rows, the red color depicts positive charges and the blue color depicts negative charges. The corresponding energy density sequences are displayed in the second and fourth rows, with the blue depicting lower density and the red depicting higher density.}
\end{figure}

The dipole CSQs are the simplest ones one can construct. As shown in \cite{Copeland:2014qra}, there are also more complex, higher multipole CSQs. In this section, we shall only briefly touch on these CSQs. In particular, we will look at the basic features and lifetimes of the quadrupole and octupole CSQs with equal charges, as shown in Fig.~\ref{fig:four_and_eight_Q_balls}. There are many possible configurations for higher multiple CSQs, a full characterization of which is certainly interesting but is beyond the scope of this paper.

Higher multiple CSQs can be prepared analogously as the dipole case. In Fig.~\ref{fig:four_and_eight_Q_balls}, we place relevant numbers of equal and opposite Q-balls next to each other with no phase differences, and then we see, with time, the positive charge will turn negative and the negative charge will turn positive. That is, the charges are swapping with neighboring charges, rather than swapping with the opposite part with respect to the origin, which are the same kind of charges in the quadrupole and octupole CSQs case. This is due to the limitation of 2 spatial dimensions, and we will see that in 3+1D, we can construct higher multipole CSQs with charges swapped between the opposite parts. We see from Fig.~\ref{fig:four_and_eight_Q_balls} that the energy density of higher multipole CSQs are mostly spherically symmetric. The charge swapping periods for the quadrupole and octupole CSQ are about $29/m$, while in the dipole case it is about $36/m$.

\begin{figure}[tbp]
\centering
\includegraphics[width=.5\textwidth,trim=90 260 110 270,clip]{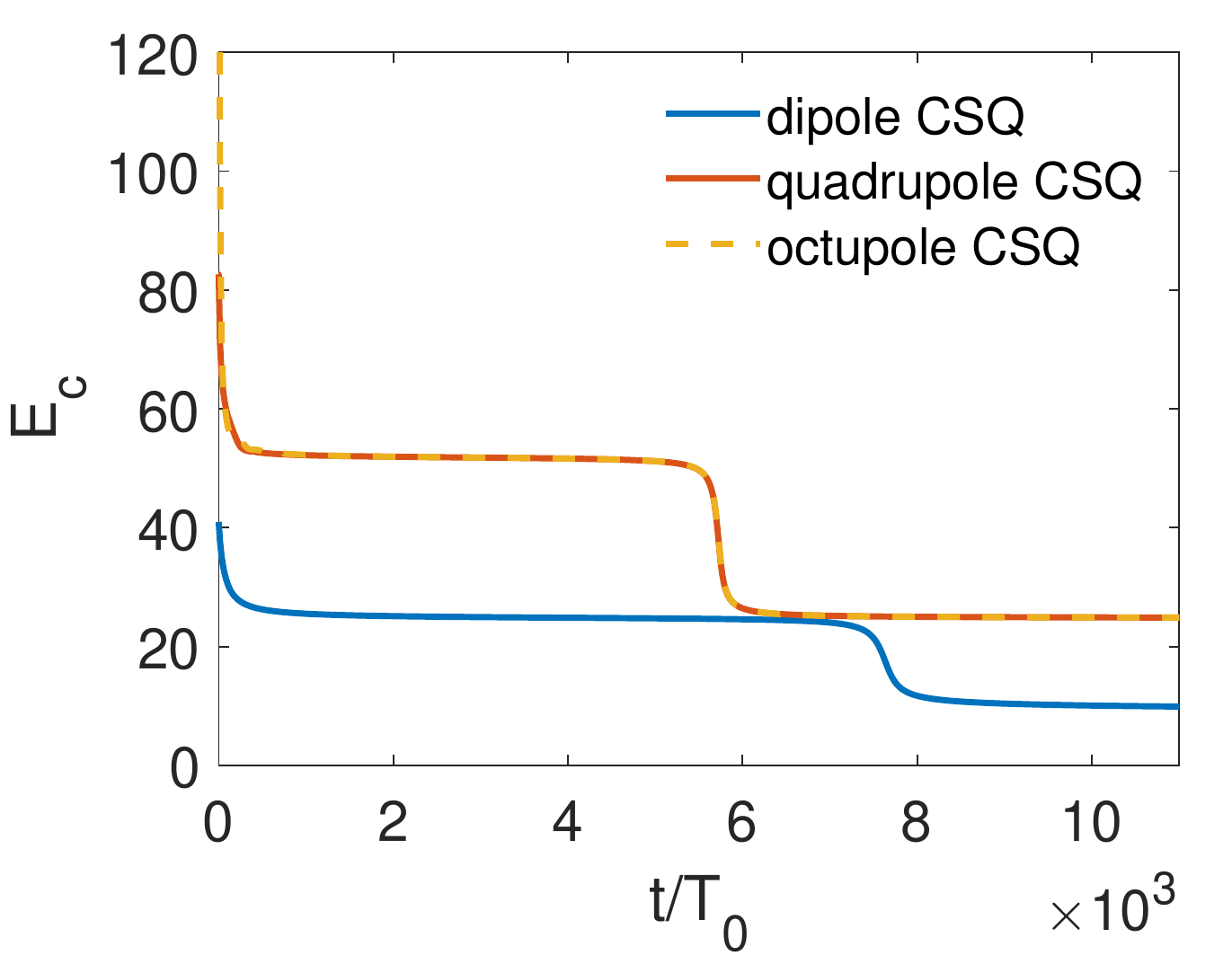}
\caption{\label{fig:energy_of_more_complex_csq}Energy evolution of different multipole CSQs. 
The initial distances between constituent Q-balls and the coordinate origin are: $2$ for the dipole CSQ, $4$ for the quadrupole CSQ and $8$ for the octupole CSQ, in units of $1/m$. The frequencies of constituent Q-balls are all chosen to be $\oi=0.84$.
}
\end{figure}

\begin{figure}[tbp]
\centering
\includegraphics[width=.45\textwidth,trim=90 260 110 280,clip]{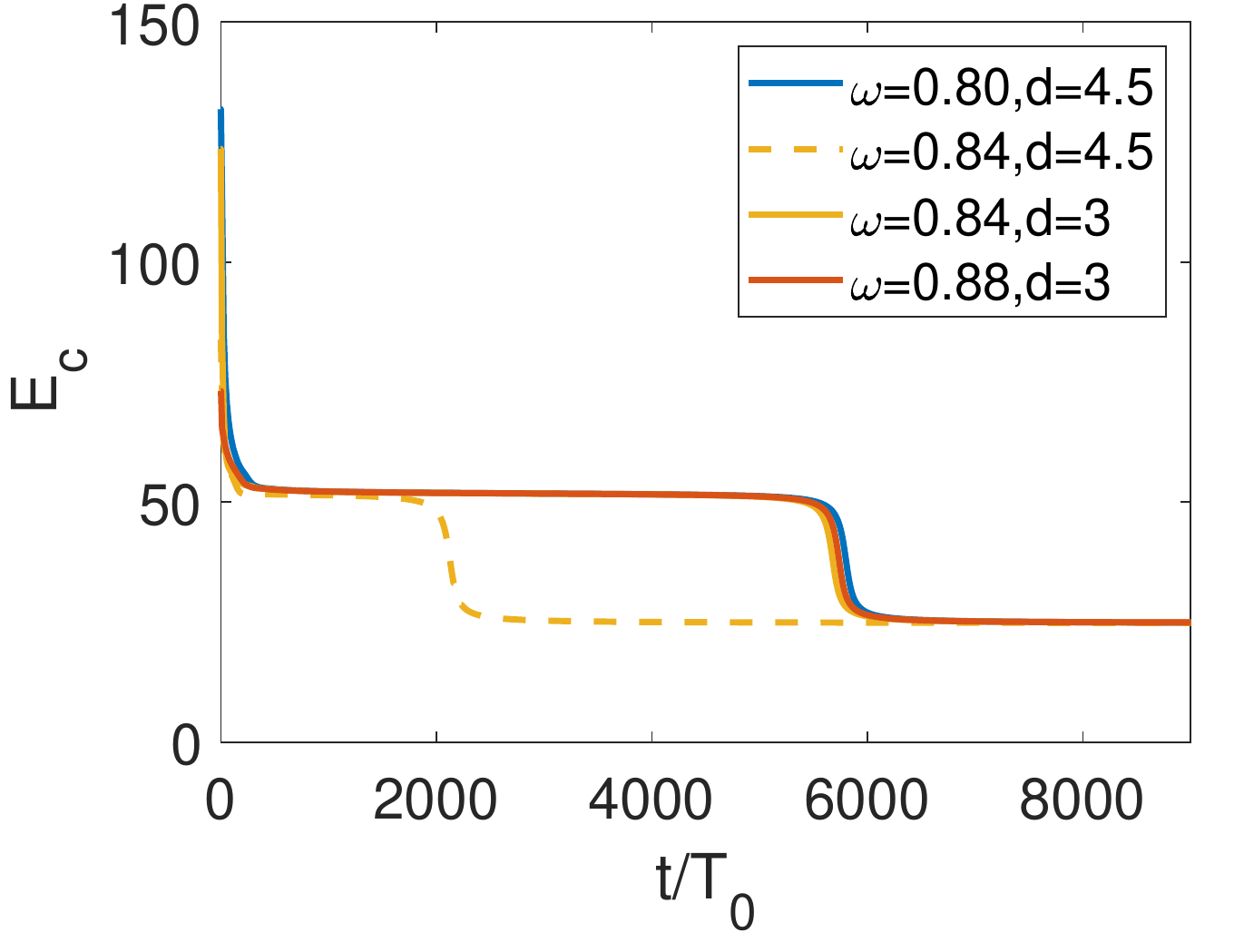}
\hfill
\includegraphics[width=.45\textwidth,trim=90 260 110 280,clip]{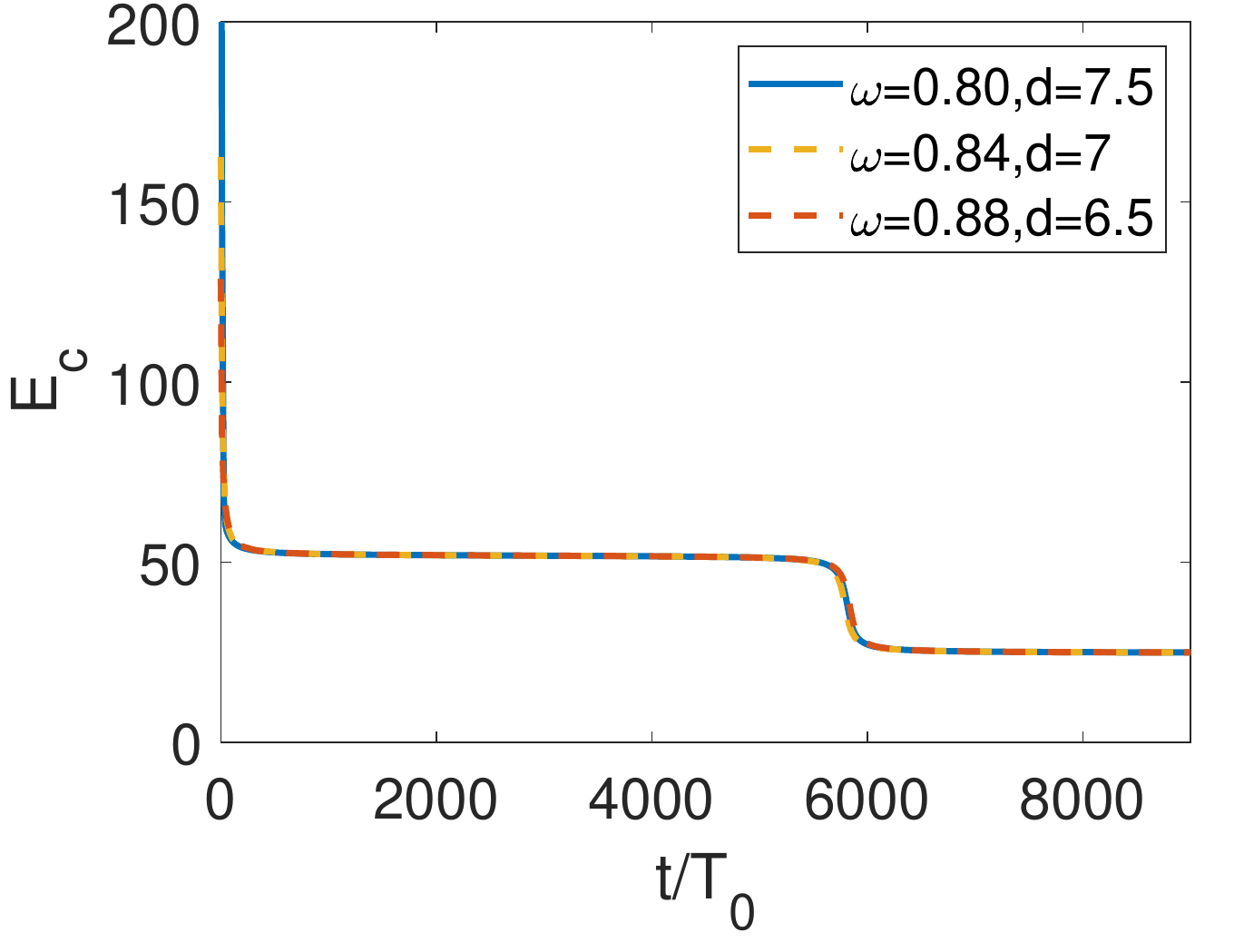}
\caption{\label{fig:wd_lifetime_complex_CSQ}
Lifetimes of quadrupole (left) and octupole (right) CSQs with different initial $\oi$ and $d$. Lifetimes are shown in units of $T_0=2\pi/m$. }
\end{figure}

The evolution of energy $E_c$ of the quadrupole and octupole CSQ are plotted, together with the dipole CSQ, in Fig.~\ref{fig:energy_of_more_complex_csq}.  Similar to the dipole case, the energy initially drops rapidly before coming to a plateau, which lasts for thousands of oscillation periods, and then quickly decays to a second plateau. The quadrupole and octupole CSQ share almost the same duration of lifetime, which is less than the lifetime of the dipole CSQ. Also, we see that the first $E_c$ plateau of the quadrupole and octupole CSQ is about twice that of the first $E_c$ of the dipole CSQ, which interestingly is around the level of the second plateau of the quadrupole and octupole CSQ. 
Fig.~\ref{fig:wd_lifetime_complex_CSQ} shows how the lifetime of CSQ changes with different initial Q-ball frequency $\oi$ and distance from the origin $d$. Roughly speaking, along the diagonal strip of Fig.~\ref{parameter space}, higher multipole CSQs share similar lifespans, while away from the diagonal strip (i.e., the case of $\oi=0.84,d=4.5$), the lifespans can be much shorter.
 
 \begin{figure}[tbp]
\centering
\includegraphics[width=.5\textwidth,trim=90 260 110 270,clip]{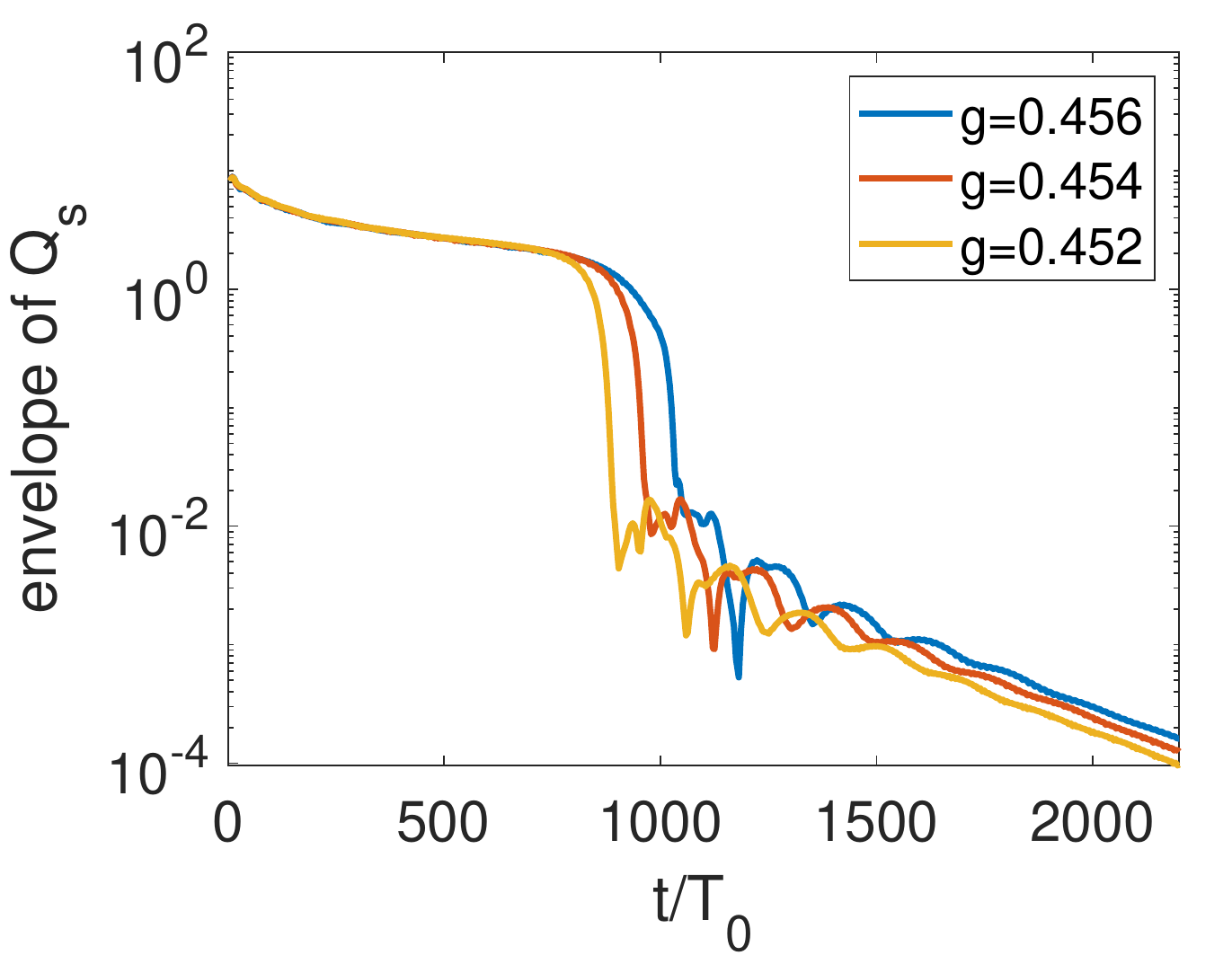}
\caption{\label{fig:plateaus_of_4_CSQ} Charge plateau of a quadrupole CSQ. 
The other parameters are set to be $d=3$ and $\oi=0.84$. 
}
\end{figure}

However, for the model with $g=1/2$, the charges of the quadrupole and octupole CSQ decrease much faster than those in the dipole CSQ, so even in the first plateau the charges in these higher multipole CSQs are very small at late times, due to the fast exponential decay. Nevertheless, fully fledged higher multipole CSQs do exist for different $g$ or with different forms of the potential \cite{HSXZ}. For example, in Fig.~\ref{fig:plateaus_of_4_CSQ}, we see that for a smaller $g$ the first charge plateau does form at least for quadrupole CSQs.  As we have seen in Section \ref{sec:lifeDiCSQ}, this is surprising as reducing $g$ increases the amplitude of the charge densities, thus increasing the charge to energy ratio. Note that here $Q_s$ is defined only in the spirit of Fig.~\ref{fig:integration range}: It is defined as the same charge in one sector of the CSQ, that is, a quadrant for the quadrupole Q-balls CSQ. 

\begin{figure}[tbp]
\centering
\includegraphics[width=.5\textwidth,trim=90 260 110 270,clip]{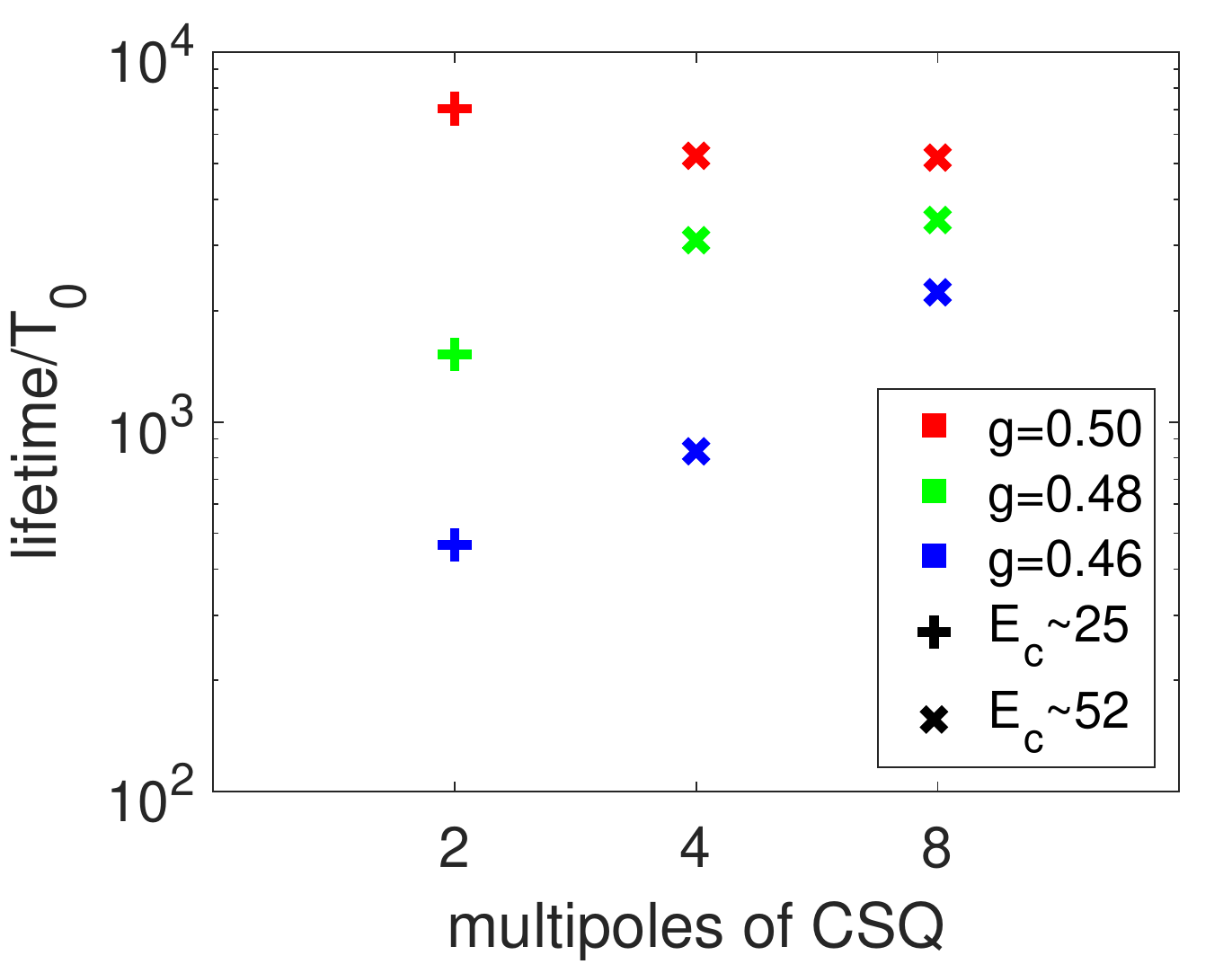}
\caption{\label{fig:g_and_complex_CSQ}
Lifetimes of CSQs for different couplings ($g$) and multipoles. Lifetimes are shown in units of $T_0=2\pi/m$.
}
\end{figure}

In Fig.~\ref{fig:g_and_complex_CSQ}, we show how the CSQ lifetime changes with coupling $g$ and  its multipole. Note that for smaller $g$ the higher multipole CSQs actually have longer lifetimes, and it appears that the lifetimes converge to the same value at high multipoles for different $g$. (Note that here the lifetime is directly extracted from the energy $E_c$ plateaus.)

\section{CSQs in 3+1D}
\label{sec:csq_in_3d}

\begin{figure}[tbp]
\centering
\includegraphics[width=.5\textwidth,trim=90 260 90 270,clip]{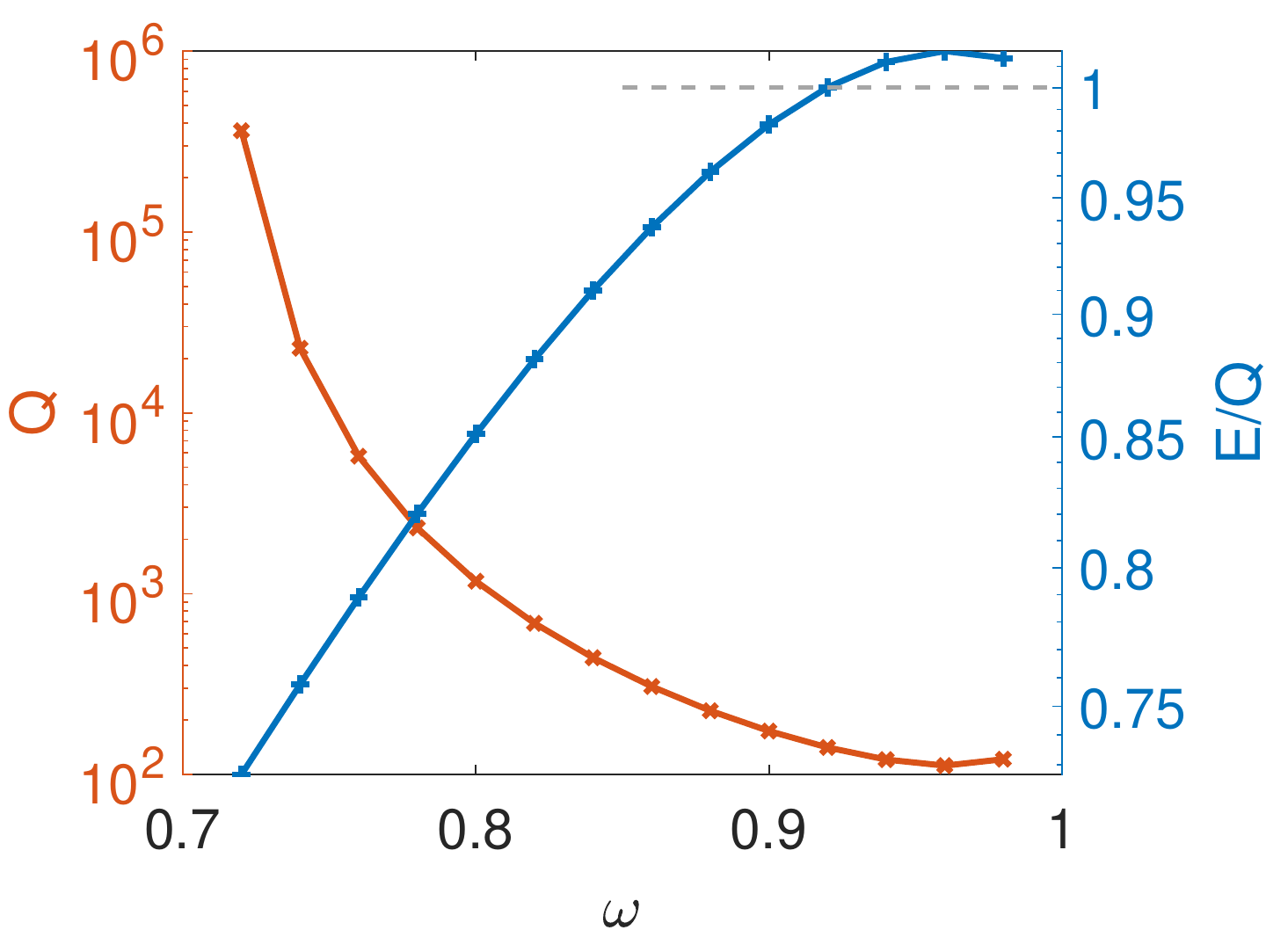}
\caption{\label{fig:PropertiesCSQin3D}
Dependence of total charge $Q$ and energy-charge ratio $E/Q$ on frequency $\oi$ of an elementary Q-ball in 3+1D in the potential (\ref{Veffphi}). 
}
\end{figure}

In this section, we shall briefly investigate CSQs and their lifetimes in 3+1D. We shall still construct the CSQs with elementary Q-balls. See Fig.~\ref{fig:PropertiesCSQin3D} for the dependence of the total charge and the ratio between the total energy and the total charge on the oscillating frequency for 3+1D elementary Q-balls in the potential (\ref{Veffphi}). From Fig.~\ref{fig:PropertiesCSQin3D}, we see that Q-balls become unstable when frequency $\oi\gtrsim 0.92$ as their $E/Q>1$.

\begin{figure}[tbp]
\centering
\includegraphics[width=.5\textwidth,trim=90 260 110 280,clip]{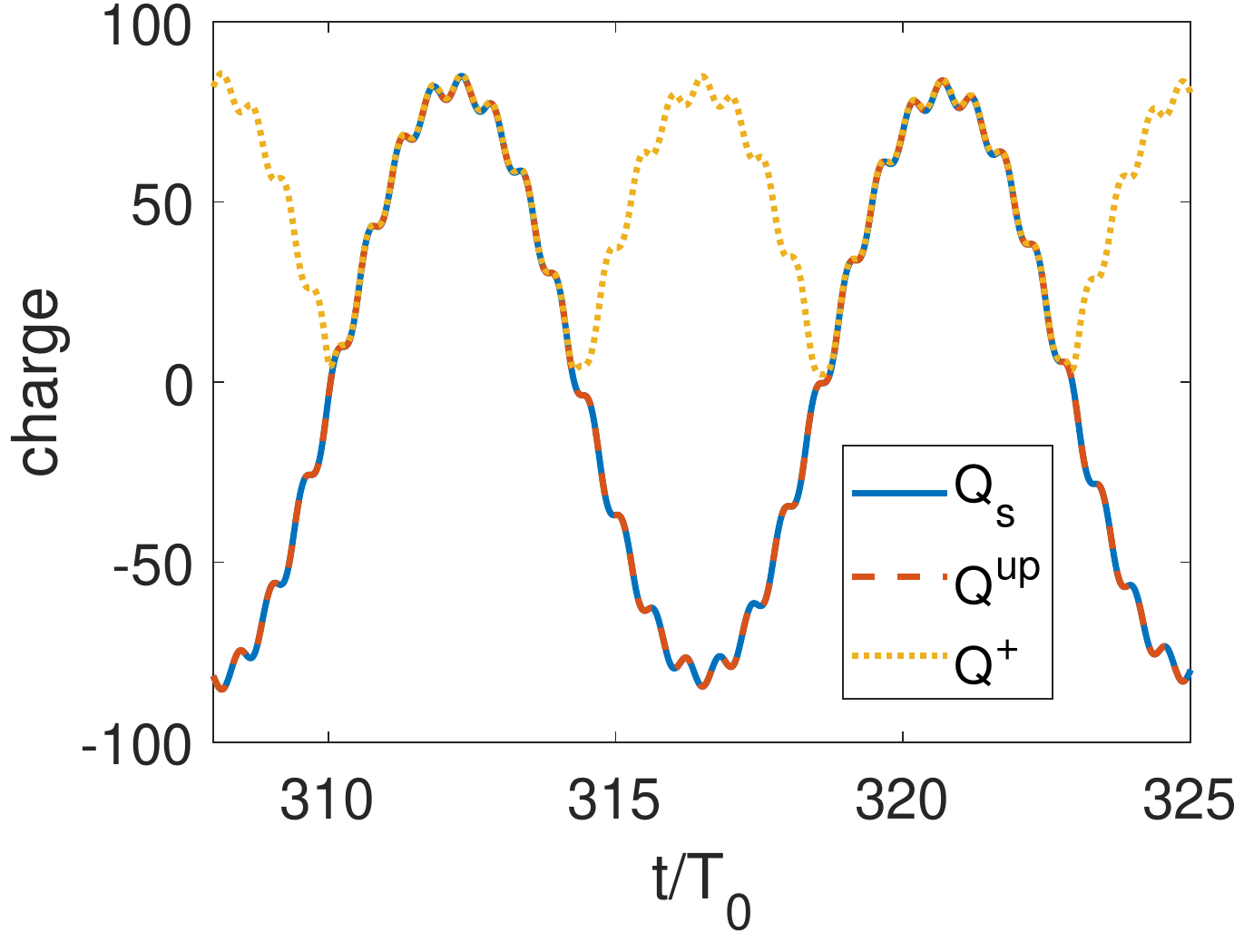}
\caption{\label{fig:swapCSQ_3D}
Charge swapping patterns of a dipole CSQ in 3+1D. $Q_s$, $Q^{\rm up}$ and $Q^+$ are defined analogously to those in Section \ref{sec:CSQs}; see Fig.~\ref{fig:integration range}. 
}
\end{figure}

\begin{figure}[tbp]
\centering
\includegraphics[width=.45\textwidth,trim=90 260 110 270,clip]{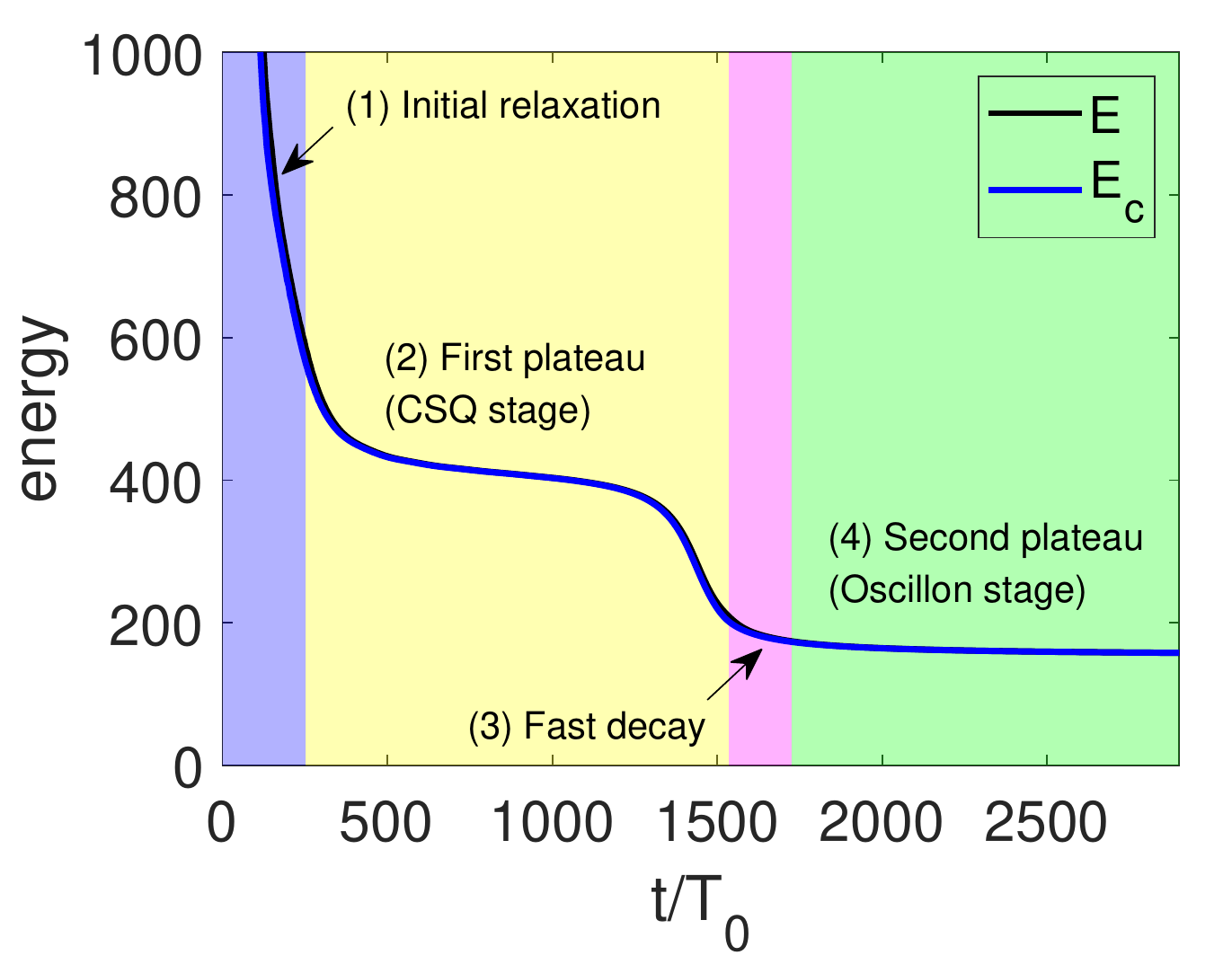}
\hfill
\includegraphics[width=.45\textwidth,trim=90 260 110 270,clip]{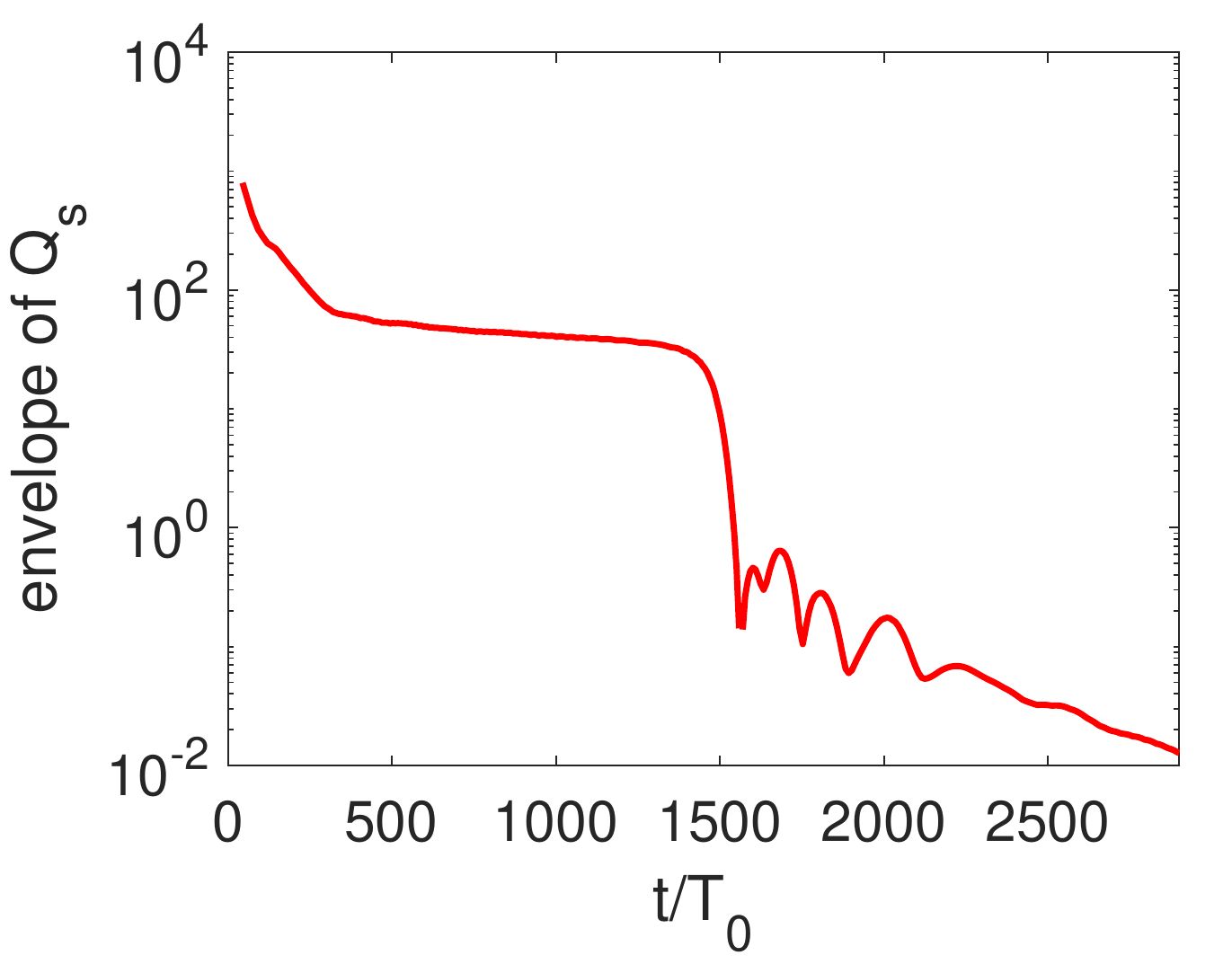}
\caption{\label{fig:EandQ_3D} Evolution of energy $E$ and $E_c$ and envelope of charge $Q_s$ for a dipole CSQ in 3+1D. The initial constituents are elementary Q-balls with frequency $\omega=\pm0.84$ and spacing $d=10$.
}
\end{figure}

\begin{figure}[tbp]
\centering
\includegraphics[width=.45\textwidth,trim=90 260 100 280,clip]{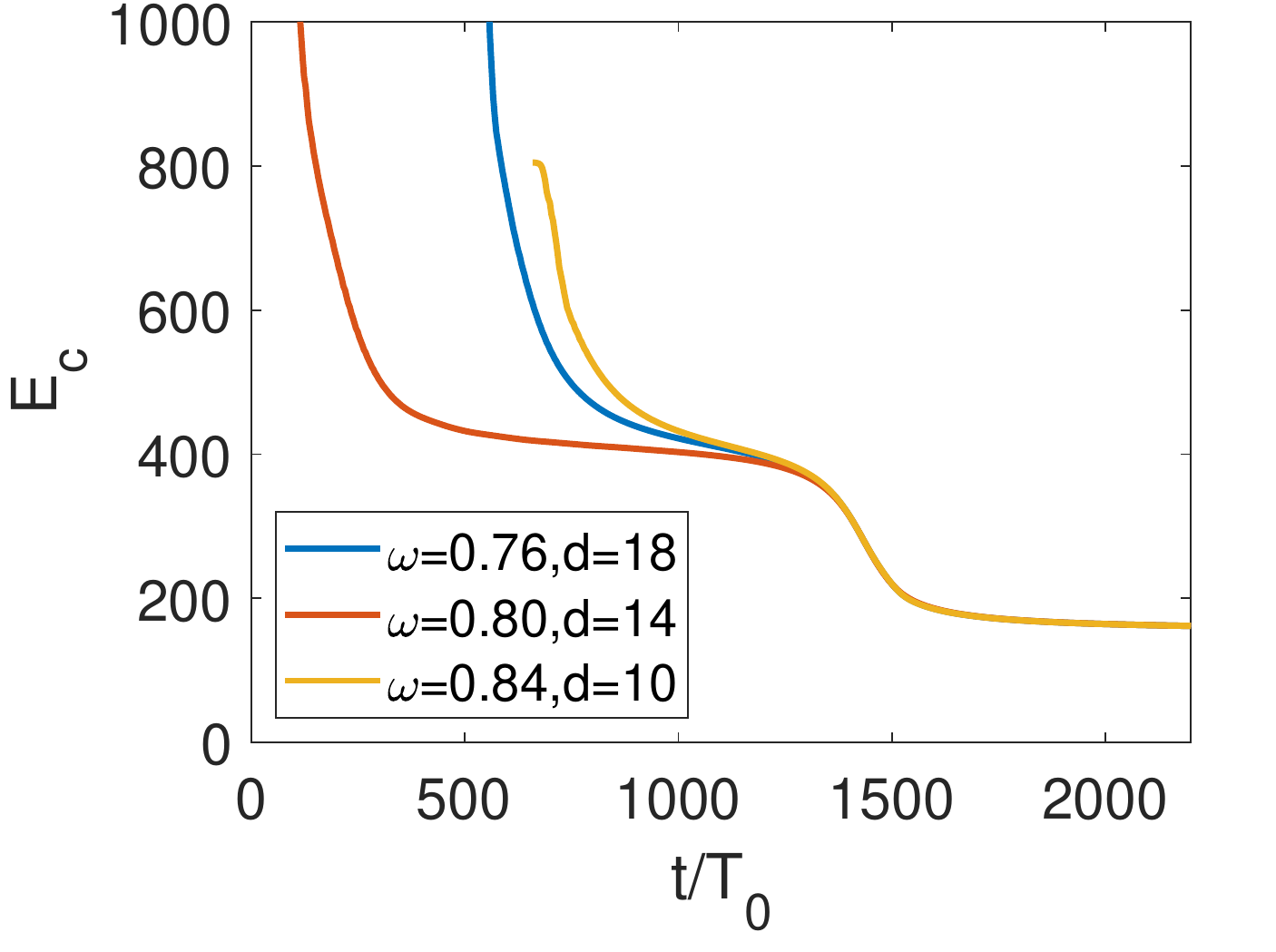}
\hfill
\includegraphics[width=.45\textwidth,trim=90 260 100 280,clip]{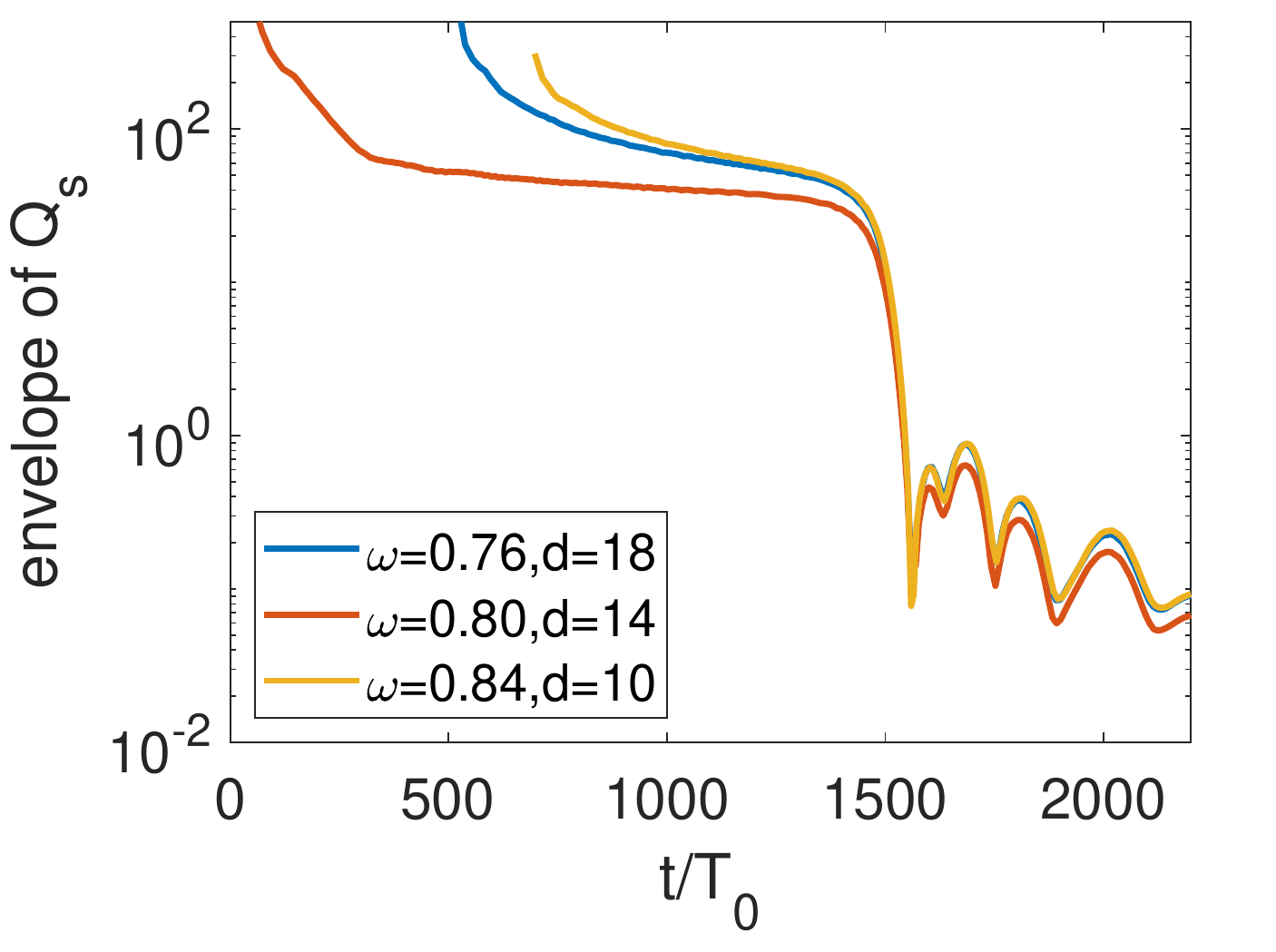}
\caption{\label{fig:attractor_3D}
Attractor behavior of the dipole CSQ in 3+1D. The initial configurations are superimposed elementary Q-balls. Initial configurations close to $\oi=0.80,d=14$ have longer lifetimes.}

\end{figure}

To construct dipole CSQs in 3+1D, we still superimpose an elementary Q-ball and an elementary anti-Q-ball closely to each other so that their nonlinear cores overlap, and then let the configuration relax. In Fig.~\ref{fig:swapCSQ_3D}, we see that the charge swapping patterns of a dipole CSQ in 3+1D is very similar to the 2+1D counterpart (cf.~Fig.~\ref{fig:charge1}), although the charge swapping period in 3+1D is about 1.5 times that in 2+1D for the dipole case. In Fig.~\ref{fig:EandQ_3D}, we see that the time evolution still has four stages: (1) Initial relaxation, (2) First plateau (CSQ stage), (3) Fast decay and (4) Second plateau (oscillon stage). The big difference is that the lifetimes of the CSQ stages are much shorter than those in 2+1D, which is in line with the fact that in 3+1D there are more ``channels'' to decay for the quasi-stable CSQs. However, it is also to do with the potential we are using. For the logarithmic potential, for example, the 3+1D CSQs are also very long lived \cite{HSXZ}, even for the quadrupole or octupole CSQs.  In Fig.~\ref{fig:attractor_3D}, we see that CSQs in 3+1D are also attractor solutions, as expected, and their lifetimes can be significantly prolonged by tuning initial $\oi$ and $d$.

\begin{figure}[tbp]
\centering
\includegraphics[width=.45\textwidth,trim=250 250 250 250,clip]{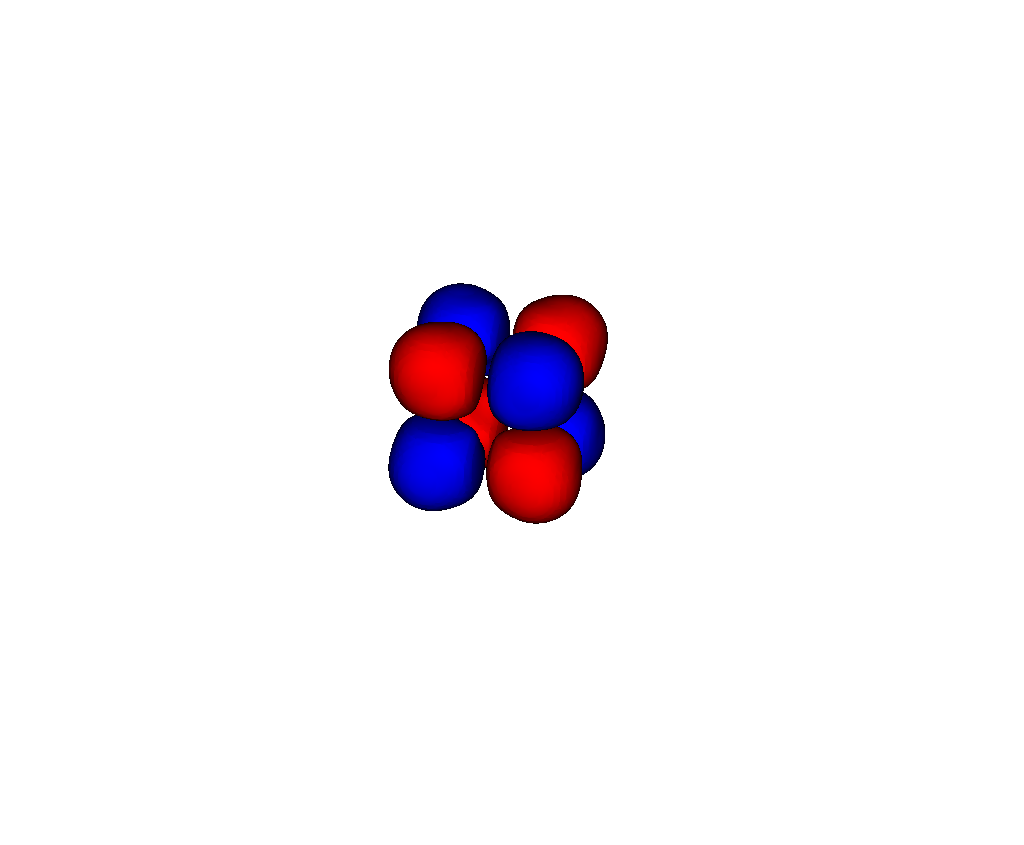}
\caption{\label{fig:octupoleCSQ}
Octupole CSQ in 3+1D.
}
\end{figure}

\begin{figure}[tbp]
\centering
\includegraphics[width=.45\textwidth,trim=90 260 110 260,clip]{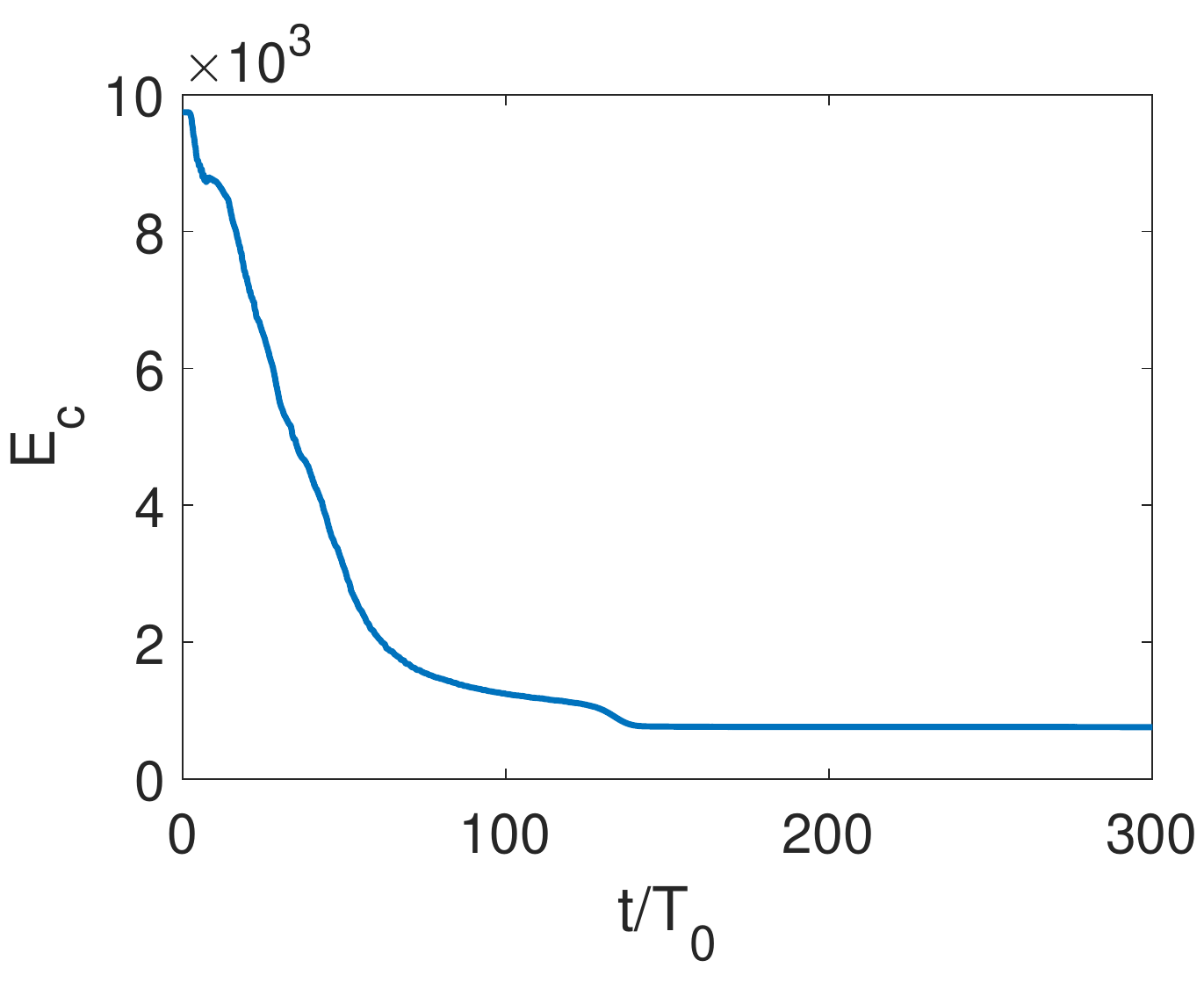}
\hfill
\includegraphics[width=.45\textwidth,trim=90 260 110 260,clip]{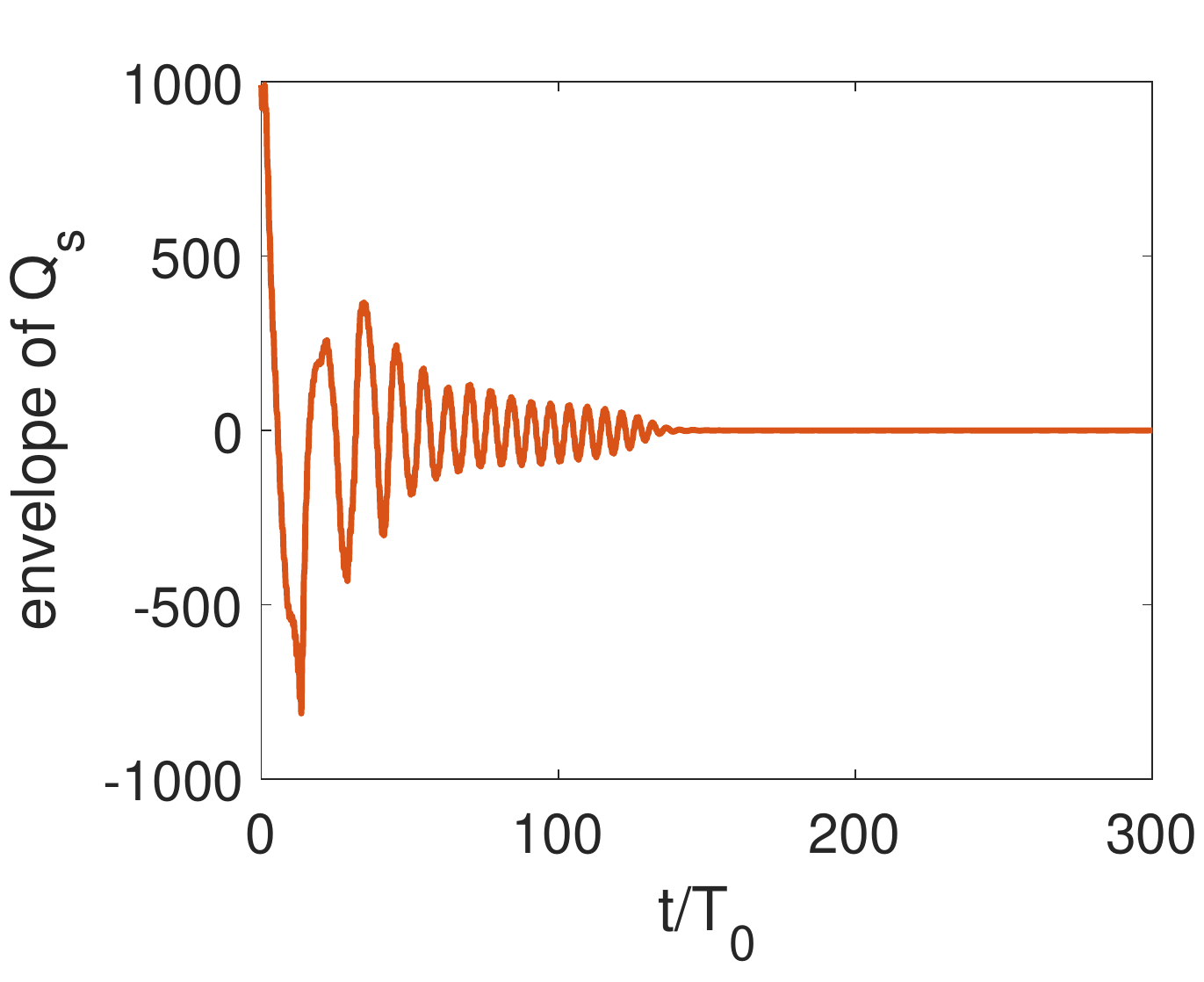}
\caption{\label{fig:octupoleCSQ_EandQ}
Energy (left) and charge (right) evolution of the octupole CSQ of Fig.~\ref{fig:octupoleCSQ}. The parameters in the potential are chosen as $g=0.6, \omega=0.84, d=3$.
}
\end{figure}

There are also higher multipole CSQs in 3+1D, although their lifetimes are even shorter. In 3+1D, while the quadrupole CSQ still has a planar configuration, the octupole CSQ could be arranged as in Fig.~\ref{fig:octupoleCSQ}, unlike the planar octupole configuration in 2+1D. In Fig.~\ref{fig:octupoleCSQ_EandQ}, we find a very short CSQ stage for the octupole CSQ, which is close to the oscillon plateau. If we were to construct the planar octupole CSQ in 3+1D, we would not be able to identity the first CSQ plateau in its evolution. 

\acknowledgments

We would like to thank Zachariah Etienne, Xiao-Xiao Kou and Chi Tian for helpful discussions. 
PMS acknowledges support from STFC grant ST/P000703/1. SYZ acknowledges support from the starting grants from University of Science and Technology of China under grant No.~KY2030000089 and GG2030040375, and is also supported by National Natural Science Foundation of China under grant No.~11947301, 12075233 and 12047502, and supported by the Fundamental Research Funds for the Central Universities under grant No.~WK2030000036. \\

~\\
\appendix

\noindent{\bf APPENDIX}

\section{Other absorbing boundary conditions}
\label{sec:otherABCs}

Here we describe other ABCs we have explored and used to cross-check the validity of some of our results. The Sommerfeld ABC is widely used in numerically relativity \cite{alcubierre2008introduction}, while the Engquist-Majda ABCs have been previously used to study stability of oscillons in spherically symmetry \cite{Salmi:2012ta}. 

\subsection{Sommerfeld's absorbing boundary condition}

The Sommerfeld boundary condition assumes that the outgoing waves have a spherical form
\be
\phi=\phi_0+\f{u(r-vt)}{r^{(d-1)/2}}+\f{h(t)}{r^n}  ,
\ee
where $d$ is the number of the spatial dimensions,  $\phi_0$ is the field value at the spatial infinity, $v$ is the wave velocity at spatial infinity, $u(r-vt)$ is an out-going spherical perturbation and $h(t)$ simulates the non-wavelike behavior that has an $n$-th power law decay.  The Sommerfeld ABC that absorbs such a wave is given by
\be
\pd_t \phi+v \pd_r \phi+\f{(d-1)v}{2r} (\phi-\phi_0)=\f{h'}{r^n}  , \label{newrad}
\ee
and can be re-cast in Cartesian coordinates as
\be
\pd_t \phi+\f{vx^i}{r}\pd_i \phi+\f{(d-1)v}{2r} (\phi-\phi_0)=\f{h'}{r^n} , \label{newrad_cart}
\ee
with $r$ replaced by $(\delta_{ij}x^ix^j)^{1/2}$. Setting $v=1$ and $\phi_0=0$, in the large $r$ limit, \eref{newrad} is just the 1st order Higdon ABC with the addition of a non-wavelike term. Practically, $h'$ is taken as only a function of $t$, and its value at fixed $t$ is evaluated at the outermost layer adjacent to the boundary by \eref{newrad_cart}, which is then used to solve the boundary conditions at the boundary.

\subsection{Engquist-Majda's absorbing boundary conditions}

The Engquist-Majda ABCs \cite{engquist1977absorbing} are designed to absorb plane waves $\phi=\phi_0 e^{i( \oi t+ k_j x^j)}$. At boundary $x^i=a$, such a wave is annihilated by the local operator
\be
\(\f{\pd}{\pd x^i} - ik_i\)\phi\rvert_{x^i=a}= \(\f{\pd}{\pd x^i} - i\,{\rm sign}(k_i) \sqrt{\oi^2-\sum_{j\neq i}{k_j k_j}-m^2}\)\phi\rvert_{x^i=a}=0  ,
\ee
which uses both the Fourier space and real space coordinates. To extract boundary conditions in real space in terms of differential operators, we expand the square root around $\sum_{j\neq i}{k_j k_j}+m^2=0$ by a Taylor series (Alternatively, one may also use a Pade expansion \cite{engquist1977absorbing}.).
The first two orders are given by
\bal
 \(\f{\pd}{\pd x^i} - {\rm sign}(k_i) i\oi \)\phi\rvert_{x^i=a}&=0 ,
 \\
 \[  - i\oi \f{\pd}{\pd x^i} + {\rm sign}(k_i)\( (i\oi)^2-\frac12 \sum_{j\neq i}{(ik_j) (ik_j)} +\frac12 m^2\)  \] \phi\rvert_{x^i=a}&=0  .
\eal
Transforming the Fourier space coordinates to real space coordinates, we have
\bal
\( \f{\pd}{\pd t}\pm\f{\pd}{\pd x^i} \)\phi|_{x^i=a} &= 0 , \label{Eng1}
\\
\[ \f{\pd}{\pd x^i \pd t}\mp \( \f{\pd^2}{\pd t^2}-\sum_{j\neq i}\f{1}{2} \f{\pd^2}{\pd x^j \pd x^j} +\frac12 m^2 \) \] \phi|_{x^i=a} &= 0 , \label{Eng2}
\eal
where the $+$ ($-$) sign is for the right (left) boundary. The 1st order ABC above is just that of the 1st order Higdon ABC with $c_1=1$,  and, as mentioned, the 2nd order ABC above is just that of the 2nd order Higdon ABC with $c_1=c_2=1$, upon using the linearized Klein-Gordon equation of motion. Engquist and Majda also generalized their absorbing boundary conditions for general curvilinear coordinates \cite{engquist1977absorbing}.

\bibliographystyle{JHEP}
\bibliography{refs}

\providecommand{\href}[2]{#2}\begingroup\raggedright\begin{thebibliography}{10}

\bibitem{Friedberg:1976me}
R.~Friedberg, T.~Lee and A.~Sirlin, \emph{{A Class of Scalar-Field Soliton
  Solutions in Three Space Dimensions}},
  \href{http://dx.doi.org/10.1103/PhysRevD.13.2739}{\emph{Phys. Rev. D}
  {\bfseries 13} (1976) 2739--2761}.

\bibitem{Coleman:1985ki}
S.~R. Coleman, \emph{{Q Balls}},
  \href{http://dx.doi.org/10.1016/0550-3213(86)90520-1}{\emph{Nucl. Phys. B}
  {\bfseries 262} (1985) 263}.

\bibitem{Safian:1987pr}
A.~M. Safian, S.~R. Coleman and M.~Axenides, \emph{{SOME NONABELIAN Q BALLS}},
  \href{http://dx.doi.org/10.1016/0550-3213(88)90315-X}{\emph{Nucl. Phys. B}
  {\bfseries 297} (1988) 498--514}.

\bibitem{Kusenko:1997ad}
A.~Kusenko, \emph{{Small Q balls}},
  \href{http://dx.doi.org/10.1016/S0370-2693(97)00582-0}{\emph{Phys. Lett. B}
  {\bfseries 404} (1997) 285},
  [\href{https://arxiv.org/abs/hep-th/9704073}{{\ttfamily hep-th/9704073}}].

\bibitem{Laine:1998rg}
M.~Laine and M.~E. Shaposhnikov, \emph{{Thermodynamics of nontopological
  solitons}},
  \href{http://dx.doi.org/10.1016/S0550-3213(98)00474-X}{\emph{Nucl. Phys. B}
  {\bfseries 532} (1998) 376--404},
  [\href{https://arxiv.org/abs/hep-ph/9804237}{{\ttfamily hep-ph/9804237}}].

\bibitem{Axenides:1999hs}
M.~Axenides, S.~Komineas, L.~Perivolaropoulos and M.~Floratos, \emph{{Dynamics
  of nontopological solitons: Q balls}},
  \href{http://dx.doi.org/10.1103/PhysRevD.61.085006}{\emph{Phys. Rev. D}
  {\bfseries 61} (2000) 085006},
  [\href{https://arxiv.org/abs/hep-ph/9910388}{{\ttfamily hep-ph/9910388}}].

\bibitem{Multamaki:1999an}
T.~Multamaki and I.~Vilja, \emph{{Analytical and numerical properties of Q
  balls}}, \href{http://dx.doi.org/10.1016/S0550-3213(99)00827-5}{\emph{Nucl.
  Phys. B} {\bfseries 574} (2000) 130--152},
  [\href{https://arxiv.org/abs/hep-ph/9908446}{{\ttfamily hep-ph/9908446}}].

\bibitem{Battye:2000qj}
R.~Battye and P.~Sutcliffe, \emph{{Q-ball dynamics}},
  \href{http://dx.doi.org/10.1016/S0550-3213(00)00506-X}{\emph{Nucl. Phys. B}
  {\bfseries 590} (2000) 329--363},
  [\href{https://arxiv.org/abs/hep-th/0003252}{{\ttfamily hep-th/0003252}}].

\bibitem{Paccetti:2001uh}
F.~Paccetti~Correia and M.~Schmidt, \emph{{Q balls: Some analytical results}},
  \href{http://dx.doi.org/10.1007/s100520100710}{\emph{Eur. Phys. J. C}
  {\bfseries 21} (2001) 181--191},
  [\href{https://arxiv.org/abs/hep-th/0103189}{{\ttfamily hep-th/0103189}}].

\bibitem{Volkov:2002aj}
M.~S. Volkov and E.~Wohnert, \emph{{Spinning Q balls}},
  \href{http://dx.doi.org/10.1103/PhysRevD.66.085003}{\emph{Phys. Rev. D}
  {\bfseries 66} (2002) 085003},
  [\href{https://arxiv.org/abs/hep-th/0205157}{{\ttfamily hep-th/0205157}}].

\bibitem{Gleiser:2005iq}
M.~Gleiser and J.~Thorarinson, \emph{{Energy landscape of d-dimensional
  Q-balls}}, \href{http://dx.doi.org/10.1103/PhysRevD.73.065008}{\emph{Phys.
  Rev. D} {\bfseries 73} (2006) 065008},
  [\href{https://arxiv.org/abs/hep-th/0505251}{{\ttfamily hep-th/0505251}}].

\bibitem{Campanelli:2007um}
L.~Campanelli and M.~Ruggieri, \emph{{Supersymmetric Q-balls: A Numerical
  study}}, \href{http://dx.doi.org/10.1103/PhysRevD.77.043504}{\emph{Phys. Rev.
  D} {\bfseries 77} (2008) 043504},
  [\href{https://arxiv.org/abs/0712.3669}{{\ttfamily 0712.3669}}].

\bibitem{Sakai:2007ft}
N.~Sakai and M.~Sasaki, \emph{{Stability of Q-balls and Catastrophe}},
  \href{http://dx.doi.org/10.1143/PTP.119.929}{\emph{Prog. Theor. Phys.}
  {\bfseries 119} (2008) 929--937},
  [\href{https://arxiv.org/abs/0712.1450}{{\ttfamily 0712.1450}}].

\bibitem{Bowcock:2008dn}
P.~Bowcock, D.~Foster and P.~Sutcliffe, \emph{{Q-balls, Integrability and
  Duality}}, \href{http://dx.doi.org/10.1088/1751-8113/42/8/085403}{\emph{J.
  Phys. A} {\bfseries 42} (2009) 085403},
  [\href{https://arxiv.org/abs/0809.3895}{{\ttfamily 0809.3895}}].

\bibitem{Tsumagari:2008bv}
M.~I. Tsumagari, E.~J. Copeland and P.~M. Saffin, \emph{{Some stationary
  properties of a Q-ball in arbitrary space dimensions}},
  \href{http://dx.doi.org/10.1103/PhysRevD.78.065021}{\emph{Phys. Rev. D}
  {\bfseries 78} (2008) 065021},
  [\href{https://arxiv.org/abs/0805.3233}{{\ttfamily 0805.3233}}].

\bibitem{Copeland:2009as}
E.~J. Copeland and M.~I. Tsumagari, \emph{{Q-balls in flat potentials}},
  \href{http://dx.doi.org/10.1103/PhysRevD.80.025016}{\emph{Phys. Rev. D}
  {\bfseries 80} (2009) 025016},
  [\href{https://arxiv.org/abs/0905.0125}{{\ttfamily 0905.0125}}].

\bibitem{Mai:2012yc}
M.~Mai and P.~Schweitzer, \emph{{Energy momentum tensor, stability, and the
  D-term of Q-balls}},
  \href{http://dx.doi.org/10.1103/PhysRevD.86.076001}{\emph{Phys. Rev. D}
  {\bfseries 86} (2012) 076001},
  [\href{https://arxiv.org/abs/1206.2632}{{\ttfamily 1206.2632}}].

\bibitem{Gulamov:2015fya}
I.~Gulamov, E.~Nugaev, A.~Panin and M.~Smolyakov, \emph{{Some properties of
  U(1) gauged Q-balls}},
  \href{http://dx.doi.org/10.1103/PhysRevD.92.045011}{\emph{Phys. Rev. D}
  {\bfseries 92} (2015) 045011},
  [\href{https://arxiv.org/abs/1506.05786}{{\ttfamily 1506.05786}}].

\bibitem{Bazeia:2015gkq}
D.~Bazeia, M.~Marques and R.~Menezes, \emph{{Exact solutions, energy and charge
  of stable Q-balls}},
  \href{http://dx.doi.org/10.1140/epjc/s10052-016-4059-z}{\emph{Eur. Phys. J.
  C} {\bfseries 76} (2016) 241},
  [\href{https://arxiv.org/abs/1512.04279}{{\ttfamily 1512.04279}}].

\bibitem{Bazeia:2016wco}
D.~Bazeia, L.~Losano, M.~Marques, R.~Menezes and R.~da~Rocha, \emph{{Compact
  Q-balls}},
  \href{http://dx.doi.org/10.1016/j.physletb.2016.04.060}{\emph{Phys. Lett. B}
  {\bfseries 758} (2016) 146--151},
  [\href{https://arxiv.org/abs/1604.08871}{{\ttfamily 1604.08871}}].

\bibitem{Levkov:2017paj}
D.~Levkov, E.~Nugaev and A.~Popescu, \emph{{The fate of small classically
  stable Q-balls}},
  \href{http://dx.doi.org/10.1007/JHEP12(2017)131}{\emph{JHEP} {\bfseries 12}
  (2017) 131}, [\href{https://arxiv.org/abs/1711.05279}{{\ttfamily
  1711.05279}}].

\bibitem{Smolyakov:2017axd}
M.~N. Smolyakov, \emph{{Perturbations against a Q-ball: Charge, energy, and
  additivity property}},
  \href{http://dx.doi.org/10.1103/PhysRevD.97.045011}{\emph{Phys. Rev. D}
  {\bfseries 97} (2018) 045011},
  [\href{https://arxiv.org/abs/1711.05730}{{\ttfamily 1711.05730}}].

\bibitem{Loiko:2018mhb}
V.~Loiko, I.~Perapechka and Y.~Shnir, \emph{{Q-balls without a potential}},
  \href{http://dx.doi.org/10.1103/PhysRevD.98.045018}{\emph{Phys. Rev. D}
  {\bfseries 98} (2018) 045018},
  [\href{https://arxiv.org/abs/1805.11929}{{\ttfamily 1805.11929}}].

\bibitem{Hasegawa:2019bbo}
F.~Hasegawa, J.-P. Hong and M.~Suzuki, \emph{{More about Q-ball with elliptical
  orbit}}, \href{http://dx.doi.org/10.1016/j.physletb.2019.135001}{\emph{Phys.
  Lett. B} {\bfseries 798} (2019) 135001},
  [\href{https://arxiv.org/abs/1903.07281}{{\ttfamily 1903.07281}}].

\bibitem{Frieman:1988ut}
J.~A. Frieman, G.~Gelmini, M.~Gleiser and E.~W. Kolb, \emph{{Solitogenesis:
  Primordial Origin of Nontopological Solitons}},
  \href{http://dx.doi.org/10.1103/PhysRevLett.60.2101}{\emph{Phys. Rev. Lett.}
  {\bfseries 60} (1988) 2101}.

\bibitem{Enqvist:1997si}
K.~Enqvist and J.~McDonald, \emph{{Q balls and baryogenesis in the MSSM}},
  \href{http://dx.doi.org/10.1016/S0370-2693(98)00271-8}{\emph{Phys. Lett. B}
  {\bfseries 425} (1998) 309--321},
  [\href{https://arxiv.org/abs/hep-ph/9711514}{{\ttfamily hep-ph/9711514}}].

\bibitem{Kusenko:1997zq}
A.~Kusenko, \emph{{Solitons in the supersymmetric extensions of the standard
  model}}, \href{http://dx.doi.org/10.1016/S0370-2693(97)00584-4}{\emph{Phys.
  Lett. B} {\bfseries 405} (1997) 108},
  [\href{https://arxiv.org/abs/hep-ph/9704273}{{\ttfamily hep-ph/9704273}}].

\bibitem{Kasuya:1999wu}
S.~Kasuya and M.~Kawasaki, \emph{{Q ball formation through Affleck-Dine
  mechanism}}, \href{http://dx.doi.org/10.1103/PhysRevD.61.041301}{\emph{Phys.
  Rev. D} {\bfseries 61} (2000) 041301},
  [\href{https://arxiv.org/abs/hep-ph/9909509}{{\ttfamily hep-ph/9909509}}].

\bibitem{Enqvist:2000cq}
K.~Enqvist, A.~Jokinen, T.~Multamaki and I.~Vilja, \emph{{Numerical simulations
  of fragmentation of the Affleck-Dine condensate}},
  \href{http://dx.doi.org/10.1103/PhysRevD.63.083501}{\emph{Phys. Rev. D}
  {\bfseries 63} (2001) 083501},
  [\href{https://arxiv.org/abs/hep-ph/0011134}{{\ttfamily hep-ph/0011134}}].

\bibitem{Fujii:2001xp}
M.~Fujii and K.~Hamaguchi, \emph{{Higgsino and wino dark matter from Q ball
  decay}}, \href{http://dx.doi.org/10.1016/S0370-2693(01)01412-5}{\emph{Phys.
  Lett. B} {\bfseries 525} (2002) 143--149},
  [\href{https://arxiv.org/abs/hep-ph/0110072}{{\ttfamily hep-ph/0110072}}].

\bibitem{Postma:2001ea}
M.~Postma, \emph{{Solitosynthesis of Q balls}},
  \href{http://dx.doi.org/10.1103/PhysRevD.65.085035}{\emph{Phys. Rev. D}
  {\bfseries 65} (2002) 085035},
  [\href{https://arxiv.org/abs/hep-ph/0110199}{{\ttfamily hep-ph/0110199}}].

\bibitem{Fujii:2002kr}
M.~Fujii and K.~Hamaguchi, \emph{{Nonthermal dark matter via Affleck-Dine
  baryogenesis and its detection possibility}},
  \href{http://dx.doi.org/10.1103/PhysRevD.66.083501}{\emph{Phys. Rev. D}
  {\bfseries 66} (2002) 083501},
  [\href{https://arxiv.org/abs/hep-ph/0205044}{{\ttfamily hep-ph/0205044}}].

\bibitem{Kawasaki:2002hq}
M.~Kawasaki, F.~Takahashi and M.~Yamaguchi, \emph{{Large lepton asymmetry from
  Q balls}}, \href{http://dx.doi.org/10.1103/PhysRevD.66.043516}{\emph{Phys.
  Rev. D} {\bfseries 66} (2002) 043516},
  [\href{https://arxiv.org/abs/hep-ph/0205101}{{\ttfamily hep-ph/0205101}}].

\bibitem{Allahverdi:2002vy}
R.~Allahverdi, A.~Mazumdar and A.~Ozpineci, \emph{{Q ball formation in the wake
  of Hubble induced radiative corrections}},
  \href{http://dx.doi.org/10.1103/PhysRevD.65.125003}{\emph{Phys. Rev. D}
  {\bfseries 65} (2002) 125003},
  [\href{https://arxiv.org/abs/hep-ph/0203062}{{\ttfamily hep-ph/0203062}}].

\bibitem{Palti:2004is}
E.~Palti, P.~Saffin and E.~Copeland, \emph{{Dynamics of Q-balls in an expanding
  universe}}, \href{http://dx.doi.org/10.1103/PhysRevD.70.083520}{\emph{Phys.
  Rev. D} {\bfseries 70} (2004) 083520},
  [\href{https://arxiv.org/abs/hep-th/0405081}{{\ttfamily hep-th/0405081}}].

\bibitem{Berkooz:2005rn}
M.~Berkooz, D.~J. Chung and T.~Volansky, \emph{{High density preheating effects
  on Q-ball decays and MSSM inflation}},
  \href{http://dx.doi.org/10.1103/PhysRevLett.96.031303}{\emph{Phys. Rev.
  Lett.} {\bfseries 96} (2006) 031303},
  [\href{https://arxiv.org/abs/hep-ph/0510186}{{\ttfamily hep-ph/0510186}}].

\bibitem{Pearce:2012jp}
L.~Pearce, \emph{{Solitosynthesis induced phase transitions}},
  \href{http://dx.doi.org/10.1103/PhysRevD.85.125022}{\emph{Phys. Rev. D}
  {\bfseries 85} (2012) 125022},
  [\href{https://arxiv.org/abs/1202.0873}{{\ttfamily 1202.0873}}].

\bibitem{Krylov:2013qe}
E.~Krylov, A.~Levin and V.~Rubakov, \emph{{Cosmological phase transition,
  baryon asymmetry and dark matter Q-balls}},
  \href{http://dx.doi.org/10.1103/PhysRevD.87.083528}{\emph{Phys. Rev. D}
  {\bfseries 87} (2013) 083528},
  [\href{https://arxiv.org/abs/1301.0354}{{\ttfamily 1301.0354}}].

\bibitem{Zhou:2015yfa}
S.-Y. Zhou, \emph{{Gravitational waves from Affleck-Dine condensate
  fragmentation}},
  \href{http://dx.doi.org/10.1088/1475-7516/2015/06/033}{\emph{JCAP} {\bfseries
  06} (2015) 033}, [\href{https://arxiv.org/abs/1501.01217}{{\ttfamily
  1501.01217}}].

\bibitem{Hasegawa:2018yuy}
F.~Hasegawa and M.~Kawasaki, \emph{{Primordial Black Holes from Affleck-Dine
  Mechanism}},
  \href{http://dx.doi.org/10.1088/1475-7516/2019/01/027}{\emph{JCAP} {\bfseries
  01} (2019) 027}, [\href{https://arxiv.org/abs/1807.00463}{{\ttfamily
  1807.00463}}].

\bibitem{Cotner:2019ykd}
E.~Cotner, A.~Kusenko, M.~Sasaki and V.~Takhistov, \emph{{Analytic Description
  of Primordial Black Hole Formation from Scalar Field Fragmentation}},
  \href{http://dx.doi.org/10.1088/1475-7516/2019/10/077}{\emph{JCAP} {\bfseries
  10} (2019) 077}, [\href{https://arxiv.org/abs/1907.10613}{{\ttfamily
  1907.10613}}].

\bibitem{Kusenko:1997si}
A.~Kusenko and M.~E. Shaposhnikov, \emph{{Supersymmetric Q balls as dark
  matter}}, \href{http://dx.doi.org/10.1016/S0370-2693(97)01375-0}{\emph{Phys.
  Lett. B} {\bfseries 418} (1998) 46--54},
  [\href{https://arxiv.org/abs/hep-ph/9709492}{{\ttfamily hep-ph/9709492}}].

\bibitem{Enqvist:1998ds}
K.~Enqvist and J.~McDonald, \emph{{D term inflation and B ball baryogenesis}},
  \href{http://dx.doi.org/10.1103/PhysRevLett.81.3071}{\emph{Phys. Rev. Lett.}
  {\bfseries 81} (1998) 3071--3074},
  [\href{https://arxiv.org/abs/hep-ph/9806213}{{\ttfamily hep-ph/9806213}}].

\bibitem{Kasuya:2000sc}
S.~Kasuya and M.~Kawasaki, \emph{{A New type of stable Q balls in the gauge
  mediated SUSY breaking}},
  \href{http://dx.doi.org/10.1103/PhysRevLett.85.2677}{\emph{Phys. Rev. Lett.}
  {\bfseries 85} (2000) 2677--2680},
  [\href{https://arxiv.org/abs/hep-ph/0006128}{{\ttfamily hep-ph/0006128}}].

\bibitem{Banerjee:2000mb}
R.~Banerjee and K.~Jedamzik, \emph{{On B-ball dark matter and baryogenesis}},
  \href{http://dx.doi.org/10.1016/S0370-2693(00)00688-2}{\emph{Phys. Lett. B}
  {\bfseries 484} (2000) 278--282},
  [\href{https://arxiv.org/abs/hep-ph/0005031}{{\ttfamily hep-ph/0005031}}].

\bibitem{Kusenko:2004yw}
A.~Kusenko, L.~Loveridge and M.~Shaposhnikov, \emph{{Supersymmetric dark matter
  Q-balls and their interactions in matter}},
  \href{http://dx.doi.org/10.1103/PhysRevD.72.025015}{\emph{Phys. Rev. D}
  {\bfseries 72} (2005) 025015},
  [\href{https://arxiv.org/abs/hep-ph/0405044}{{\ttfamily hep-ph/0405044}}].

\bibitem{Roszkowski:2006kw}
L.~Roszkowski and O.~Seto, \emph{{Axino dark matter from Q-balls in
  Affleck-Dine baryogenesis and the Omega(b) - Omega(DM) coincidence problem}},
  \href{http://dx.doi.org/10.1103/PhysRevLett.98.161304}{\emph{Phys. Rev.
  Lett.} {\bfseries 98} (2007) 161304},
  [\href{https://arxiv.org/abs/hep-ph/0608013}{{\ttfamily hep-ph/0608013}}].

\bibitem{Kasuya:2011ix}
S.~Kasuya and M.~Kawasaki, \emph{{Gravitino dark matter and baryon asymmetry
  from Q-ball decay in gauge mediation}},
  \href{http://dx.doi.org/10.1103/PhysRevD.84.123528}{\emph{Phys. Rev. D}
  {\bfseries 84} (2011) 123528},
  [\href{https://arxiv.org/abs/1107.0403}{{\ttfamily 1107.0403}}].

\bibitem{Kasuya:2014ofa}
S.~Kasuya and M.~Kawasaki, \emph{{Baryogenesis from the gauge-mediation type
  Q-ball and the new type of Q-ball as the dark matter}},
  \href{http://dx.doi.org/10.1103/PhysRevD.89.103534}{\emph{Phys. Rev. D}
  {\bfseries 89} (2014) 103534},
  [\href{https://arxiv.org/abs/1402.4546}{{\ttfamily 1402.4546}}].

\bibitem{Kawasaki:2019ywz}
M.~Kawasaki and H.~Nakatsuka, \emph{{Q-ball decay through A-term in the
  gauge-mediated SUSY breaking scenario}},
  \href{http://dx.doi.org/10.1088/1475-7516/2020/04/017}{\emph{JCAP} {\bfseries
  04} (2020) 017}, [\href{https://arxiv.org/abs/1912.06993}{{\ttfamily
  1912.06993}}].

\bibitem{Enqvist:2003zb}
K.~Enqvist and M.~Laine, \emph{{Q-ball dynamics from atomic Bose-Einstein
  condensates}},
  \href{http://dx.doi.org/10.1088/1475-7516/2003/08/003}{\emph{JCAP} {\bfseries
  08} (2003) 003}, [\href{https://arxiv.org/abs/cond-mat/0304355}{{\ttfamily
  cond-mat/0304355}}].

\bibitem{Bunkov:2007fe}
Y.~Bunkov and G.~Volovik, \emph{{Magnons condensation into Q-ball in He-3 -
  B}}, \href{http://dx.doi.org/10.1103/PhysRevLett.98.265302}{\emph{Phys. Rev.
  Lett.} {\bfseries 98} (2007) 265302},
  [\href{https://arxiv.org/abs/cond-mat/0703183}{{\ttfamily
  cond-mat/0703183}}].

\bibitem{Copeland:2014qra}
E.~J. Copeland, P.~M. Saffin and S.-Y. Zhou, \emph{{Charge-Swapping Q-balls}},
  \href{http://dx.doi.org/10.1103/PhysRevLett.113.231603}{\emph{Phys. Rev.
  Lett.} {\bfseries 113} (2014) 231603},
  [\href{https://arxiv.org/abs/1409.3232}{{\ttfamily 1409.3232}}].

\bibitem{Bogolyubsky:1976yu}
I.~Bogolyubsky and V.~Makhankov, \emph{{Lifetime of Pulsating Solitons in Some
  Classical Models}}, {\emph{Pisma Zh. Eksp. Teor. Fiz.} {\bfseries 24} (1976)
  15--18}.

\bibitem{Copeland:1995fq}
E.~J. Copeland, M.~Gleiser and H.-R. Muller, \emph{{Oscillons: Resonant
  configurations during bubble collapse}},
  \href{http://dx.doi.org/10.1103/PhysRevD.52.1920}{\emph{Phys. Rev. D}
  {\bfseries 52} (1995) 1920--1933},
  [\href{https://arxiv.org/abs/hep-ph/9503217}{{\ttfamily hep-ph/9503217}}].

\bibitem{Honda:2001xg}
E.~P. Honda and M.~W. Choptuik, \emph{{Fine structure of oscillons in the
  spherically symmetric phi**4 Klein-Gordon model}},
  \href{http://dx.doi.org/10.1103/PhysRevD.65.084037}{\emph{Phys. Rev. D}
  {\bfseries 65} (2002) 084037},
  [\href{https://arxiv.org/abs/hep-ph/0110065}{{\ttfamily hep-ph/0110065}}].

\bibitem{Adib:2002ff}
A.~B. Adib, M.~Gleiser and C.~A. Almeida, \emph{{Long lived oscillons from
  asymmetric bubbles: Existence and stability}},
  \href{http://dx.doi.org/10.1103/PhysRevD.66.085011}{\emph{Phys. Rev. D}
  {\bfseries 66} (2002) 085011},
  [\href{https://arxiv.org/abs/hep-th/0203072}{{\ttfamily hep-th/0203072}}].

\bibitem{Fodor:2006zs}
G.~Fodor, P.~Forgacs, P.~Grandclement and I.~Racz, \emph{{Oscillons and
  Quasi-breathers in the phi**4 Klein-Gordon model}},
  \href{http://dx.doi.org/10.1103/PhysRevD.74.124003}{\emph{Phys. Rev. D}
  {\bfseries 74} (2006) 124003},
  [\href{https://arxiv.org/abs/hep-th/0609023}{{\ttfamily hep-th/0609023}}].

\bibitem{Saffin:2006yk}
P.~M. Saffin and A.~Tranberg, \emph{{Oscillons and quasi-breathers in D+1
  dimensions}},
  \href{http://dx.doi.org/10.1088/1126-6708/2007/01/030}{\emph{JHEP} {\bfseries
  01} (2007) 030}, [\href{https://arxiv.org/abs/hep-th/0610191}{{\ttfamily
  hep-th/0610191}}].

\bibitem{Farhi:2007wj}
E.~Farhi, N.~Graham, A.~H. Guth, N.~Iqbal, R.~Rosales and N.~Stamatopoulos,
  \emph{{Emergence of Oscillons in an Expanding Background}},
  \href{http://dx.doi.org/10.1103/PhysRevD.77.085019}{\emph{Phys. Rev. D}
  {\bfseries 77} (2008) 085019},
  [\href{https://arxiv.org/abs/0712.3034}{{\ttfamily 0712.3034}}].

\bibitem{Gleiser:2008ty}
M.~Gleiser and D.~Sicilia, \emph{{Analytical Characterization of Oscillon
  Energy and Lifetime}},
  \href{http://dx.doi.org/10.1103/PhysRevLett.101.011602}{\emph{Phys. Rev.
  Lett.} {\bfseries 101} (2008) 011602},
  [\href{https://arxiv.org/abs/0804.0791}{{\ttfamily 0804.0791}}].

\bibitem{Fodor:2009kf}
G.~Fodor, P.~Forgacs, Z.~Horvath and M.~Mezei, \emph{{Radiation of scalar
  oscillons in 2 and 3 dimensions}},
  \href{http://dx.doi.org/10.1016/j.physletb.2009.03.054}{\emph{Phys. Lett. B}
  {\bfseries 674} (2009) 319--324},
  [\href{https://arxiv.org/abs/0903.0953}{{\ttfamily 0903.0953}}].

\bibitem{Gleiser:2009ys}
M.~Gleiser and D.~Sicilia, \emph{{A General Theory of Oscillon Dynamics}},
  \href{http://dx.doi.org/10.1103/PhysRevD.80.125037}{\emph{Phys. Rev. D}
  {\bfseries 80} (2009) 125037},
  [\href{https://arxiv.org/abs/0910.5922}{{\ttfamily 0910.5922}}].

\bibitem{Amin:2010dc}
M.~A. Amin, R.~Easther and H.~Finkel, \emph{{Inflaton Fragmentation and
  Oscillon Formation in Three Dimensions}},
  \href{http://dx.doi.org/10.1088/1475-7516/2010/12/001}{\emph{JCAP} {\bfseries
  12} (2010) 001}, [\href{https://arxiv.org/abs/1009.2505}{{\ttfamily
  1009.2505}}].

\bibitem{Amin:2011hj}
M.~A. Amin, R.~Easther, H.~Finkel, R.~Flauger and M.~P. Hertzberg,
  \emph{{Oscillons After Inflation}},
  \href{http://dx.doi.org/10.1103/PhysRevLett.108.241302}{\emph{Phys. Rev.
  Lett.} {\bfseries 108} (2012) 241302},
  [\href{https://arxiv.org/abs/1106.3335}{{\ttfamily 1106.3335}}].

\bibitem{Salmi:2012ta}
P.~Salmi and M.~Hindmarsh, \emph{{Radiation and Relaxation of Oscillons}},
  \href{http://dx.doi.org/10.1103/PhysRevD.85.085033}{\emph{Phys. Rev. D}
  {\bfseries 85} (2012) 085033},
  [\href{https://arxiv.org/abs/1201.1934}{{\ttfamily 1201.1934}}].

\bibitem{Amin:2013ika}
M.~A. Amin, \emph{{K-oscillons: Oscillons with noncanonical kinetic terms}},
  \href{http://dx.doi.org/10.1103/PhysRevD.87.123505}{\emph{Phys. Rev. D}
  {\bfseries 87} (2013) 123505},
  [\href{https://arxiv.org/abs/1303.1102}{{\ttfamily 1303.1102}}].

\bibitem{Zhou:2013tsa}
S.-Y. Zhou, E.~J. Copeland, R.~Easther, H.~Finkel, Z.-G. Mou and P.~M. Saffin,
  \emph{{Gravitational Waves from Oscillon Preheating}},
  \href{http://dx.doi.org/10.1007/JHEP10(2013)026}{\emph{JHEP} {\bfseries 10}
  (2013) 026}, [\href{https://arxiv.org/abs/1304.6094}{{\ttfamily 1304.6094}}].

\bibitem{Mukaida:2016hwd}
K.~Mukaida, M.~Takimoto and M.~Yamada, \emph{{On Longevity of
  I-ball/Oscillon}},
  \href{http://dx.doi.org/10.1007/JHEP03(2017)122}{\emph{JHEP} {\bfseries 03}
  (2017) 122}, [\href{https://arxiv.org/abs/1612.07750}{{\ttfamily
  1612.07750}}].

\bibitem{Lozanov:2017hjm}
K.~D. Lozanov and M.~A. Amin, \emph{{Self-resonance after inflation: oscillons,
  transients and radiation domination}},
  \href{http://dx.doi.org/10.1103/PhysRevD.97.023533}{\emph{Phys. Rev. D}
  {\bfseries 97} (2018) 023533},
  [\href{https://arxiv.org/abs/1710.06851}{{\ttfamily 1710.06851}}].

\bibitem{Gleiser:2018kbq}
M.~Gleiser, M.~Stephens and D.~Sowinski, \emph{{Configurational entropy as a
  lifetime predictor and pattern discriminator for oscillons}},
  \href{http://dx.doi.org/10.1103/PhysRevD.97.096007}{\emph{Phys. Rev. D}
  {\bfseries 97} (2018) 096007},
  [\href{https://arxiv.org/abs/1803.08550}{{\ttfamily 1803.08550}}].

\bibitem{Amin:2018xfe}
M.~A. Amin, J.~Braden, E.~J. Copeland, J.~T. Giblin, C.~Solorio, Z.~J. Weiner
  et~al., \emph{{Gravitational waves from asymmetric oscillon dynamics?}},
  \href{http://dx.doi.org/10.1103/PhysRevD.98.024040}{\emph{Phys. Rev. D}
  {\bfseries 98} (2018) 024040},
  [\href{https://arxiv.org/abs/1803.08047}{{\ttfamily 1803.08047}}].

\bibitem{Ibe:2019vyo}
M.~Ibe, M.~Kawasaki, W.~Nakano and E.~Sonomoto, \emph{{Decay of I-ball/Oscillon
  in Classical Field Theory}},
  \href{http://dx.doi.org/10.1007/JHEP04(2019)030}{\emph{JHEP} {\bfseries 04}
  (2019) 030}, [\href{https://arxiv.org/abs/1901.06130}{{\ttfamily
  1901.06130}}].

\bibitem{Kou:2019bbc}
X.-X. Kou, C.~Tian and S.-Y. Zhou, \emph{{Oscillon Preheating in Full General
  Relativity}},  \href{https://arxiv.org/abs/1912.09658}{{\ttfamily
  1912.09658}}.

\bibitem{Zhang:2020bec}
H.-Y. Zhang, M.~A. Amin, E.~J. Copeland, P.~M. Saffin and K.~D. Lozanov,
  \emph{{Classical Decay Rates of Oscillons}},
  \href{http://dx.doi.org/10.1088/1475-7516/2020/07/055}{\emph{JCAP} {\bfseries
  07} (2020) 055}, [\href{https://arxiv.org/abs/2004.01202}{{\ttfamily
  2004.01202}}].

\bibitem{alcubierre2008introduction}
M.~Alcubierre, \emph{Introduction to 3+ 1 numerical relativity}, vol.~140.
\newblock Oxford University Press, 2008.

\bibitem{engquist1977absorbing}
B.~Engquist and A.~Majda, \emph{Absorbing boundary conditions for numerical
  simulation of waves}, {\emph{Proceedings of the National Academy of Sciences}
  {\bfseries 74} (1977) 1765--1766}.

\bibitem{higdon1994radiation}
R.~L. Higdon, \emph{Radiation boundary conditions for dispersive waves},
  {\emph{SIAM Journal on Numerical Analysis} {\bfseries 31} (1994) 64--100}.

\bibitem{higdon1986absorbing}
R.~L. Higdon, \emph{Absorbing boundary conditions for difference approximations
  to the multidimensional wave equation}, {\emph{Mathematics of computation}
  {\bfseries 47} (1986) 437--459}.

\bibitem{HSXZ}
S.-Y. Hou, P.~M. Saffin, Q.-X. Xie and S.-Y. Zhou, \emph{{in preparation}}, .

\bibitem{David:2015eya}
D.~Daverio, M.~Hindmarsh and N.~Bevis, \emph{{Latfield2: A c++ library for
  classical lattice field theory}},
  \href{https://arxiv.org/abs/1508.05610}{{\ttfamily 1508.05610}}.

\end{thebibliography}\endgroup

\end{document}